\documentclass[aps,prb,reprint,superscriptaddress,floatfix]{revtex4-2}
\usepackage{lineno}
\usepackage{graphicx}
\usepackage{amsmath}
\usepackage{amssymb}
\usepackage{algorithm}
\usepackage{algpseudocode}
\usepackage[version=4]{mhchem}
\usepackage{tabularx}
\usepackage{booktabs}
\usepackage{multirow}

\algblockdefx[Try]{Try}{EndTry}{\textbf{try}}{}
\algblockdefx[Catch]{Catch}{EndTry}[1]{\textbf{catch} #1}{\textbf{end try}}

\begin{document}
\title[Short title]{Better accuracy with fewer qubits: Single-particle basis set optimization for electronic structure calculations on quantum computers} 

\author{Subimal Deb} 
\affiliation{Centre for Quantum Engineering, Research and Education, TCG CREST, Kolkata 700091, India}
\email{subimal.deb@gmail.com}
\author{V. S. Prasannaa}
\affiliation{Centre for Quantum Engineering, Research and Education, TCG CREST, Kolkata 700091, India}
\affiliation{Academy of Scientific and Innovative Research (AcSIR), Ghaziabad- 201002, India} 
\email{srinivasaprasannaa@gmail.com}

\date{\today}

\begin{abstract} 
In spite of recent advances, quantum computers are expected to be sufficiently noisy in the coming few years to the extent of limiting atomic and molecular/quantum chemical calculations to a relatively small number of orbitals. However, even with reasonable quality single particle basis sets, small active spaces with limited orbitals (fewer qubits) can result in a significant fraction of correlation energy being lost, motivating the design of moderate quality qubit-efficient single particle basis sets with as few orbitals as possible for quantum algorithms. We begin by reoptimizing the existing minimal basis sets using a genetic algorithm-inspired approach in conjunction with aggressive refinement strategies, and generate modified minimal basis sets (which we call MSTO-$k$G basis, for cardinal number $k=2$ through $11$, that is, $k$ Gaussian primitives per contraction) for atoms from \ce{H} through \ce{F} (excluding \ce{He}). The minimization is performed at the configuration interaction singles and doubles level of theory. Our open-source codes are available at \url{https://github.com/subimal/MSTO-kG}. The MSTO basis sets consistently improve upon the FCI ground state energies obtained using the conventional STO-$k$G bases. For H, Li, and Be, the lowest FCI energies obtained among our MSTO-$k$G bases lie below the corresponding 6-31G values. For Li, MSTO-10G also lies 3.846 mHa below the cc-pVQZ value. For the remaining atoms, the lowest FCI energies obtained with our basis sets are higher than those from the 6-31G basis. These improvements over the traditional minimal bases are achieved without increasing the number of contracted orbitals and hence the qubit counts. We also investigate the transferability of our atom-optimized basis sets to molecular systems through potential energy curve calculations, obtaining favourable results for systems such as Li\(_2\) and LiH, while observing limitations for other systems, notably H\(_2\) and C\(_2\) near their equilibrium geometries. Finally, taking the Li atom as a best-case example, we compare illustrative resource estimates for the variational quantum eigensolver with the unitary coupled cluster singles and doubles ansatz, the quantum phase estimation algorithm applied to complete active space configuration interaction, and the Harrow-Hassidim-Lloyd algorithm applied to the linearized coupled cluster singles and doubles method. We find that the MSTO bases yield lower classically computed FCI reference energies than the 6-31G and cc-pVDZ bases, while requiring fewer logical qubits and substantially fewer CX gates. Our analysis also illustrates the dependence of logical T counts on the choice of basis set. Overall, our work demonstrates the potential of reoptimized minimal basis sets for improving the accuracy of quantum chemical calculations within a restricted orbital count and quantum resource budget. 
\end{abstract} 

\maketitle

\tableofcontents 

\section{Introduction} 

Atomic and molecular physics/quantum chemistry calculations on quantum computers has been a fast-growing field, owing to the promise of speed-up from quantum algorithms such as quantum phase estimation (QPE) \cite{kitaev1995quantum,aspuru2005simulated,abrams1999quantum} and more recently the Harrow-Hassidim-Lloyd (HHL) \cite{harrow2009quantum,baskaran2023adapting,tsemo2025enhancing, peniel2026} algorithms. Long-term implications from such a quantum advantage could range from finding new materials for several real-world applications \cite{motlagh2025quantum,verma2026multireference} to drug discovery \cite{zhou2026quantum,santagati2024drug}. In the NISQ era, the variational quantum eigensolver (VQE) algorithm \cite{peruzzo2014variational} has been the workhorse for quantum chemical calculations on noisy quantum computers. 

    \begin{figure*}[t]
        \centering
        \includegraphics[scale=0.4]{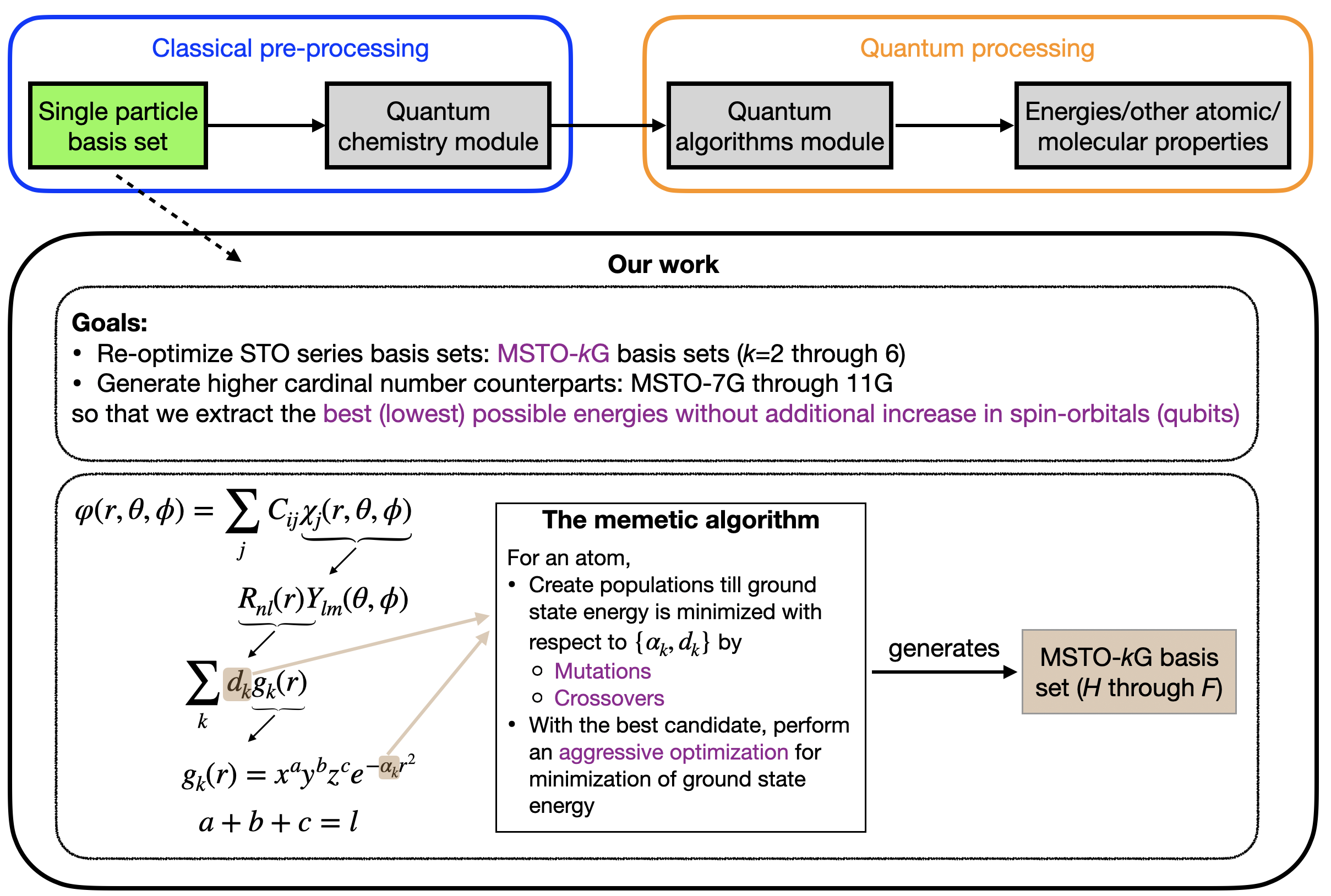}
        \caption{Schematic of our workflow. A typical quantum chemical calculation in the quantum computing framework involves classical (or quantum-classical hybrid) pre-processing and quantum processing. While there have been many efforts on the processing and the quantum chemical part (Hartree-Fock and integral evaluation) of the pre-processing steps, little work has been done on reducing required quantum resources via basis set optimization. We develop our memetic algorithm to optimize exponents and contraction coefficients of the STO basis sets, extend it to cardinal number 11, and show that the resulting modified STO bases yield lower atomic ground state energies than the conventional minimal bases without increasing the number of spin orbitals. For some atoms, they also yield lower FCI energies than the split-valence 6-31G basis. }
        \label{fig:schematic}
    \end{figure*}

All of these algorithms have been extended in their ability and scope in several directions. In the case of VQE, the modifications typically occur in the ansatz choice (for example, see Ref. \cite{kandala2017hardware} that introduced hardware efficient ans\"atze, and Ref. \cite{grimsley2019adaptive}, which introduced adaptive ans\"atze via their ADAPT-VQE algorithm) or the optimizer module (for instance, see the recent work by authors in Ref. \cite{mohammad2025hopso}), to improve its performance in some directions with their corresponding trade-offs. Works in literature also focus on resource reduction techniques including finding equivalent unitaries but with cheaper decomposition and tapering off qubits with little loss in results, which enable executing the algorithm on current noisy hardware and yield fairly accurate results \cite{chawla2025relativistic,chawla2025trapped,guo2024experimental}. Of late, the problem of executing the QPE algorithm on quantum computers has also been receiving attention owing to the tremendous advances on the quantum hardware and quantum error correction fronts \cite{yamamoto2025quantum}. Several variants of QPE exist that reduce resource requirements (for example, see \cite{li2024iterative,schiffer2025hardware,wang2025resource}). The authors of Ref. \cite{baskaran2023adapting} propose application of the HHL algorithm to quantum chemistry, and in the same work, variants of the algorithm improve it. A subsequent work in 2026 shows that one may realize an exponential advantage in the HHL-chemistry framework \cite{peniel2026}. Both of these fault-tolerant era algorithms, QPE and HHL, have been executed on current quantum hardware, notably the former was executed with quantum error correction partially baked into their workflow on the Quantinuum quantum computer with reasonable error margins given the scale of the resources required \cite{yamamoto2025quantum}. 

Existing demonstrations of VQE, QPE, and the HHL algorithms for chemistry problems on real quantum computers indicate that despite the rapid advances on the hardware front, practical computations in the near-term are likely to be restricted to relatively small active spaces (limited orbitals, and therefore a small number of qubits), primarily owing to the result fidelity deteriorating exponentially as we go to deeper circuits for a given gate fidelity. A small active space could result from a low quality basis set such as a minimal basis or picking a subset of orbitals from a high quality basis set such as Dunning's basis for post-Hartree-Fock (HF) calculations. In the former, the quality of ground state energies is not very good due to the basis size being insufficient. In the latter, a significant fraction of dynamical correlation effects in predicting ground state energies may be lost, and we also need to rely upon careful selection of orbitals to pick the right ones so that a good part of static correlation can be captured. These considerations motivate the search for approaches to improve achievable accuracy with a fixed (and limited) quantum resource budget. While techniques such as tapering that were mentioned in the earlier paragraph lower qubit count, where the goal is then to achieve comparable accuracy with fewer quantum resources, they introduce additional classical overheads. An alternative to obtaining comparable accuracies with fewer qubits could be indirect resource reduction through \emph{classical} pre-processing routines; for example, use of frozen natural orbitals to lower spin orbital and hence qubit count in VQE (see Ref. \cite{verma2021scaling}). A not-so-pursued approach in the context of indirect resource reduction is to optimize the single particle basis itself; if one could improve upon the contraction coefficients and exponents of existing contracted minimal basis sets and also systematically develop new contracted basis sets of higher cardinal numbers ($k>6$), we naturally could reach better energies with the same number of spin orbitals/ qubits. We would still be limited by the small size of such bases, but we do the best one possibly can via re-optimization and reaching higher cardinal numbers so that we can improve the accuracy at a fixed qubit count. This is the primary goal of our work; we develop a genetic algorithm-based approach in conjunction with aggressive refinement strategies to carry out this exercise for atoms Hydrogen (\ce{H}) through Fluorine (\ce{F}) excluding Helium. This exclusion is implied in the remainder of the manuscript. We now expand upon some of the points of this paragraph in the subsequent ones. 

We developed our own memetic algorithm to carry out the extension of basis sets to higher cardinal numbers. It is worth emphasizing at this point that although our heuristic algorithms can scale poorly with system size, basis set generation is a one time task, and therefore worth the computational effort.  The possibility of using the same number of qubits with better (lower) energy values motivates our search for optimized basis sets and extension to higher cardinal numbers. 

We first describe the memetic algorithm that improves upon the exponents and contraction coefficients of Pople's minimal bases (contracted) from cardinal numbers ($k$) 2 through 6, that is, the Slater type orbitals (STO) approximated by $k$ weighted and zero-centered Gaussians. We then extend the basis sets to higher cardinal numbers by using the fact that in their contracted form, the number of spin orbitals does not change with the cardinal number. Increasing $k$ therefore allows us to improve the atomic energies while retaining the same number of qubits as the parent minimal basis. We pick the minimal basis sets in particular, not only because they possess the aforementioned desirable property, but also because given the current two-qubit gate fidelities on quantum computers, it is sufficient to optimize these bases. We generate the modified STO-$k$G (MSTO-$k$G) basis sets for $k=2$ through $11$ for atoms Hydrogen through Fluorine, and benchmark full configuration interaction (FCI) ground state energies for these atoms obtained using our basis sets with FCI values obtained using STO-$k$G ($k=2$ through $6$), 6-31G, and cc-pVDZ basis sets. Throughout this work, the cardinal number $k$ in STO-$k$G and MSTO-$k$G denotes the number of Gaussian primitives per contraction. The index, $X$, in cc-pV$X$Z denotes the valence zeta quality. A schematic of our workflow and its place in an end-to-end quantum computing workflow for a chemistry problem is presented in Fig. \ref{fig:schematic}. 

We note that the STO series bases are of `single zeta' quality and are historically very important, but they have been superseded by modern basis sets. However, they made a comeback in quantum computing, given the field is still nascent, and we expect them to stay for at least a few years given the current status of quantum technology. As an aside, the STO-$k$G series bases are really not STOs but a linear combination of Gaussian type orbitals (GTOs) that are meant to mimic the behaviour of STOs. For a review on the work carried out using STOs themselves, interested readers could refer to Ref. \cite{hoggan2011molecular}. The authors of Ref. \cite{nagy2017basis} provide a good review on basis sets for chemistry in general. 

One could always choose a higher quality basis set and work in a small active space, but it can sometimes have the pitfall of yielding very small correlation energies (see Table \ref{tab:Ecor_elements} of Appendix B), in which case the noisy quantum computer can fail to capture the correlation energy accurately. With Pople's minimal basis sets, the correlation energies within a reasonable active space size are found to be in general larger than those with other basis sets such as Dyall's or Sapporo's, or for that matter, even the split-valence basis sets (we will expand upon this in Section \ref{sec:results}). Having said that, our memetic algorithm for improving upon exponents and contraction coefficients is general in nature, and can be employed to higher quality basis set families in the future. 

The article is organized as follows: Section \ref{sec:theory} briefly discusses minimal basis sets, while in Section \ref{sec:opt}, we explain our memetic algorithm and its role in optimizing STO-$k$G basis sets as well as in generating higher cardinal number MSTO-$k$G basis sets. Section \ref{sec:results} presents our results for the ground state energies of atomic systems, Hydrogen through Fluorine, using MSTO-$k$G basis sets, and their comparison with energies obtained using Pople's minimal and split valence 6-31G as well as Dunning's basis sets. This is followed by results for ground state energies and potential energy curves (PECs) for some molecular systems built using the atoms for which we optimized the basis sets. We then discuss in Section \ref{scaling} the quantum resources (number of qubits and CX gates) required for VQE, QPE, and HHL algorithms for chemistry calculations, followed by how single particle basis choice affects these resources in relation to ground state energies for the representative example of the \ce{Li} atom. We then compare our work with other related ones in literature, in Section \ref{lit}. We finally conclude in Section \ref{sec:conclusion}. 

\section{Theory and methodology}\label{sec:theory} 

We can compute the energy of the 1s orbital as follows \cite{magalhaes2014gaussian}: 

\begin{equation}
    E_{1s} = \dfrac{\int \chi_{\mu}^{*}\hat{H}\chi_{\mu} d\tau}{\int \chi_{\mu}^{*}\chi_{\mu} d\tau}, \label{eq:rayleighratio}
\end{equation} 

where $g_{1s}(\alpha) =  (2\alpha/\pi )^{3/4} \exp{(-\alpha r^2)}$ and $\chi_{\mu} = \sum_{i=1}^{k}\limits d_{i} g_{1s}(\alpha_{i})$. In the above equation, the integrals are over all space, $d\tau$ being an infinitesimal volume element. Optimized basis sets $\{ d_{i}, \alpha_{i} \} ~\forall i $, where $d_i$ refers to the $i^{th}$ contraction coefficient whereas $\alpha_i$ to the $i^{th}$ exponent, can be found either by finding the least squares fit, minimizing the error $\epsilon_{1s} = \int {(\chi_{1s}^{STO} - \chi_{\mu})^2}d\tau$ (the superscript $STO$ refers to the Slater-type orbital) 
or by minimizing the Rayleigh ratio (Eq. (\ref{eq:rayleighratio})). We chose to minimize the energy of an atom by gradual perturbation of the atomic basis exponents and contraction coefficients. The basis set parameters can, in principle, be optimized by minimizing the Hartree–Fock (HF), full configuration interaction (FCI), or configuration interaction singles and doubles (CISD) ground state energies of the atom of interest. In this work, we employ the CISD energy as the objective function. We note that minimizing the CISD energy does not necessarily result in a simultaneous minimization of the HF and FCI energies. Pople's basis sets are available for  $2\le k \le 6$. Further, we generate optimized basis sets with cardinal numbers up to $k=11$, with the objective of improving the ground state energies without increasing the number of contracted orbitals. However, owing to the heuristic nature of the optimization procedure, the resulting energies are not guaranteed to improve monotonically with increasing $k$. 

The use of Gaussian-type functions in electronic structure calculations was established in the pioneering work of Boys \cite{boys1950}, who demonstrated that the molecular integrals could be evaluated analytically using such functions. Their computational benefits subsequently motivated the development of contracted Gaussian basis sets, including the STO-$k$G family constructed by Hehre et al. \cite{hehre1969self} using a linear combination of atomic orbitals self-consistent-field (LCAO SCF) method for a minimal basis set of STOs for atoms not heavier than Fluorine. The representation of the STOs was in terms of GTOs - a sum of $k$ zero-centered Gaussians with a scaling parameter that accounted for the screening effect of each orbital. The parameters of the Gaussian basis sets, ($\{ d_k, \alpha_k\}$), were determined by a least square fit of the GTOs to the corresponding STOs for each principal quantum number. The volume element for the 1s orbital in the least square fit results in using a weight factor $w(r)$ proportional to $r^2$ for estimating $\{ d_i, \alpha_i\}$. As noted by Shavitt in Ref. \cite{shavitt1963the}, such a choice of $w(r)$ is appropriate for a three-dimensional integration. It leads to a better fit far away from the origin at the expense of accuracy in the near-nuclear region. 

Direct STO-based methods have also been developed for molecular electronic structure calculations, including the SMILES implementation of Fernandez Rico et al \cite{sto1}. Several studies have also developed and optimized STO basis sets for light elements, including the polarized valence and core-valence bases of Ema et al \cite{sto2}, the even-tempered STO bases of Chong et al \cite{sto5}, and the optimized STO bases of van Lenthe and Baerends \cite{sto6}. These studies demonstrate the continued development of STO-based approaches. Our work, however, focuses specifically on reoptimizing the existing STO-$k$G family of contracted Gaussian minimal basis sets to improve their electronic energies without increasing the number of contracted orbitals. We add that the choice of contracted Gaussian basis functions in this work is motivated by their widespread use in electronic structure packages and the computational convenience associated with Gaussian integral evaluation. Furthermore, STO-$k$G basis sets have been widely employed in electronic structure calculations on quantum computers. Our objective is therefore not to establish the superiority of Gaussian-type orbitals over STOs, but to investigate improvements that can be achieved within this established family of contracted Gaussian minimal basis sets. 

\section{Optimizing basis sets with a memetic algorithm}\label{sec:opt} 

For the case of the hydrogen atom, it is straightforward to compute the expression for the ground state energy by computing the integrals in Eq. (\ref{eq:rayleighratio}) as

\begin{eqnarray}
    E_{1s} &=& \left[\sum_{i=1}^{k}\limits \sum_{j=1}^{k}\limits \pi d_{i} d_{j} \left( \dfrac{4\alpha_i\alpha_j}{\pi^2} \right)^{\frac{3}{4}} \dfrac{1}{s_{i,j}} \times\left( 3\sqrt{\dfrac{\pi\alpha_{i}^{2}}{s_{i,j}}} \right. \right. \nonumber\\
    &&\left.\left.- 3\sqrt{\dfrac{\pi\alpha_{i}^{4}}{s_{i,j}^{3}}} -2 \right)\right] \left[\sum_{i=1}^{k}\limits \sum_{j=1}^{k}\limits d_{i} d_{j} \left( \dfrac{4\alpha_i\alpha_j}{s_{i,j}^{2}} \right)^{\frac{3}{4}}\right]^{-1} \label{eq:E1s}\\
    \nonumber
\end{eqnarray} 

where  $s_{i,j} = \alpha_i + \alpha_j$. This expression should be minimized with respect to the parameters $\{ d_{i}, \alpha_{i} \}$, $i=1,2,\cdots ,k$. While using the above expression for minimization of energy, it is to be noted that the contraction coefficients have to be rescaled to normalize the orbital wave functions. From the above expression, it is easy to see that  $\alpha_{1}=8/(9\pi)$ for $k=1$  (for example, see Ref. \cite{magalhaes2014gaussian}) leaving $d_{1}$ constrained only by the normalization condition on the wavefunction for the 1s orbital in the STO-$k$G basis. Eq. (\ref{eq:E1s}) clearly would not yield simple closed form expressions for atomic basis sets with higher orbitals included. We have therefore resorted to numerical optimization of the ground state energy of the atoms of interest. We next describe our metaheuristic approach to generate the MSTO-$k$G ($2 \le k \le 11$) basis. 

We use a memetic algorithm for optimizing the basis sets. A genetic algorithm (GA)-based approach \cite{holland1992genetic,forrest1996genetic} (Algorithm \ref{alg:ga}) involving mutations (Algorithm \ref{alg:mutate}) and crossovers  (Algorithm \ref{alg:crossover}) of candidate basis sets for a given atom in a minimal basis set, and selection of the best candidates (where the objective function is  the ground state electronic energy, computed using the Configuration Interaction Singles and Doubles (CISD) method) heuristically optimizes the basis set parameters, often reducing the ground state energy significantly. This is followed by an aggressive gradient-free search (or hill climbing optimization, for example \cite{walton2011modified,chinnasamy2022review}) for a lower ground state energy in the space of $k$ exponents and $k$ contraction coefficients (Algorithm \ref{alg:aggref}) in the modified basis set of STO-$k$G (MSTO-$k$G). The hybrid approach combines a global exploration by GA and a local exploration via the gradient-free aggressive search. The adopted method aims to combine accuracy with computational efficiency. 

For a given element, we start with the STO-$k$G basis sets for $2\le k \le 6$ and apply the memetic algorithm to refine the set for each $k$. For $k>6$ we use the MSTO-($k-1$)G basis set and add an additional exponent and contraction coefficient (initialized to arbitrary numbers, for example, 1.0 for both the quantities) as a starting point for an MSTO-$k$G basis. This transfer optimization (see \cite{gupta2017insights} for example) leads to a faster convergence for the higher $k$ values. The energy tolerance employed in the genetic algorithm part of the optimization was $10^{-10}$ Ha. This tolerance serves as a stopping criterion and does not provide a measure of the deviation from the global minimum. Given the stochastic nature of the approach and the finite population size and number of iterations employed, convergence to the global optimum is not guaranteed. Multiple passes of the memetic algorithm are used to improve the quality of the resulting basis sets. Further, for the memetic algorithm part, $N=10$ candidate basis sets were prepared from the bootstrapped basis which can limit the search space for large $k$. Although the value chosen  for \ce{N} can reduce the search space, a higher value may make the algorithm very resource intensive. The trade-off has to be decided by the user generating the basis sets based on the available hardware resources. Nevertheless, we obtain significantly improved atomic ground state energies relative to the conventional STO-$k$G bases, with lower FCI energies than those obtained with the 6-31G basis for H, Li, and Be. 

We use these basis sets and compute the ground state energy of a molecule from combinations of the standard minimal basis set and atom-optimized MSTO-$k$G bases, and compare with the 6-31G basis. 

The exponents and contraction coefficients for the MSTO-$k$G basis sets for $k$ from 2 through 11, and for atoms \ce{H} through \ce{F}, are given in Appendix \ref{app:basis_data}. Owing to the stochastic nature of the optimization procedure, independent executions of the memetic algorithm may lead to different sets of optimized basis parameters. Nevertheless, the specific MSTO-$k$G basis sets reported in this work are defined by the exponents and contraction coefficients provided in Appendix \ref{app:basis_data}, permitting their independent implementation without repeating the optimization procedure. 

\begin{algorithm}[H]
\caption{Genetic Algorithm Part of the Hybrid Memetic Optimization for MSTO-$k$G Basis}
\label{alg:ga}
\begin{algorithmic}[1]
\renewcommand{\algorithmicrequire}{\textbf{Input:}}
\renewcommand{\algorithmicensure}{\textbf{Output:}}
\Require Atom $M$, Population size $N$, Max trials $T$, Tolerance $\tau$
\Ensure Optimized basis set parameters $\Theta_{best}$

\State $Pop \gets \text{Generate\_Initial\_Population}(M, N)$ \Comment{Generate candidates by mutations of $M$.}
\State Sort $Pop$ by $E$ (CISD energy) in ascending order

\For{$t = 1$ \textbf{to} $T$}
    \State $f \gets \text{Dynamic\_Random\_Factor}(t)$ \Comment{Factor based on current trial}
    \State $Muts \gets \text{Mutate}(Pop, f)$ \Comment{Interpolative mutation}
    \State $Pop \gets \text{Keep\_Best}(Pop \cup Muts, N)$ \Comment{Elitism}
    
    \For{$j = 1$ \textbf{to} $N_{co}$} \Comment{$N_{co}$ is number of crossovers}
        \State $Cross \gets \text{Cross\_Overs}(Pop)$ \Comment{Discrete parameter swapping $d_{i} \leftrightarrow d_{j}$ or $\alpha_{i} \leftrightarrow \alpha_{j}$}
        \State $Pop \gets \text{Keep\_Best}(Pop \cup Cross, N)$
    \EndFor
    
    \State $E_{best} \gets Pop[0].E$, $E_{worst} \gets Pop[N-1].E$
    \If{$|E_{best} - E_{worst}| < \tau$} \Comment{Check for convergence}
        \State \textbf{break}
    \EndIf
\EndFor

\State \Return $Pop[0]$ \Comment{The best candidate after an iteration of mutations and crossovers is the new solution.}
\end{algorithmic}
\end{algorithm}

\begin{algorithm}[H]
\caption{Interpolative Mutation (Mutate)}
\label{alg:mutate}
\begin{algorithmic}[1]
\Require Population $Pop$, Dynamic random factor $f$
\Ensure List of mutated candidates $MList$

\State $N \gets \text{length}(Pop)$
\State $MList \gets [Pop[0]]$ \Comment{Retain best candidate}
\Comment{Generate new candidates by tweaking parameters from candidate pairs to intermediate values.}
\For{$i = 0$ \textbf{to} $N - 2$}
    \For{$j = i + 1$ \textbf{to} $N - 1$}
        \State $\Theta_i, \Theta_j \gets \text{Parameters of } Pop[i], Pop[j]$
        \State $\Delta \Theta \gets (\Theta_j - \Theta_i) \times f$
        \State $\Theta_{new} \gets \Theta_i + \Delta \Theta$
        \Try
            \Comment{Evaluate $E$ of the new candidate}
            \State $E_{new} \gets \text{getEnergies}(\text{Build}(\Theta_{new}))$
            \Comment{Add this candidate to the list if it is better than the best candidate in the existing list. Keep it sorted by $E$.}
            \If{$E_{new} < E \text{ of } (MList[0])$} 
                \State $MList.\text{append}([E_{new}, \Theta_{new}])$
                \State Sort $MList$ and truncate to size $N$
            \EndIf
        \EndTry
    \EndFor
\EndFor
\State \Return $MList$
\end{algorithmic}
\end{algorithm}

\begin{algorithm}[H]
\caption{Discrete Parameter Swapping (Cross\_Overs) }
\label{alg:crossover}
\begin{algorithmic}[1]
\Require Population $Pop$
\Ensure Crossover offspring $CList$

\State $N \gets \text{length}(Pop)$
\State $CList \gets [\,]$ 
\If{$N \bmod 2 = 1$}
    \State $N \gets N-1$
\EndIf 
\State $Idx \gets \Call{Shuffle}{[0,1,\ldots,N-1]}$ 
\State $N_{limit} \gets \lfloor N \times 0.75 \rfloor$ \Comment{75\% crossover rate}

\For{$k = 1$ \textbf{to} $N_{limit}$}
    \State $Idx \gets \Call{Shuffle}{[0,1,\ldots,N-1]}$ 
    \State $p1, p2 \gets Idx[0], Idx[1]$ \Comment{Select two parents}
    \State $\Theta_1, \Theta_2 \gets \text{Parameters of } Pop[p1], Pop[p2]$
    \State $T \gets \text{Random\_Binary\_Template}(\text{shape of } \Theta)$
    \State $\Theta_{off} \gets (T \times \Theta_1) + ((1 - T) \times \Theta_2)$ \Comment{Exchange a subset of the parameters to generate one offspring. }
    \Try
        \State $E_{off} \gets \text{getEnergies}(\text{Build}(\Theta_{off}))$
        \State $CList.\text{append}([E_{off}, \Theta_{off}])$
    \EndTry
\EndFor
\State \Return $CList$
\end{algorithmic}
\end{algorithm}



\begin{algorithm}[H]
\caption{Parallel Aggressive Refinement}
\label{alg:aggref}
\begin{algorithmic}[1]
\Require Candidate basis $C$, passes $P$, precision order $n$, shots $S$, sample size $\mu$, max iterations $\text{iter}_{\max}$, tolerance $\tau$, cores $n_{\text{cores}}$, output file $\text{fname}$
\Ensure Refined candidate $C_{\text{best}}$

\State $C_{\text{best}} \gets C$
\State $E_{\text{best}} \gets \text{GetCISDEnergy}(C_{\text{best}})$
\State $E_{\text{prev}} \gets E_{\text{best}}$
\State $F \gets \{10^{-0}, 10^{-1}, \dots, 10^{-n}\}$ \Comment{Precision scaling factors}
\State $\text{err} \gets 10^{9}$

\For{$\text{pass} = 0$ \textbf{to} $P - 1$}
    \For{\textbf{each} $f \in F$}
        \For{$\text{shot} = 0$ \textbf{to} $S - 1$}
            \State $\text{Improved} \gets \textbf{True}$
            \State $\text{iters} \gets 0$
            \While{$\text{Improved}$ \textbf{and} $\text{err} > \tau$ \textbf{and} $\text{iters} < \text{iter}_{\max}$}
                \State $\text{Improved} \gets \textbf{False}$
                \State $f_i \gets f \times \text{Rand}(0, 1)$ \Comment{Randomized step scaling}
                \State $\text{Stream} \gets \text{ParallelGenerator}(\text{GetOneCandidate}(C_{\text{best}}, f_i), \text{tasks}=\mu, \text{jobs}=n_{\text{cores}})$ \Comment{Joblib stream}
                \For{\textbf{each} $C_{\text{cand}} \in \text{Stream}$}
                    \If{$C_{\text{cand}} \neq \textbf{None}$}
                        \State $E_{\text{cand}} \gets \text{GetCISDEnergy}(C_{\text{cand}})$
                        \If{$E_{\text{cand}} < E_{\text{best}}$}
                            \State $C_{\text{best}} \gets C_{\text{cand}}$
                            \State $E_{\text{best}} \gets E_{\text{cand}}$
                            \State $\text{Improved} \gets \textbf{True}$
                            \State $\text{err} \gets |E_{\text{prev}} - E_{\text{best}}|$
                            \State $E_{\text{prev}} \gets E_{\text{best}}$
                            \State $\text{SaveFCIResults}(C_{\text{best}}, \text{fname})$ \Comment{Evaluate FCI and dump to disk}
                        \EndIf
                    \EndIf
                \EndFor
                \State $\text{iters} \gets \text{iters} + 1$
            \EndWhile
        \EndFor
    \EndFor
\EndFor
\State \Return $C_{\text{best}}$
\end{algorithmic}
\end{algorithm}

\begin{figure}[htbp]
    \centering
    \includegraphics[width=0.8\linewidth]{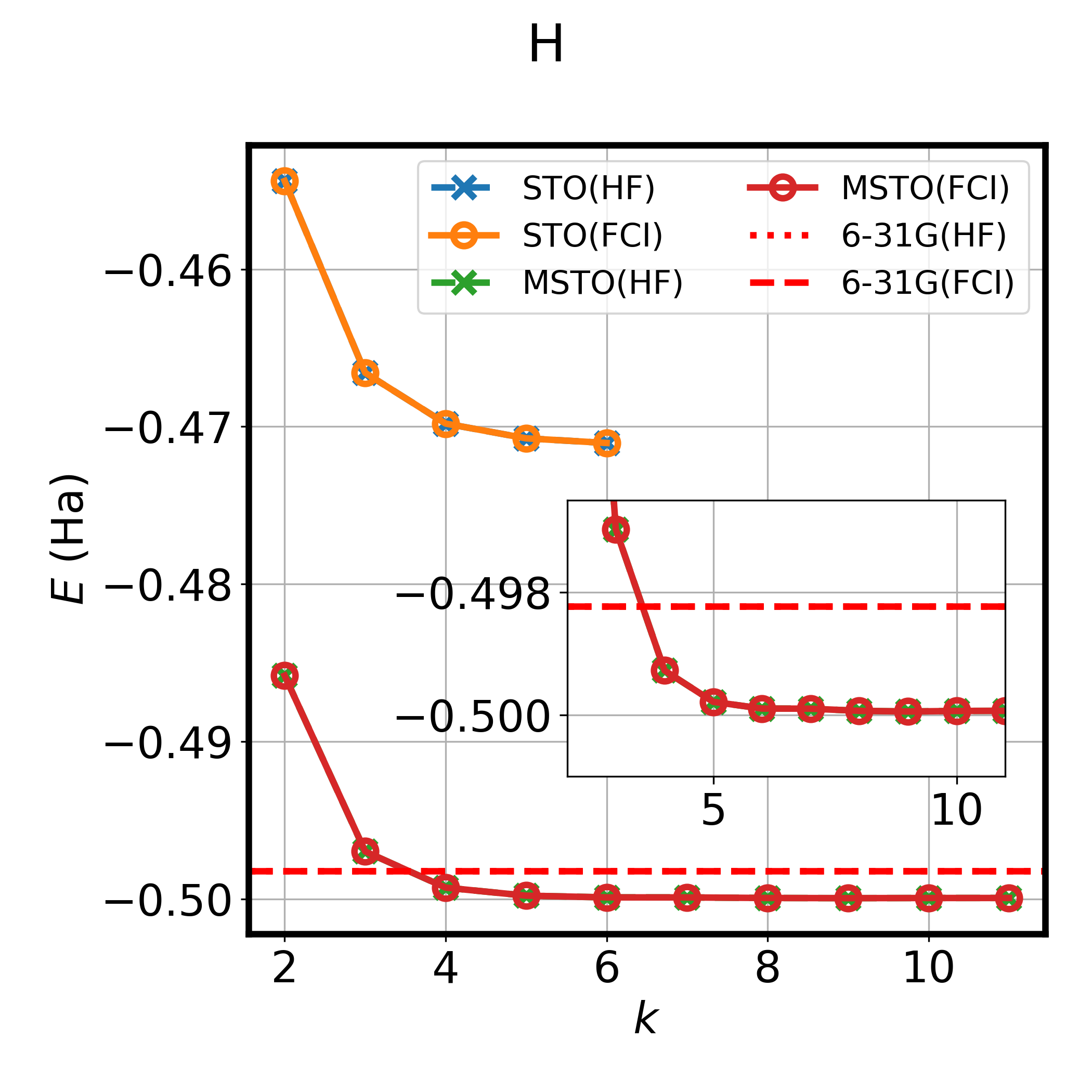}
    \caption{HF and FCI ground state energies of the hydrogen atom from MSTO-$k$G basis, compared with minimal basis sets and split valence basis sets (6-31G). } 
    \label{fig:EvsK_H}
\end{figure}

\begin{figure*}
    \includegraphics[width=0.96\textwidth]{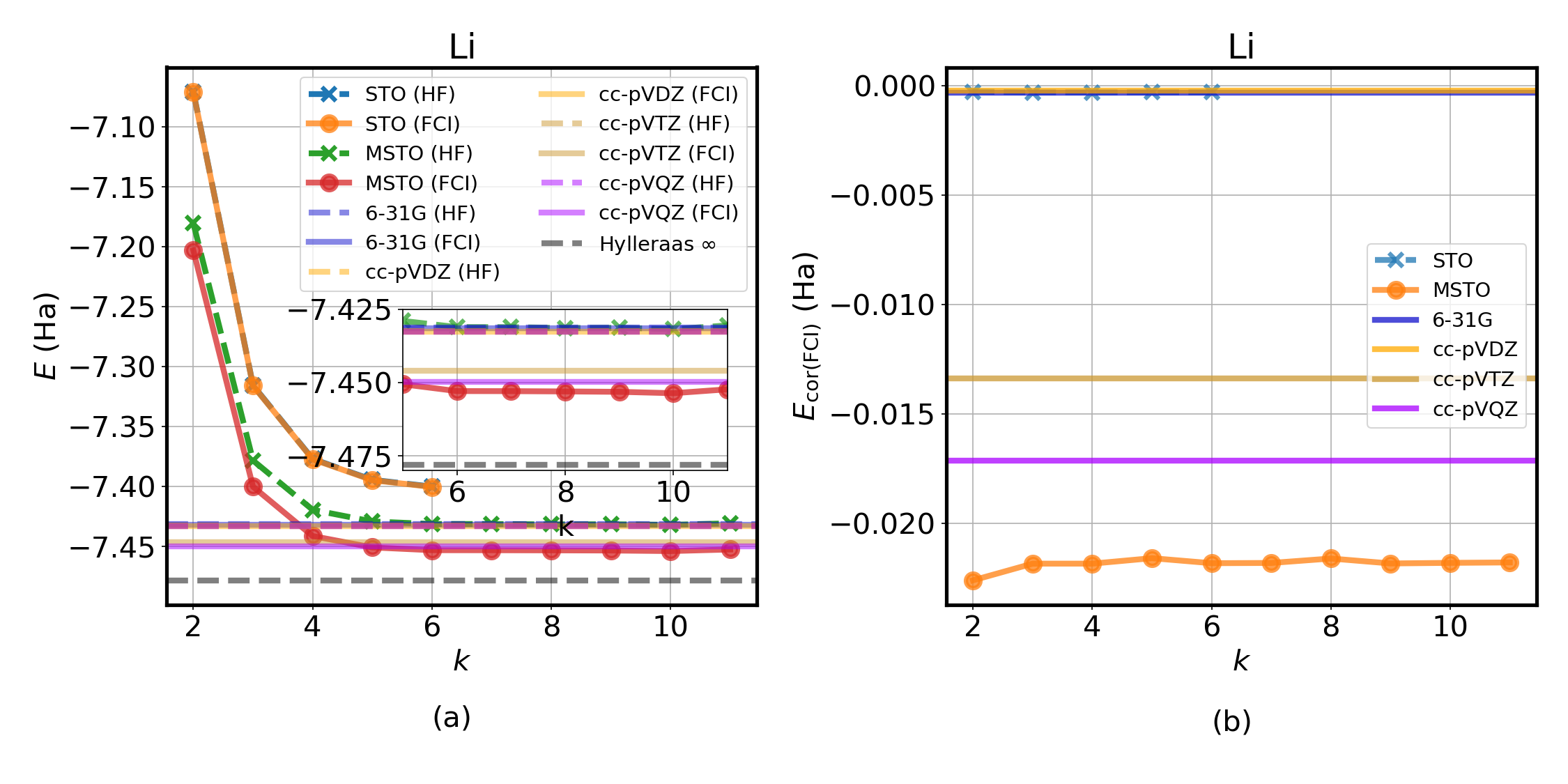} \\
    \includegraphics[width=0.96\textwidth]{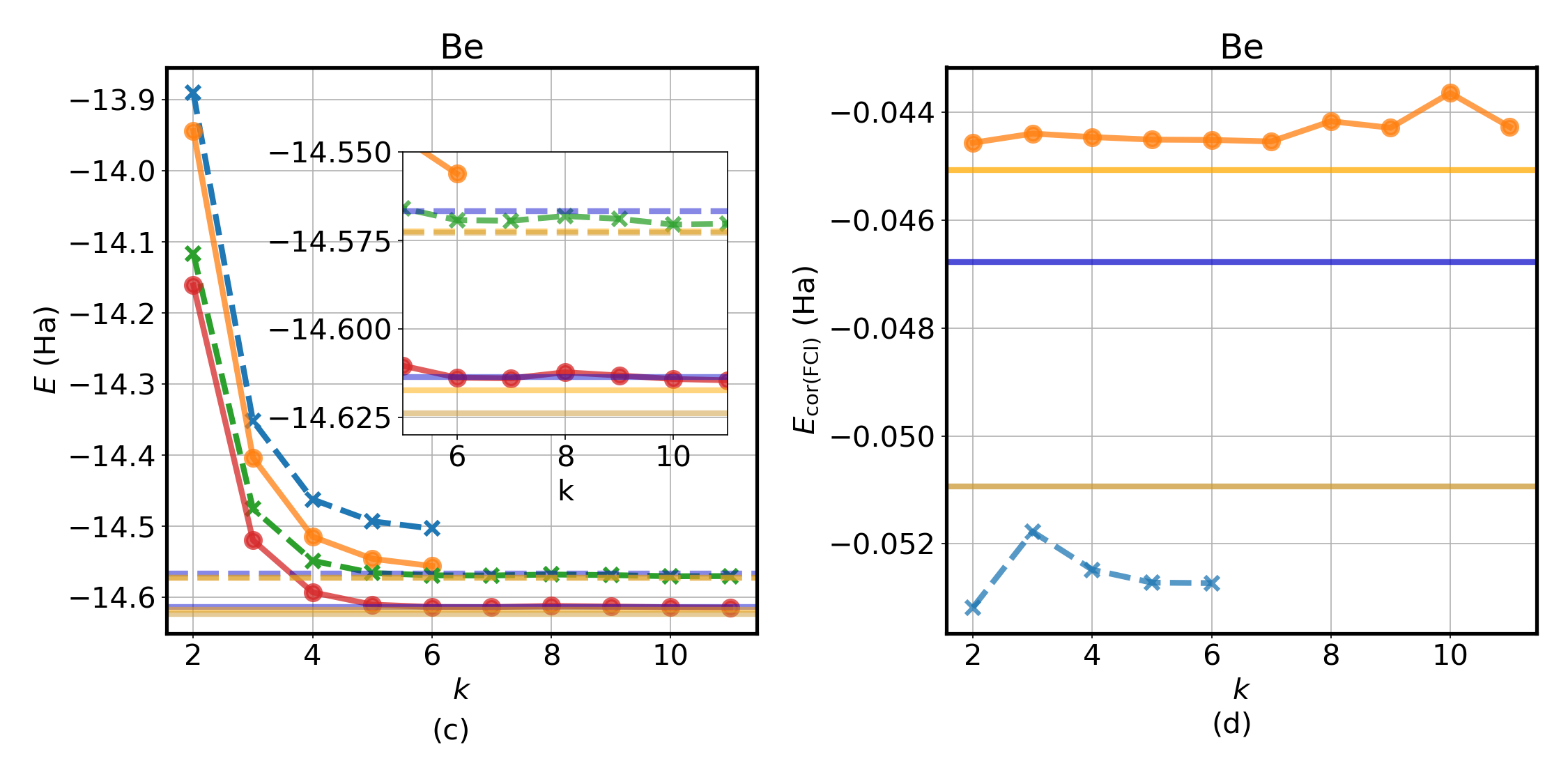} \\
    \caption{HF and FCI ground state energies of (a) Li and (c) Be obtained with our MSTO-$k$G basis sets, compared with other basis sets as indicated in the legends. The corresponding FCI correlation energies for Li and Be are presented in (b) and (d), respectively. Relative to the STO-$k$G basis sets, the MSTO bases yield larger correlation energy magnitudes for Li but smaller magnitudes for Be, despite yielding lower total FCI energies in both cases (see Tables \ref{tab:En_Li} and \ref{tab:En_Be}). The Be panels use the same symbols and line styles as the Li panels for the series shown. The Be cc-pVQZ results are given in Table \ref{tab:En_Be}. } 
    \label{fig:EvsK_LiBe}
\end{figure*}

\begin{figure*}
\includegraphics[width=0.96\textwidth]{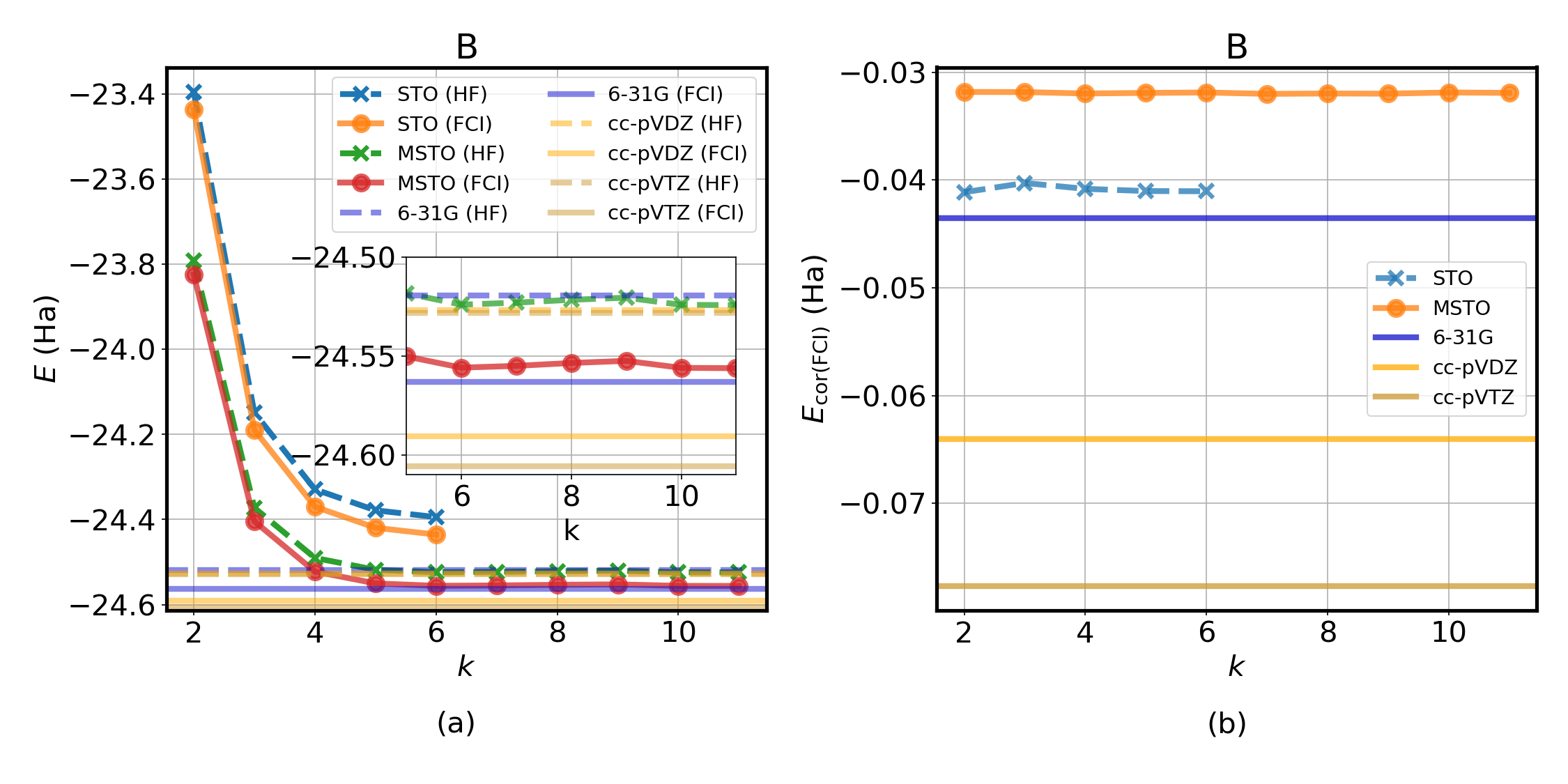} \\
    \includegraphics[width=0.96\textwidth]{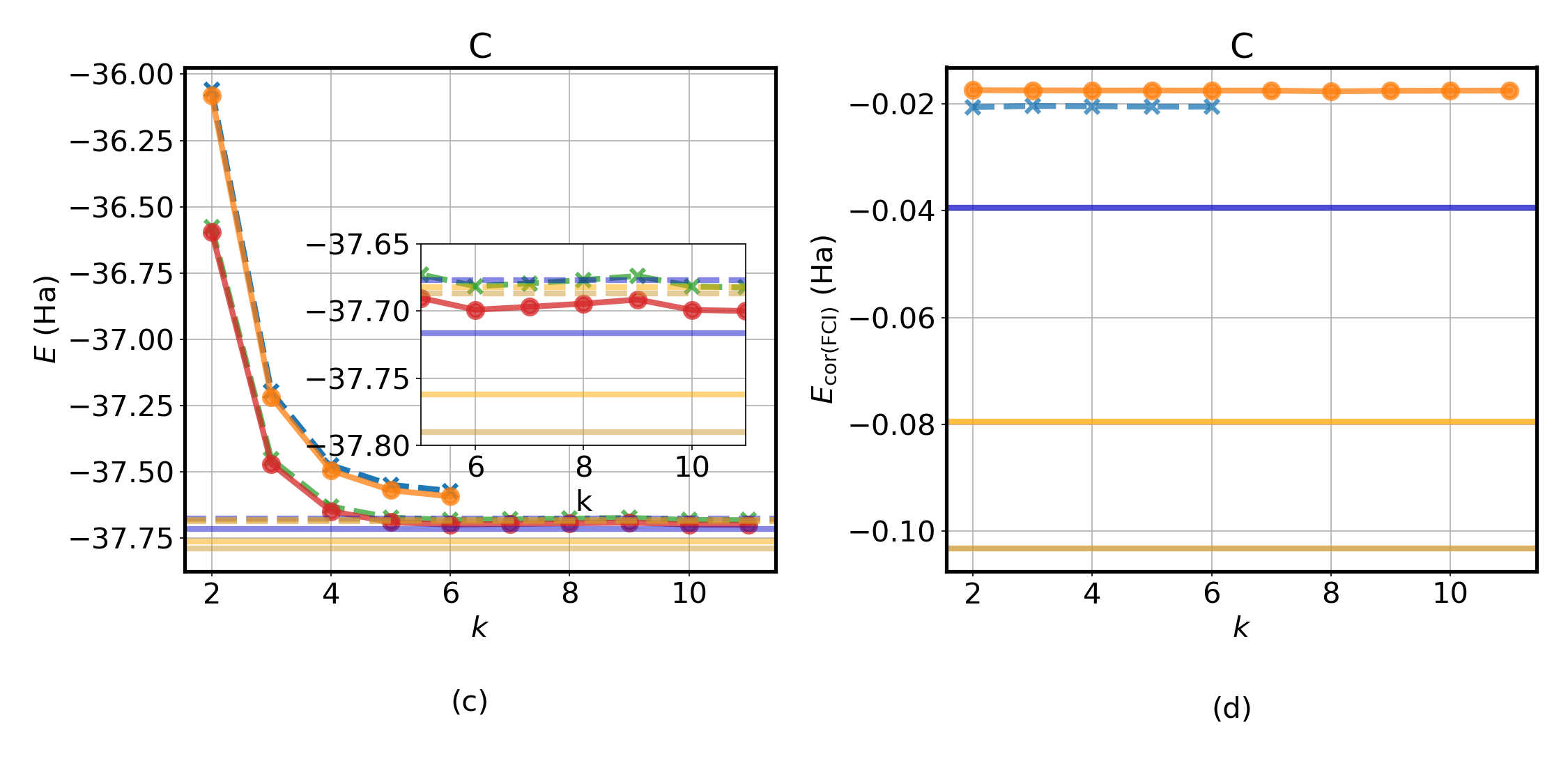} \\
    \caption{HF and FCI ground state energies of (a) B and (c) C obtained with our MSTO-$k$G basis sets, compared with the conventional basis sets indicated in the legends. The corresponding FCI correlation energies for B and C are presented in (b) and (d), respectively. Although the MSTO basis sets yield lower total FCI energies than the STO-$k$G bases, the magnitudes of their correlation energies are smaller (see Tables \ref{tab:En_B} and \ref{tab:En_C}). Panels (c) and (d) share the same legends as (a) and (b), respectively.} 
    \label{fig:EvsK_BC}
\end{figure*} 

\begin{figure}
    \centering
    \begin{tabular}{c}
    \includegraphics[width=0.8\linewidth]{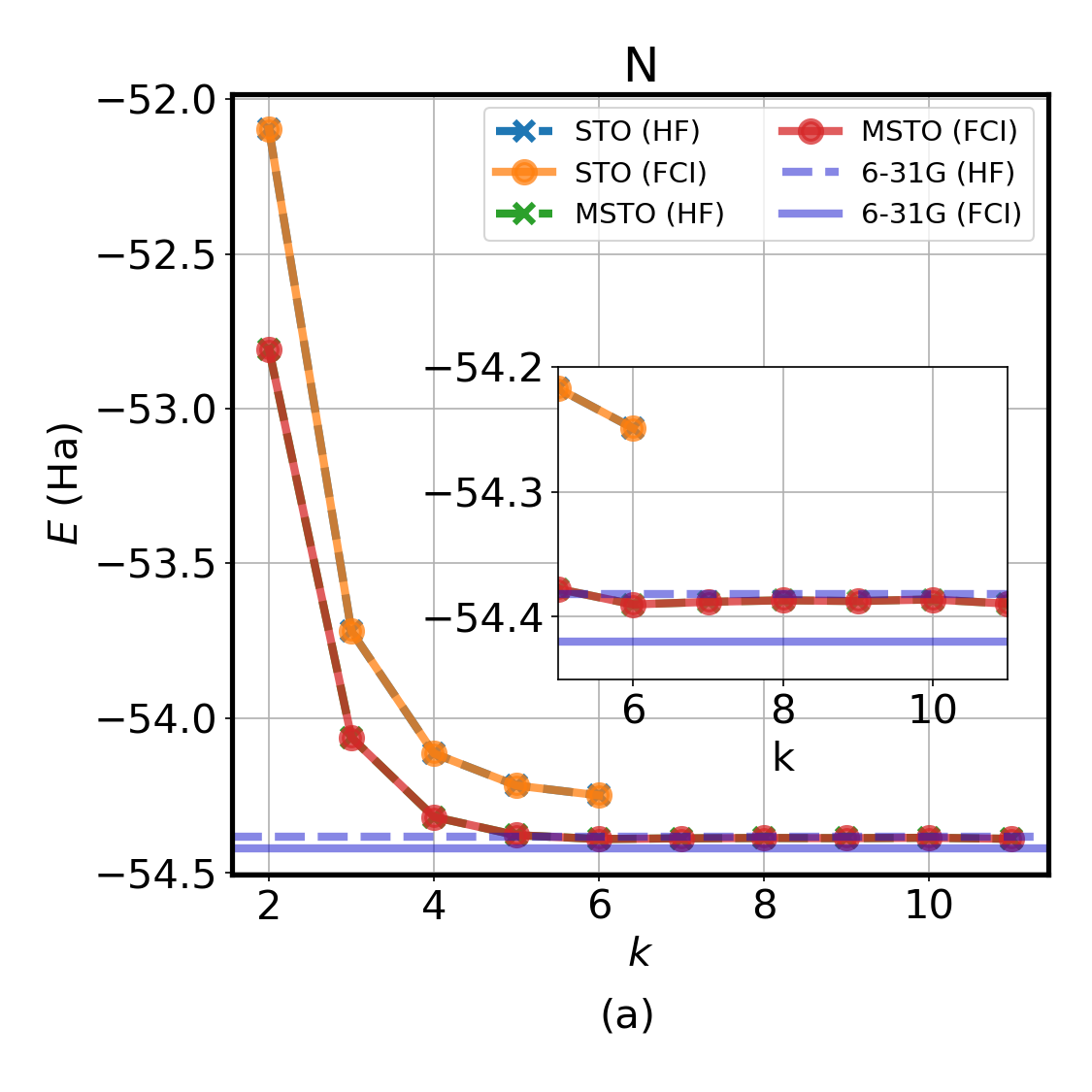} \\
    \includegraphics[width=0.8\linewidth]{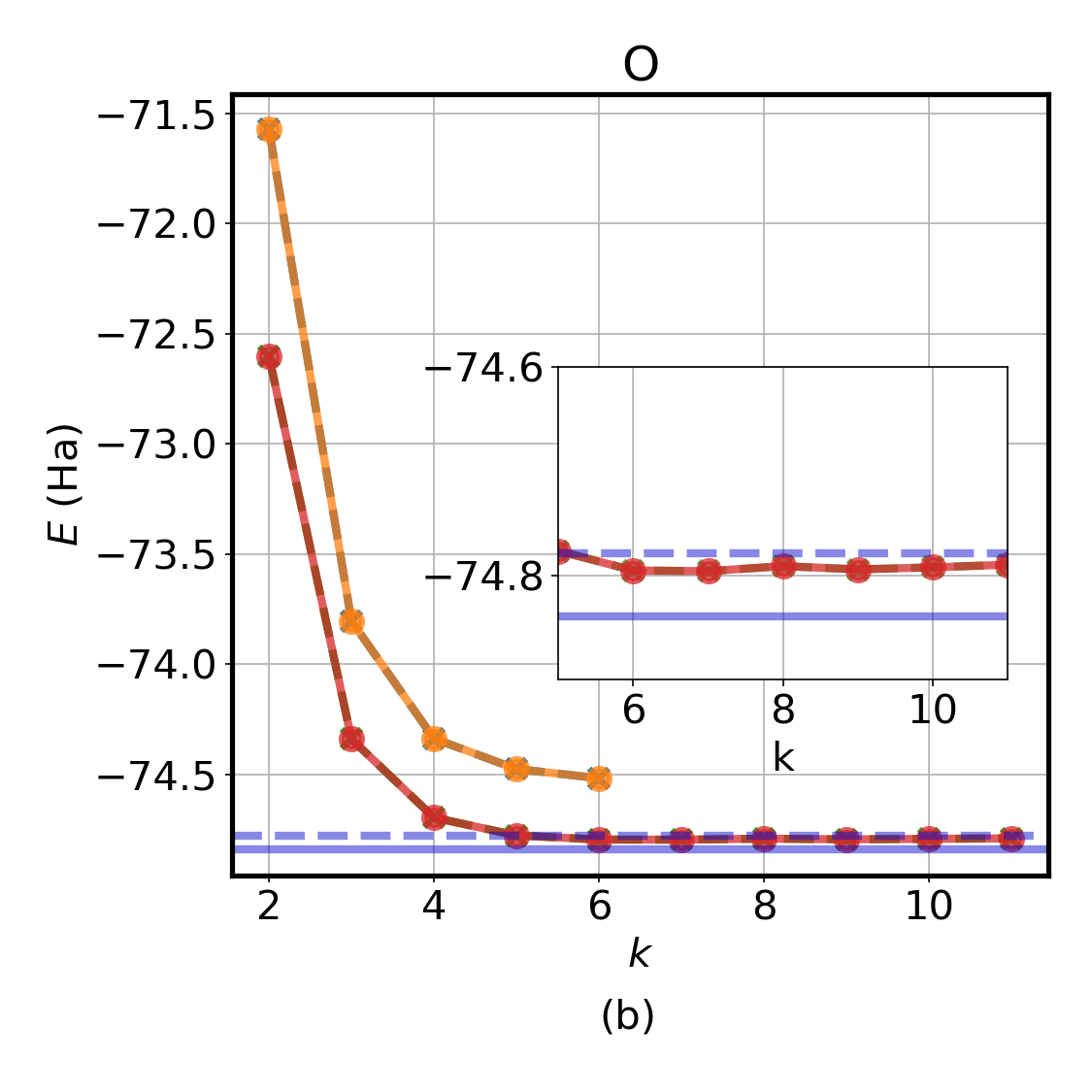} \\
    \includegraphics[width=0.8\linewidth]{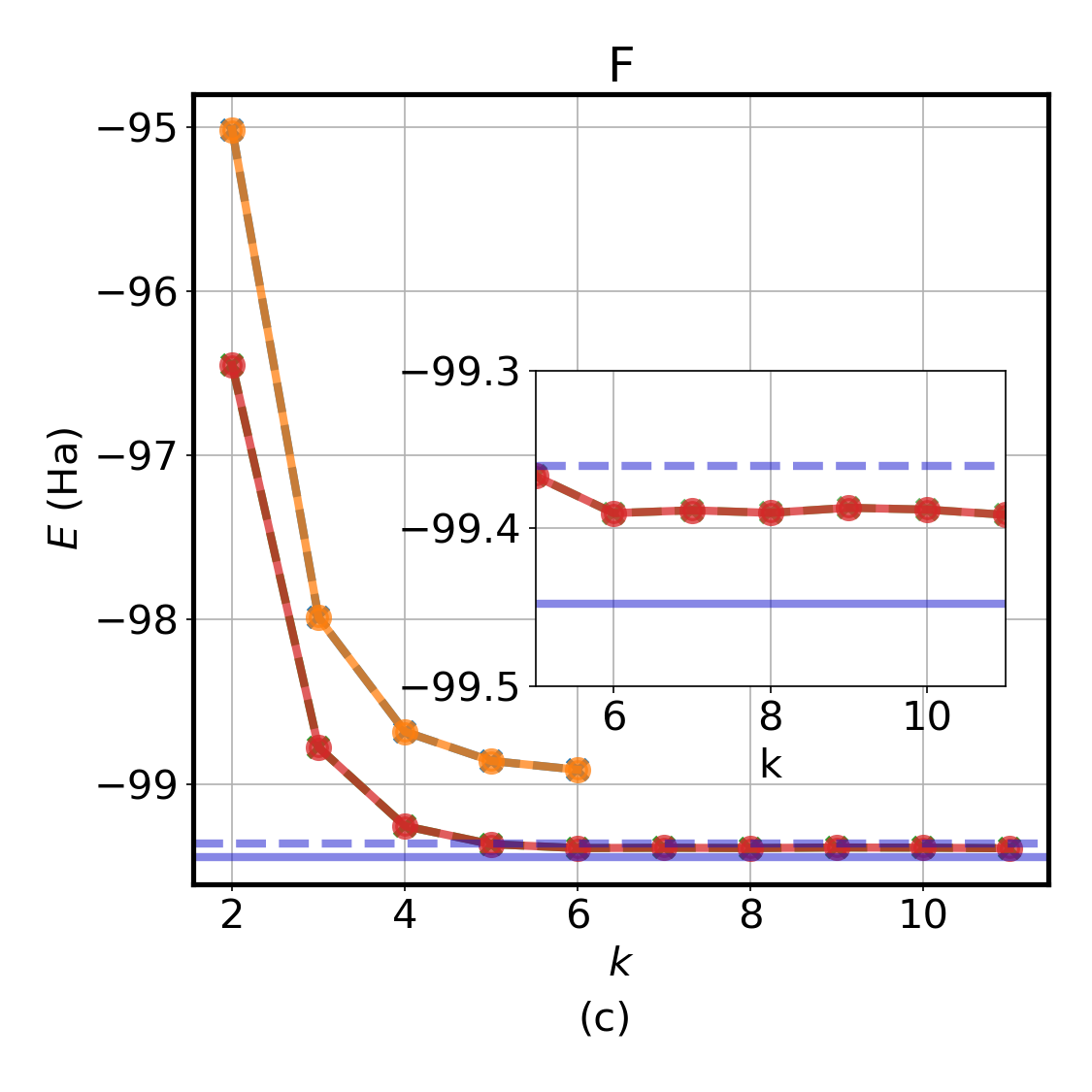} 
    \end{tabular}
    \caption{HF and FCI ground state energies of (a) Nitrogen, (b) Oxygen, and (c) Fluorine from MSTO-$k$G basis, compared with minimal basis sets and split valence basis sets (6-31G). The FCI correlation energies obtained with STO and MSTO bases are near-zero (see Tables \ref{tab:En_N}, \ref{tab:En_O} and \ref{tab:En_F}). }
    \label{fig:EvsK_NOF}
\end{figure} 

\section{Results}\label{sec:results} 

\subsection{Atomic ground state energies with our basis sets} 

Figures \ref{fig:EvsK_H} to \ref{fig:EvsK_NOF} present HF and FCI results using: 

\begin{enumerate}
\item the minimal basis sets ($k$ of 2 through 6), 
\item our MSTO-$k$G ($k$ of 2 through 11), 
\item the split-valence 6-31G, and 
\item cc-pVXZ (X=D, T, Q, where computationally
feasible, in Fig. \ref{fig:EvsK_LiBe}), 
\end{enumerate}

with the accompanying data in Tables \ref{tab:En_H} through \ref{tab:En_F} of Appendix B. Owing to the rapidly increasing computational cost of full-space FCI calculations, we restrict the cc-pVQZ FCI comparisons to H, Li, Be, and B, and the cc-pVTZ comparisons to these atoms and C. The cc-pVDZ results are reported for all the atoms considered. The tables in the Appendix also provide energies from the optimized STO-$k$G ($k=3, 6$) basis sets optimized by generalized simulated annealing \cite{de2008atomic}, STO-$k$G ($3 \le k \le 6$) basis sets by Hehre \textit{et al} and the exact values for STOs (the last two are taken from \cite{hehre1969self}). For the case of \ce{Li}, since our basis performed comparably to Dunning's cc-pVQZ basis in predicting ground state energies, we have added the results from the basis too (Fig. \ref{fig:EvsK_LiBe}(a)). We exclude He since by construction, noble gas atoms have no virtuals and hence no correlation effects in the minimal basis. Our MSTO-$k$G basis inherits the same trait. The PySCF package \cite{sun2018pyscf,sun2020recent} was used for all the computations. For the atomic benchmark calculations, we employed an all-electron restricted Hartree–Fock (RHF) reference for Be and restricted open-shell Hartree–Fock (ROHF) references for the remaining atoms. The numbers of alpha and beta electrons, respectively, were (1,0), (2,1), (2,2), (3,2), (4,2), (5,2), (5,3), and (5,4) for H, Li, Be, B, C, N, O, and F. The FCI calculations were carried out considering all electrons and all spatial orbitals of the respective basis sets, without invoking the frozen-core approximation. Throughout this work, FCI refers to full configuration interaction within the finite single-particle orbital space corresponding to a single-particle basis, and does not imply that the complete basis set limit has been attained. 

Fig. \ref{fig:EvsK_H} shows that for Hydrogen, the ground state energy using the MSTO-$k$G basis practically reaches the theoretical value of $-0.5$ Ha. The case of the H atom is exactly solvable and this exercise helps in assessing the correctness of our code. At $k=4$, our MSTO-$k$G basis surpasses the ground state energy from the 6-31G basis. Interestingly, the MSTO-9G basis, with only
one spatial orbital, yields a ground state
energy comparable to that obtained with
the cc-pVQZ basis (see Table \ref{tab:En_H}). The correlation energy was found to be consistent with zero (within a numerical error of about $10^{-16}$ Ha) in STO and MSTO bases, with the reason being the absence of electron-electron interaction in the Hamiltonian of the one-electron atom. 

For Lithium, Fig.~\ref{fig:EvsK_LiBe}(a) shows that the ground state energies from the MSTO bases at higher $k$ surpass even that obtained with the cc-pVQZ basis. Furthermore, our MSTO bases capture larger correlation energies than those obtained with the STO, 6-31G, and cc-pVQZ bases (Fig.~\ref{fig:EvsK_LiBe}(b)). Our best result, obtained with the MSTO-10G basis ($-7.453678$ Ha), is within 24.382 mHa of the benchmark ground state energy ($-7.478060$ Ha) obtained with the Hylleraas basis (see Table \ref{tab:En_Li} of Appendix B). 

For Beryllium, Fig.~\ref{fig:EvsK_LiBe} (c) shows that the MSTO basis sets yield ground state energies comparable to those from the split-valence basis. The
MSTO-11G basis yields a lower FCI energy than
the 6-31G basis by approximately 1 mHa, but
falls short of the cc-pVDZ, cc-pVTZ, and cc-pVQZ results by approximately 3, 9, and 26 mHa, respectively (see Table \ref{tab:En_Be} of Appendix B). 

Fig.~\ref{fig:EvsK_BC}(a) and (c) present our results for Boron and Carbon, respectively. While the MSTO bases consistently outperform the corresponding STO
bases, the lowest FCI energies obtained with our bases fall short of the 6-31G results by approximately 7 and 16 mHa, respectively
(see Tables \ref{tab:En_B} and \ref{tab:En_C} of Appendix B). We also observe that for these two systems and for Be, although the MSTO basis sets yield lower total FCI energies than the STO basis sets, the magnitudes of their correlation energies are smaller. This indicates that an improvement in the total ground state energy need not be accompanied by an increase in the magnitude of the correlation energy. This can impact the difficulty of resolving it on a quantum computer, because for a given uncertainty in the energy estimates, smaller correlation energies can be more difficult to resolve accurately on noisy quantum computers. For example, restricting the cc-pVDZ calculation to the first 10 spin orbitals recovers only 29.4\% and 14.9\% of the
full-space cc-pVDZ correlation energy for B and C, respectively (Table~\ref{tab:Ecor_elements} of Appendix B). We note that this is only for illustration; in practice, one invokes chemical intuition to pick the relevant orbitals for an active space, but it is not always feasible to recover most of the correlation energy with the right choice of orbitals as one would still miss dynamical correlation with a small active space. Thus, for some of the atoms considered, the MSTO bases offer a reasonable trade-off between the quality of the total ground state energies and the magnitude of the correlation energies. 

\begin{figure*}[t]
    \centering
    \begin{tabular}{cc}
       (a) & \includegraphics[width=0.7\linewidth]{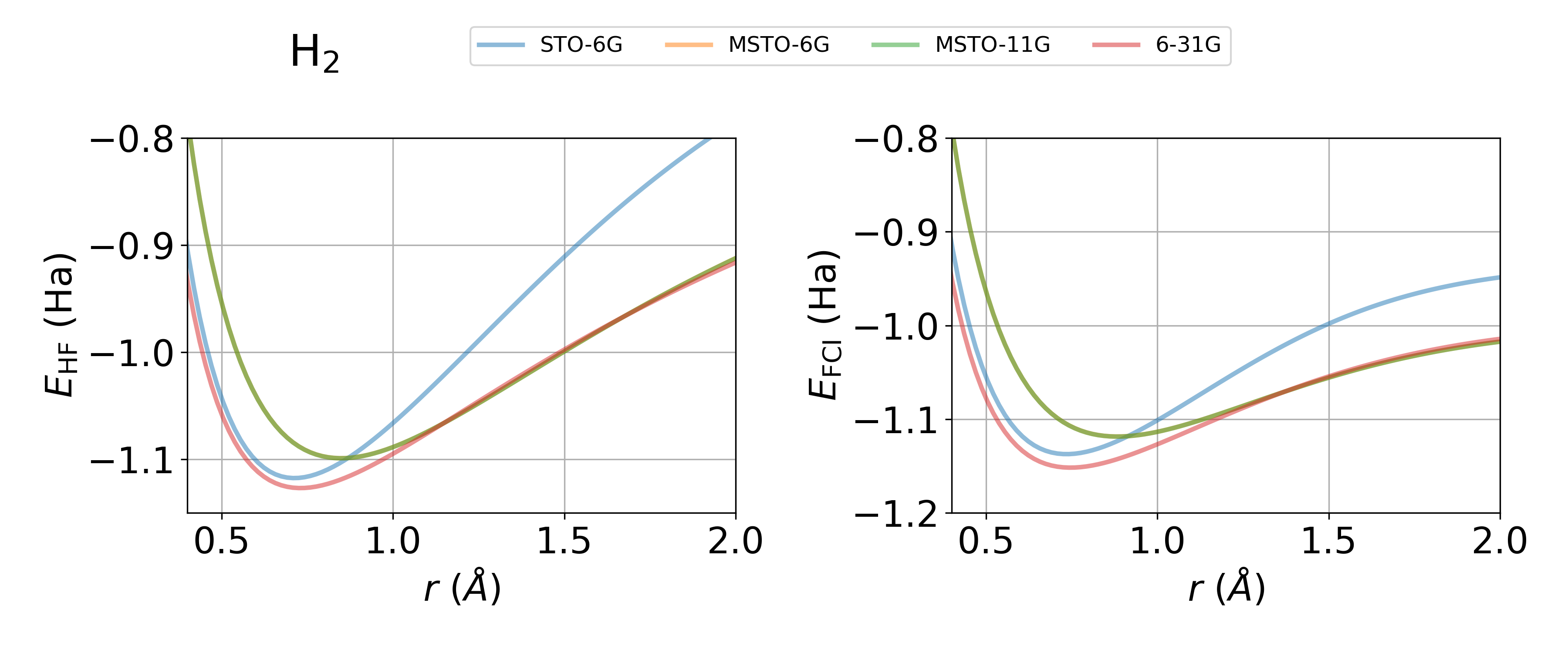} \\
       (b) & \includegraphics[width=0.7\linewidth]{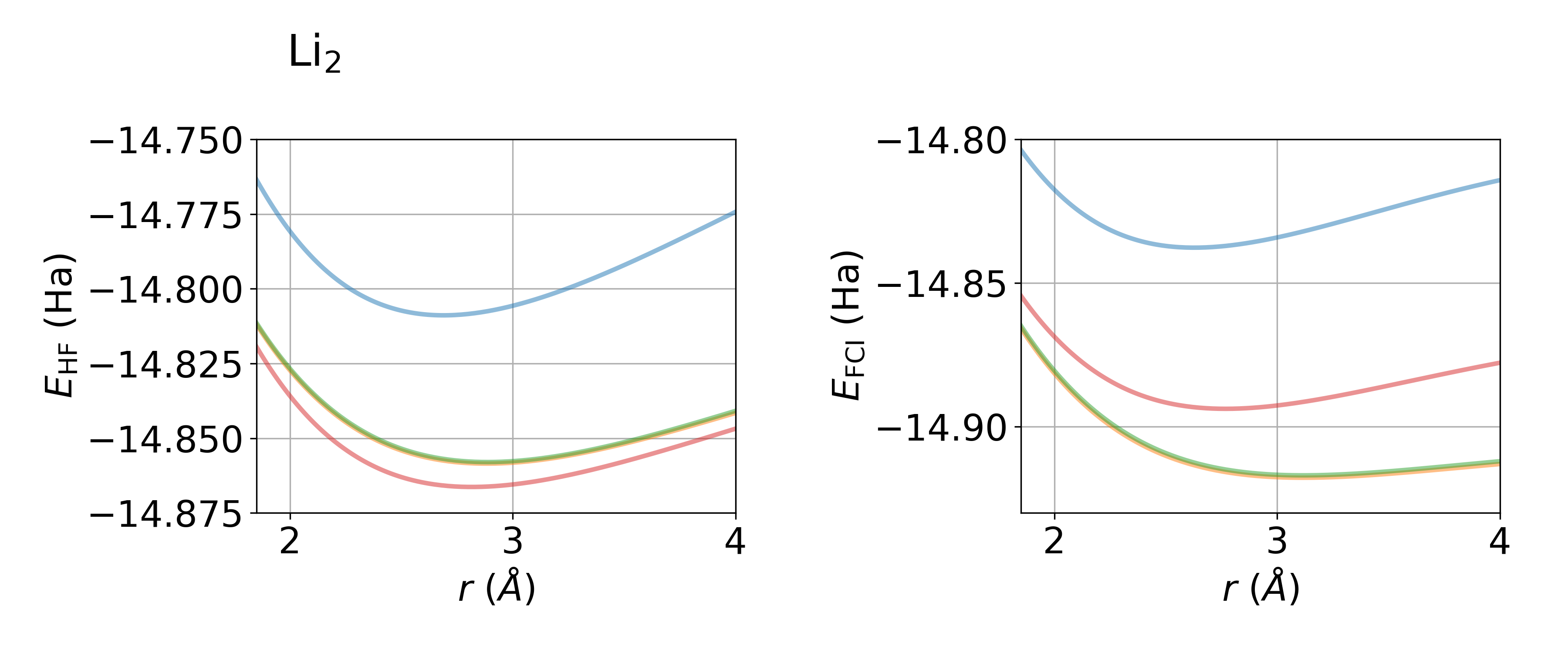}\\
       (c) & \includegraphics[width=0.7\linewidth]{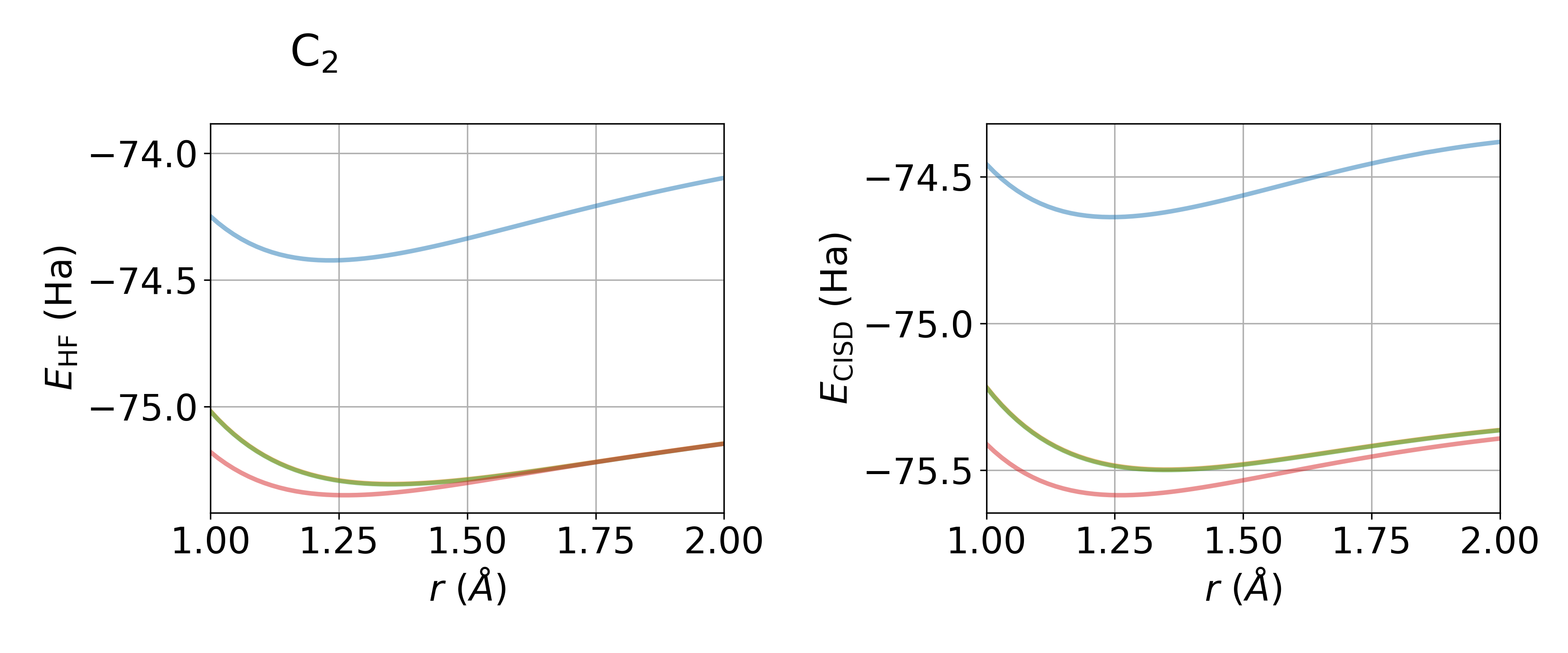}
 \end{tabular}
\caption{The HF (left panels), FCI (right panels) PECs for (a) H$_2$ and (b) Li$_2$ and CISD (right panel) for (c) C$_2$. The PECs for the atom optimized MSTO-6G and MSTO-11G bases are compared with the standard basis sets, namely STO-6G and 6-31G bases.  All plots share the same legend. In all the plots the curves for MSTO-6G and MSTO-11G coincide.}
    \label{fig:EvsrH2andLi2}
\end{figure*} 

\begin{figure*}[t]
    \centering
    \begin{tabular}{cc}
       (a) & \includegraphics[width=0.7\linewidth]{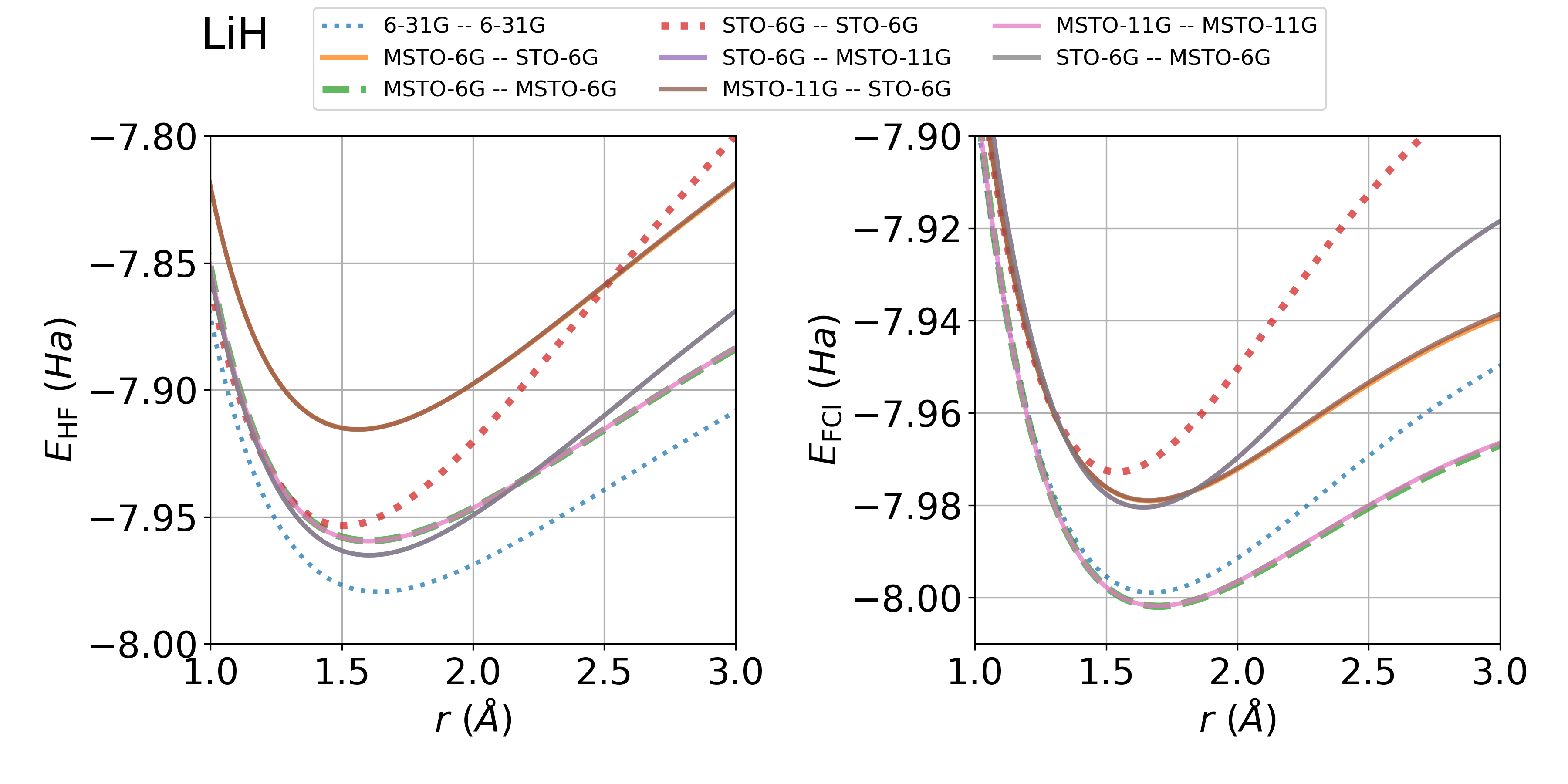}\\
       (b) & \includegraphics[width=0.7\linewidth]{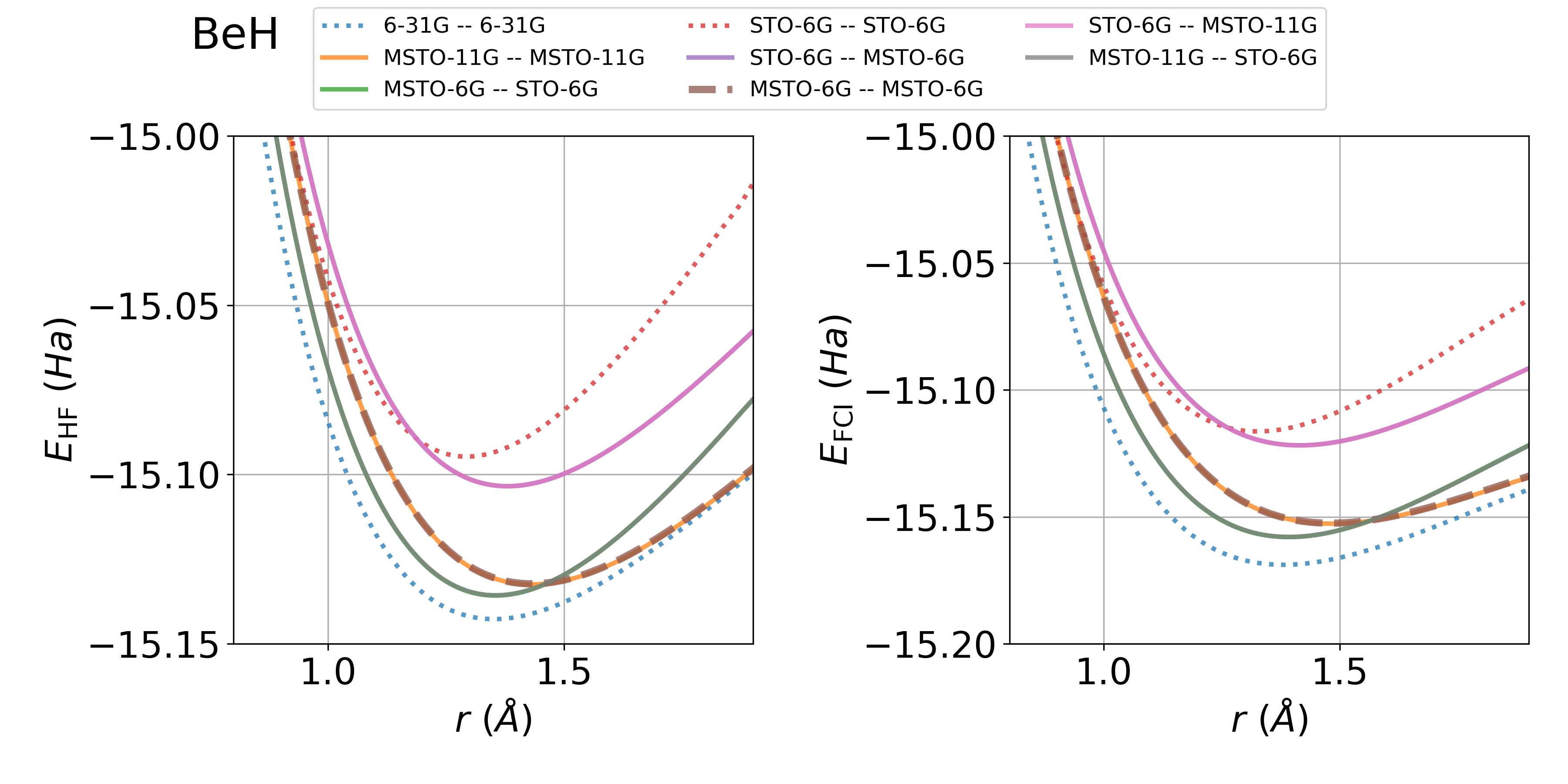}\end{tabular}
\caption{The HF (left panels) and FCI (right panels) PECs for (a) LiH and (b) BeH. The PECs for the atom optimized MSTO-6G and MSTO-11G bases are compared with the standard basis sets, namely STO-6G and 6-31G bases. The order of basis for the atoms are in the same order as the atoms in the formula. The HF and FCI panels for each molecule share the legend shown above that row. In all the plots the following pairs of curves are coincident: (i) MSTO-6G--STO-6G with MSTO-11G--STO-6G, (ii) MSTO-6G--MSTO-6G with MSTO-11G--MSTO-11G, and (iii) STO-6G--MSTO-6G with STO-6G--MSTO-11G.}
    \label{fig:EvsrLiHandBeH}
\end{figure*} 

\begin{figure*}[t]
    \centering
    \begin{tabular}{cc}
       & \includegraphics[width=0.7\linewidth]{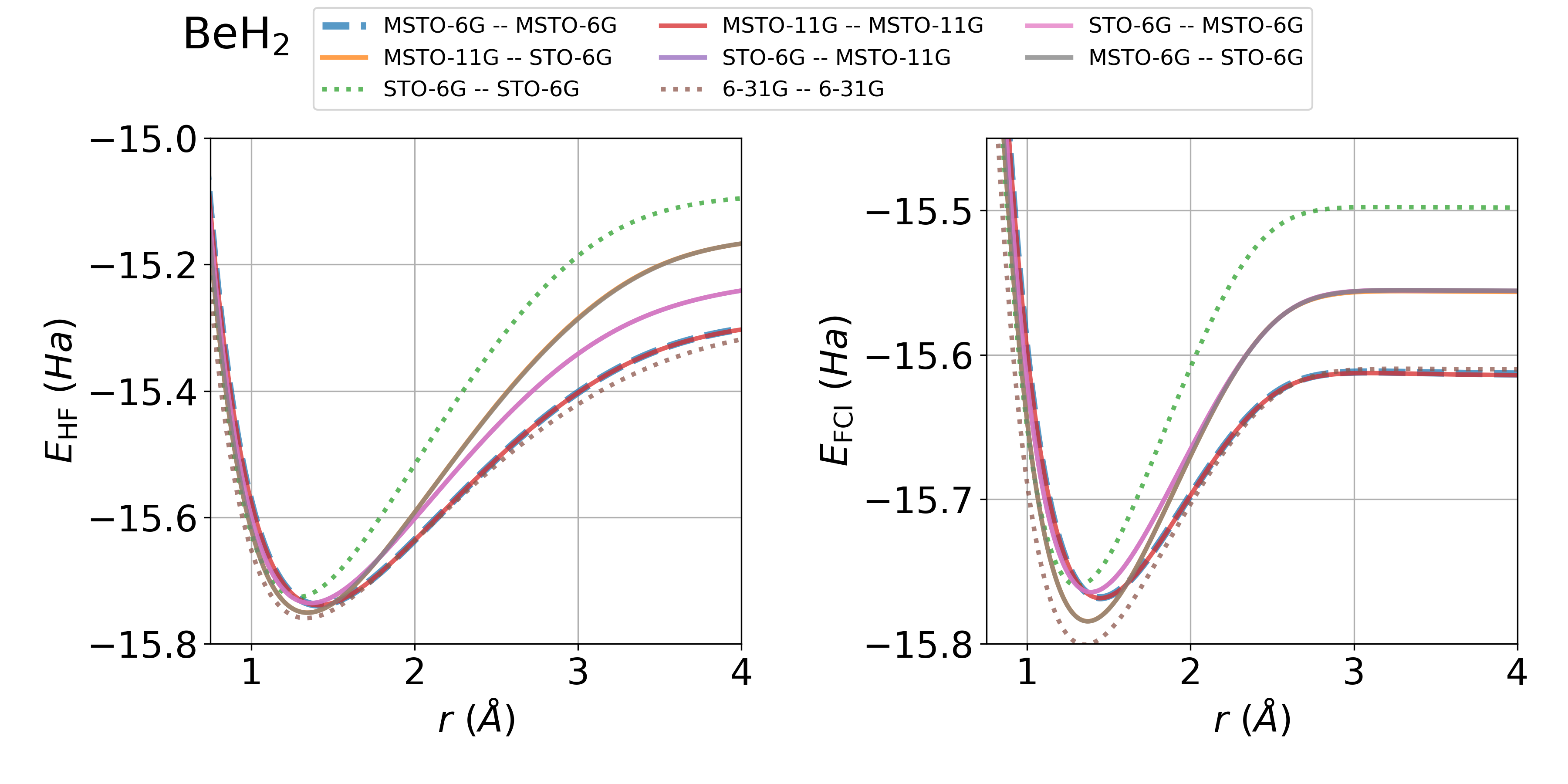}\\\end{tabular}
\caption{The HF (left panel) and FCI (right panel) PECs for \ce{BeH2}. The PECs for the atom optimized MSTO-6G and MSTO-11G bases are compared with the standard basis sets, namely STO-6G and 6-31G bases. The order of basis for the atoms are in the same order as the atoms in the chemical formula. Both the plots share the same legend. In all the plots, the following pairs of curves are coincident: (i) MSTO-6G--STO-6G with MSTO-11G--STO-6G, (ii) MSTO-6G--MSTO-6G with MSTO-11G--MSTO-11G, and (iii) STO-6G--MSTO-6G with STO-6G--MSTO-11G.}
    \label{fig:EvsrBeH2andCHplus}
\end{figure*} 

For the atoms considered, the improvements in ground state energies generally become smaller beyond $k=6$, indicating an apparent saturation within the range of $k$ explored and the optimization strategy employed. Nevertheless, the energies do not decrease monotonically with increasing $k$, and further improvements through alternative or more extensive optimization strategies cannot be ruled out. 

While we observe the superior performance of our MSTO basis sets over the STO ones in predicting HF energies for N, O, and F, it is worth noting that our bases inherit the same limitation as the STO ones in that for these 3 atoms, they do not have sufficient number of virtuals to generate any noticeable contribution to correlation energies. Improving this limitation by possibly adding diffuse and/or higher angular momenta functions is beyond the scope of the current work, and we defer it to a future study. We also note that among the MSTO-$k$G basis sets generated in this work, the lowest FCI ground state energies for H, Li, Be, B, C, N, O, and F are obtained for $k=9,10,11,11,11,6,7$, and $11$, respectively. Thus, increasing the number of primitives per contraction does not necessarily result in lower energies for the basis sets generated by our heuristic optimization procedure. Furthermore, the choice of the objective function is important. For example, among our MSTO basis sets for Be, the lowest HF and FCI energies are obtained with the MSTO-10G and MSTO-11G basis sets, respectively. 

\subsection{Molecular ground state energies with our basis sets} 

We pick the simple \ce{H2}, \ce{Li2}, and \ce{C2} molecules for computing their potential energy curves (PECs) with the STO-6G, MSTO-6G, MSTO-11G, and the 6-31G basis sets. Near equilibrium, the MSTO-6G and MSTO-11G curves nearly coincide. For H$_2$, their HF and FCI energies lie above the corresponding STO-6G and 6-31G values (Fig. \ref{fig:EvsrH2andLi2}(a)). For C$_2$, their HF and CISD energies lie below STO-6G and above 6-31G (Fig. \ref{fig:EvsrH2andLi2}(c)). This observation is consistent with that from Andrade \textit{et al} \cite{de2008atomic}. Furthermore, the equilibrium bond length from MSTO bases itself is slightly longer than the ones predicted by STO-6G and 6-31G bases. The relative performance of the basis sets changes with internuclear distance. For example, at stretched geometries of H$_2$, the MSTO basis sets yield lower FCI energies than those obtained with STO-6G. We add that the incorrect dissociation behaviour of H$_2$ at the restricted Hartree–Fock level is a known limitation of the single-determinant description, which cannot correctly describe the dissociation into two neutral open-shell hydrogen atoms. This limitation is not specific to the MSTO-$k$G basis sets and should be distinguished from errors arising from the incompleteness of the single-particle basis. The FCI wave function, being a linear combination of determinants, can describe the static correlation associated with bond dissociation, although it remains limited by the finite single-particle basis employed. 

For the case of Li$_2$ (Fig. \ref{fig:EvsrH2andLi2} (b)), we find that the FCI energies obtained using MSTO-6G and MSTO-11G bases consistently outperform the 6-31G results. Just as in the case of the \ce{H2} molecule, the MSTO bases predict slightly longer bond lengths than the minimal and the split-valence basis sets. 

We study the performance of STO and MSTO bases as well as hybrid ones (MSTO for one atom and STO for the other) for three heteronuclear molecules: \ce{LiH}, \ce{BeH} (Fig. \ref{fig:EvsrLiHandBeH}), and \ce{BeH2} (Fig. \ref{fig:EvsrBeH2andCHplus}). For LiH near equilibrium, using MSTO bases on both atoms gives a slightly lower FCI energy than 6-31G, whereas 6-31G gives the lowest HF energy (see Fig. \ref{fig:EvsrLiHandBeH}(a)). We also observe that the MSTO-MSTO, 6-31G, and hybrid bases all predict comparable equilibrium bond lengths at the FCI level, all slightly longer than the one predicted using STO-STO basis sets. This observation is consistent with that from an earlier work \cite{de2008atomic}. 

It is important to note that the MSTO-$k$G basis sets employed in these molecular calculations are optimized for isolated atoms. Improvements in atomic ground state energies do not necessarily translate into improvements in molecular energy differences, equilibrium bond lengths, or dissociation energies. The molecular results presented here therefore serve to assess the transferability, as well as the limitations, of the atom-optimized basis sets. 

\begin{figure*}[t]
\centering
\setlength{\tabcolsep}{6pt}

\begin{tabular}{cc}

\includegraphics[width=0.45\textwidth]{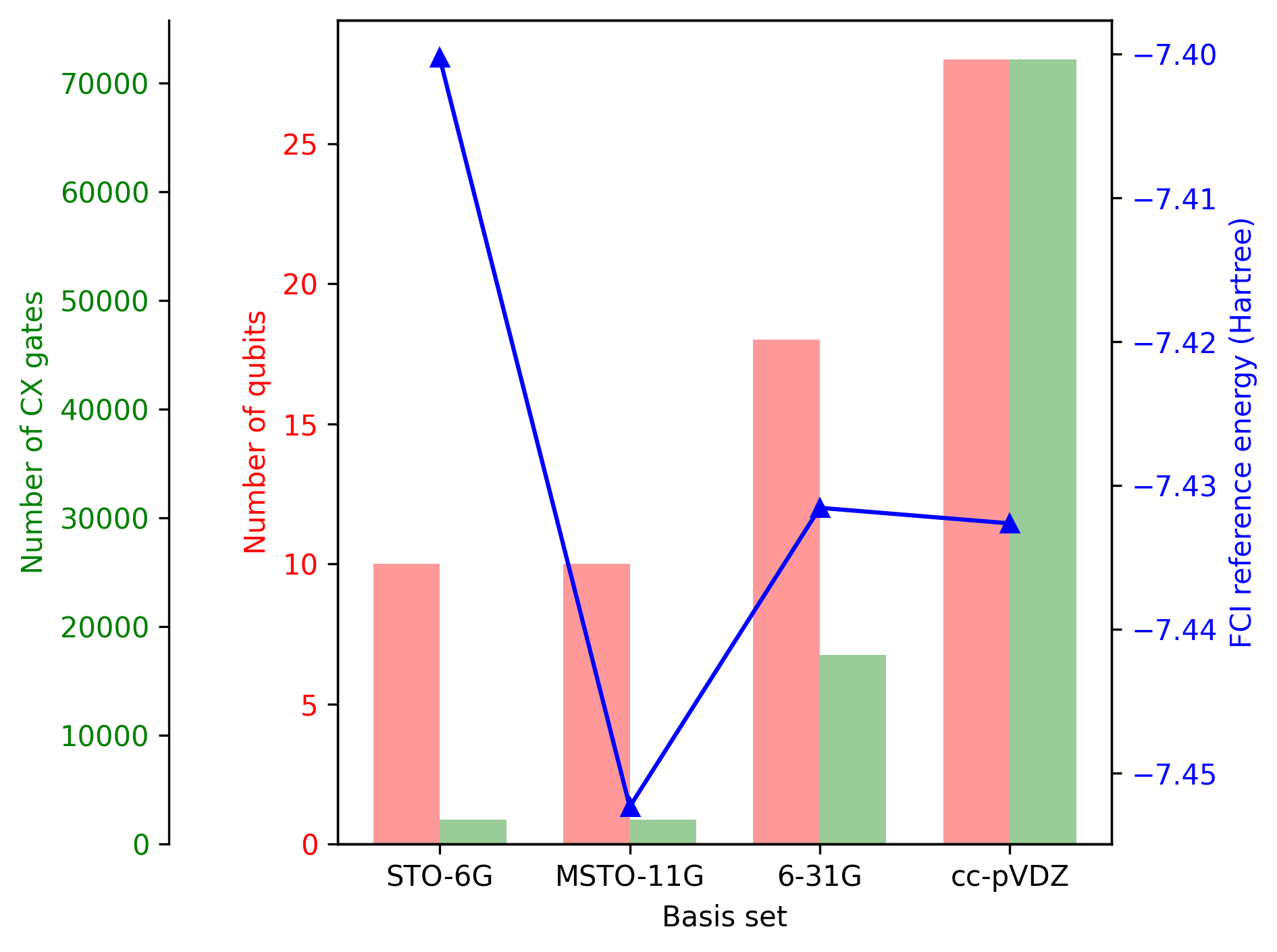}&
\includegraphics[width=0.45\textwidth]{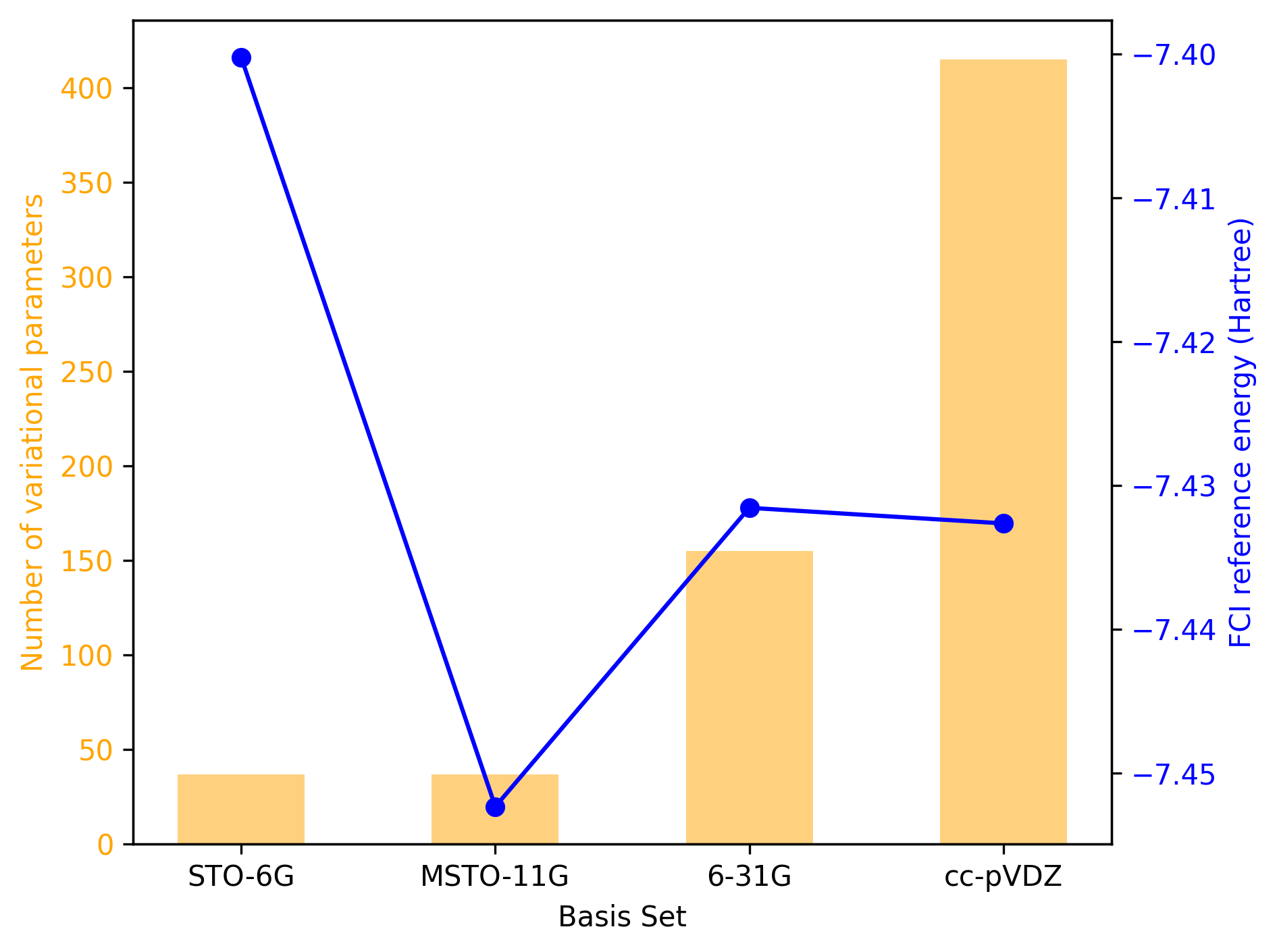}\\
(a)&(b) \\ 
\includegraphics[width=0.45\textwidth]{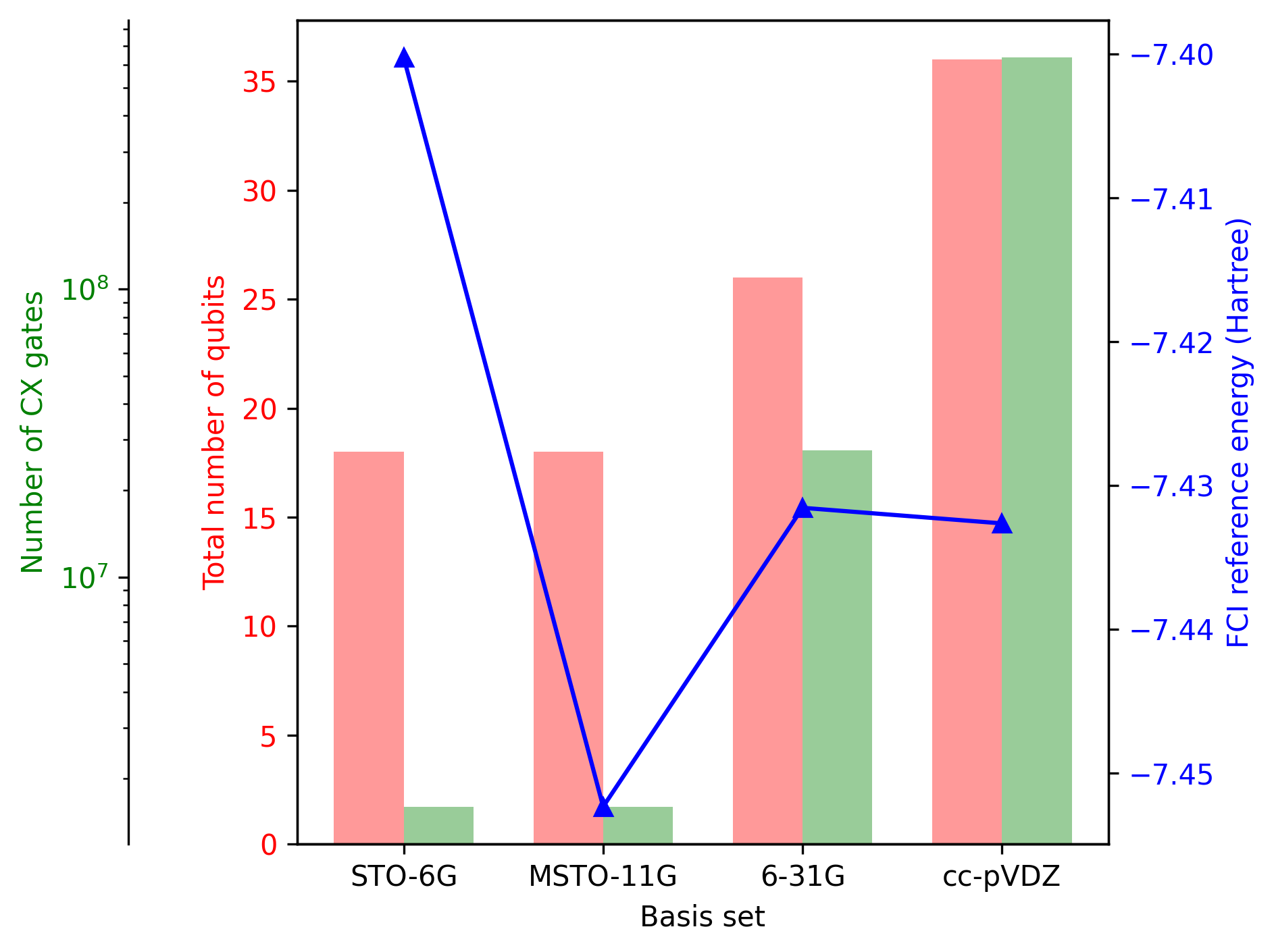}&
\includegraphics[width=0.45\textwidth]{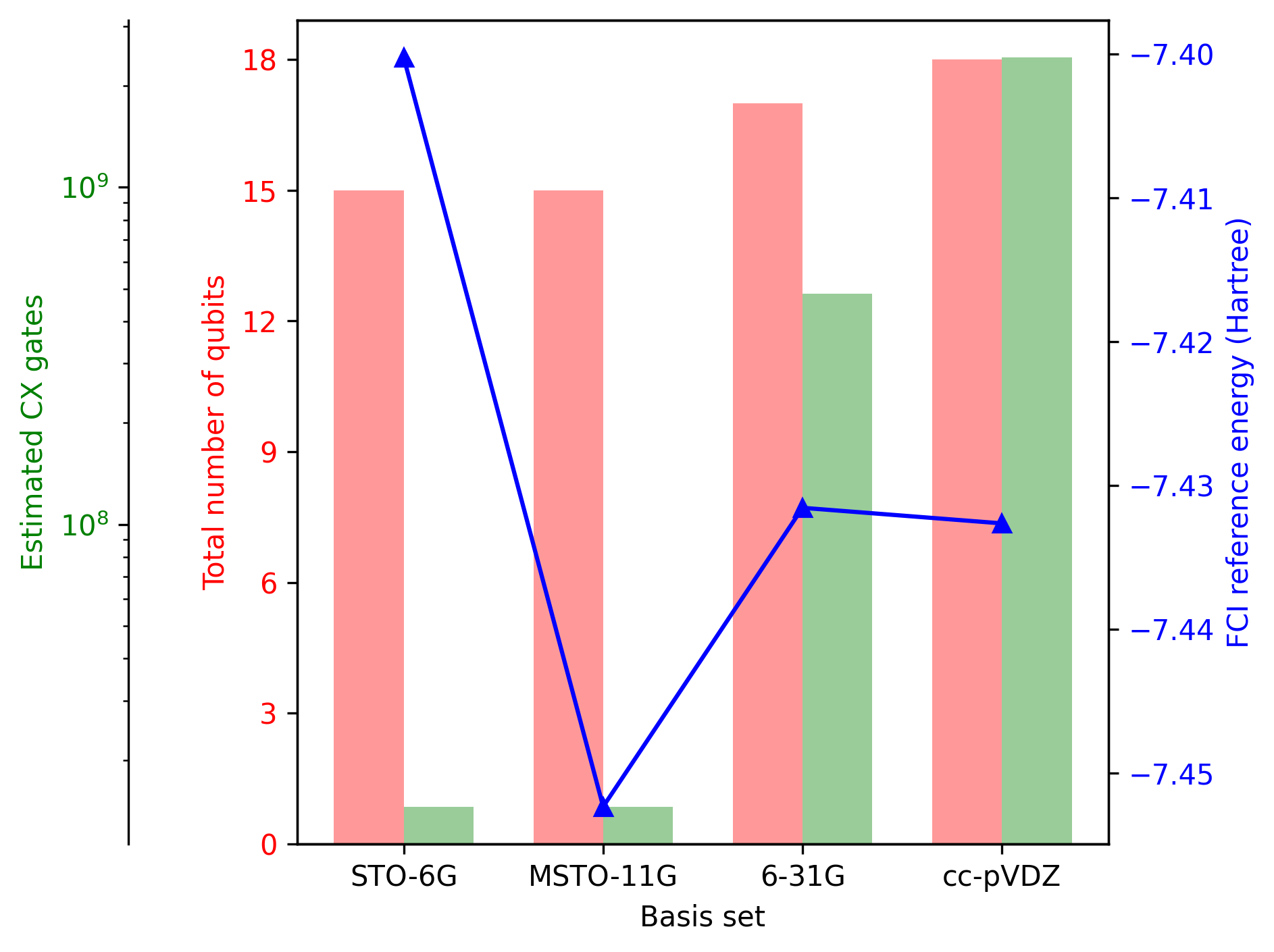}\\
(c)&(d) \\ 

\end{tabular}

\caption{Comparison of illustrative circuit resource
estimates across different basis sets for the Li atom.
The top row presents our results for VQE (circuit
resources and the number of variational parameters), while the bottom row shows the QPE-CASCI (sub-figure (c)) and HHL-LCCSD (sub-figure (d)) results
(total number of qubits and CX gates). The energies presented in all panels are classically computed FCI reference energies and not the outputs of the respective quantum algorithms. } 
\label{fig:combined_resources}

\end{figure*} 

\section{Resource estimation in quantum algorithms with our basis sets} \label{scaling}

We now discuss estimates of the quantum resources (number of CX gates and number of qubits) required for carrying out a VQE-UCCSD ground state energy calculation. We then perform the same exercise for the fault-tolerant era algorithms, QPE-CASCI and HHL-LCCSD approaches. Carrying out a detailed resource estimate (such as the number of physical qubits and its trade-off with runtime, etc) for these two algorithms is outside the scope of our work, and we defer it for a future study. We restrict ourselves only to estimates of the number of logical qubits and the number of CX gates (and logical $T$ gates) and comment on how choice of single-particle basis can affect the estimates in relation to the predicted quality of energies. 

Following estimates for the number of logical qubits and the CX gate count, we pick \ce{Li} as a representative system (which gave the best energies with our MSTO bases) and compare these two quantities against the FCI ground state energies obtained using STO-6G, MSTO-11G, 6-31G, and cc-pVDZ bases. We recall that ideally, we would want a single particle basis set to require relatively fewer qubits and incur fewer CX gates while yielding better (lower) energies. 

We note a few points about the scope of this analysis: 

\begin{enumerate} 
\item For our analysis, we only consider the quantum algorithms in their vanilla form. This is for simplicity, as often, reduction in quantum resources that occur in different variants of an algorithm either incurs classical computing overheads that complicate estimating costs (for example, a recently introduced quantum-information inspired ansatz for VQE lowers gate and parameter counts but requires knowing the logarithm of an approximately computed density matrix \cite{Kalam2026}) or the improved variant is at the level of showing that such an algorithm exists and for which circuit designs may not be obvious (the Childs-Kothari-Somma improvement over HHL \cite{CKS2017}). 
\item For the \ce{Li} atom example, we use the $\{RZ, X, SX, CX\}$ basis as the native gate set. Using other choices would change the gate counts but not the overall observed trends. 
\item We do not compute energies using the quantum
algorithms themselves. Instead, we use the classically
computed FCI energies as common reference values
to assess the quality of the respective basis sets
alongside the estimated circuit resources. We also
note that a small correlation energy relative to the
total energy does not necessarily imply a small error
relative to the energy differences being compared. 
\end{enumerate} 

\subsection{VQE-UCCSD} 

The number of qubits in VQE ($n$) is the number of spin orbitals, $n_s$. We use the UCCSD ansatz, and hence the number of parameters scales in general as $\sim n^4$. In practice, the number is lesser as we shall see in our numerical example. Upon Trotterizing (first order and first step, chosen for the purposes of simplicity), each of the $n^4$ terms of the form $e^{i\theta_i P_i}$, where $\theta_i$ is the $i^{\mathrm{th}}$ parameter/amplitude (usually up to a factor of 2 or its integral multiple) and $P_i$ is an $n-$qubit Pauli word, can be compactly represented using a Pauli gadget that has $\sim 2(n-1)$ CX gates. This gives an $\mathcal{O}(n^5)$ estimate for the number of CX gates in one preparation of the UCCSD ansatz state under the assumptions above. The CX counts for this are reported in Table~\ref{tab:VQEresources} of Appendix C. Estimating the energy at a given set of variational parameters requires repeated state preparations for the measurement shots. 

For Li, STO-6G and MSTO-11G give the same UCCSD circuit size: 10 qubits, 37 variational parameters, and 2246 CX gates per ansatz preparation (Fig.~\ref{fig:combined_resources}(a) and (b), as well as Table~\ref{tab:VQEresources} of Appendix C). The classically computed FCI reference energy is 52.107 mHa lower with MSTO-11G than with STO-6G. For comparison, 6-31G requires 18 qubits, 155 parameters, and 17,420 CX gates, while cc-pVDZ requires 28 qubits, 415 parameters, and 72,185 CX gates. For this atom, we recall that the MSTO-11G FCI reference energy is also lower than the corresponding 6-31G and cc-pVDZ values. 

\subsection{QPE-CASCI} 

We now move to gate counts and energies from QPE. The number of qubits ($n$) here is the sum of the number of spin orbitals ($n_s$) and the number of clock register qubits ($n_r$) to capture the eigenvalues to within $n_r$ digits of precision in their binary representation. We arrive at the estimate for CX gate count by: 

\begin{itemize}
\item expanding $U=e^{iHt}=e^{i \sum_{i=1}^{\ell_H}H_it} \approx \prod_{i=1}^{\ell_H}e^{iH_it}$ (we assume first order Trotterization with Trotter step set to 1 for simplicity), 
\item estimating the CX count for each Pauli gadget representing $e^{iH_i t}$ as $2(n_s-1)$, assuming that it acts non-trivially on all $n_s$ qubits \cite{whitfield2011simulation}. Thus, $\ell_H$ such gadgets give an estimate of $2\ell_H(n_s-1)$ CX gates for $U$, 
\item using the decomposition in
Ref.~\cite{blunt2024compilation}, in which each controlled-Pauli rotation is represented using two Pauli gadgets on the enlarged register of $n_s+1$ qubits, giving an estimate of $4\ell_H n_s$ CX gates per controlled-$U$, and finally 
\item we implement the controlled powers $U^{2^j}$ by repeated applications of controlled-$U$. For $n_r$ clock qubits, this requires $\sum_{j=0}^{n_r-1} 2^j=2^{n_r}-1$ applications. In the numerical resource estimates, we approximate
this count by $2^{n_r}$, giving approximately $2^{n_r}\,4\ell_H n_s$ CX gates for controlled evolution. 
\end{itemize} 

The QPE algorithm also has an inverse quantum Fourier transform (IQFT) module, which comprises $n_r(n_r-1)/2$ controlled-phase gates and $\lfloor n_r/2\rfloor$ SWAP gates. Decomposing each controlled-phase gate into two CX gates and each SWAP gate into three CX gates, the CX gate count from the IQFT module is given by $C_{\rm IQFT}=n_r(n_r-1)+3\lfloor n_r/2\rfloor$. 

Therefore, the CX gate estimate for QPE is $
C_{\rm QPE}\simeq 2^{n_r}4\ell_H n_s+
n_r(n_r-1)+3\lfloor n_r/2\rfloor
$. We make a few remarks at this juncture: 

\begin{itemize}
    \item For a fixed number of clock register qubits $n_r$, the CX gate count is governed primarily by the number of spin orbitals $n_s$ and the number of Pauli terms $\ell_H$. However, maintaining a fixed target energy accuracy may require increasing $n_r$ with system size. 
    \item Typically, for molecular Hamiltonians, $\ell_H = \mathcal{O}(n_s^4)$, due to the Hamiltonian containing at most pairwise Coulomb interactions. Upon Jordan-Wigner transforming the Hamiltonian, one typically retains this scaling. In practice, the number of terms are fewer than this worst case estimate. 
    \item Although we focus on CX gate count, since all the circuits would be executed at the logical level, the $T$ gate count is also very important to estimate, as they cannot be realized transversally in popular codes such as the surface code \cite{fowler2012surface} and the Steane code \cite{steane1996error}. We note that each Pauli gadget has one arbitrary angle rotation gate ($RZ(\theta)$), and thus the entire circuit has $\sim \ell_H/\epsilon$ such rotation gates. Here, $\epsilon$ is the error due to inadequate representation of eigenvalues of the Hamiltonian via limited $n_r$, and goes as $\sim 2^{-n_r}$. For an admissible error of $\delta$ in transpiling an arbitrary angle $Z$ rotation into one-qubit Cliffords and $T$ gates, one requires $\sim \mathrm{log}(1/\delta)$ $T$ gates. Thus, the $T$ count for QPE-CASCI is $\sim \frac{\ell_H}{\epsilon}\mathrm{log}(\frac{1}{\delta})$. 
\end{itemize} 

We now turn our attention to resource estimates from QPE-CASCI, presented in Fig. \ref{fig:combined_resources}(c), where we observe trends similar to those with the VQE algorithm, where the requirement on the number of logical qubits and the number of CX gates are fewer for MSTO bases than for the 6-31G (about 70\% of the qubits and about 6\% of the CX gates) or cc-pVDZ (about 50\% of the qubits and about 0.25\% of the CX gates) bases (also see supporting data in Table \ref{tab:QPEresources} of Appendix C). 

\subsection{HHL-LCCSD} 

Finally, we take a look at the CX gate counts and energies from the HHL-LCCSD algorithm. The algorithm itself requires two QPE modules and a controlled-rotation module (which we shall call $CR$ for brevity), and requires 3 qubit registers, the state register with $n_b$ qubits, the clock register containing $n_r$ qubits, and the HHL ancilla register with a single qubit. We now estimate the qubit count. Let $d_A$ denote the dimension of the SRLCCSD $A$ matrix. For the Li atom, which we pick as a representative example, with $N_\alpha=2$ and $N_\beta=1$ where $N_\alpha$ is the number of $\alpha$ electrons and $N_\beta$ beta electron count, the number of independent spin-conserving single and double excitation amplitudes is given by $d_A =
2(m-2)+(m-1)+\binom{m-2}{2}+2(m-2)(m-1)$, where $m=n_s/2$ is the number of spatial orbitals. The resulting dimensions are $d_A=37,155$, and $415$ for the minimal, 6-31G, and cc-pVDZ bases, respectively. Padding the matrix dimension to the next nearest power of two gives $D_A=2^{n_b}$, with $n_b=\log_2 D_A=\lceil\log_2 d_A\rceil$. In the above equation, $n_b$ is expressed as a logarithm of $D_A$ due to amplitude encoding. 

In general, the dimension of the LCCSD matrix is bounded by $d_A\leq n_s^4$. The corresponding padded matrix dimension therefore satisfies $
D_A=2^{\lceil\log_2 d_A\rceil}
\leq 2^{\lceil\log_2(n_s^4)\rceil}$. Therefore, the number of state register qubits satisfies $n_b=\lceil\log_2d_A\rceil \leq\lceil4\log_2n_s\rceil
=\mathcal{O}(\log n_s)$. 

We now move to the gate counts. As seen earlier, the CX count for one QPE module is estimated as $
C_{\rm QPE}\simeq 2^{n_r}4\ell_A n_b+
n_r(n_r-1)+3\lfloor n_r/2\rfloor
$, where $\ell_A$ denotes the number of Pauli terms in the decomposition of the $A$ matrix. Since HHL involves a QPE module and its inverse, the CX count from them is twice the above estimate. Adding the controlled-rotation module, the total CX gate count is estimated to be 

\begin{equation}
C_{\rm HHL}\simeq
2\left[
2^{n_r}4\ell_A n_b+
n_r(n_r-1)+3\lfloor n_r/2\rfloor
\right]+C_{\rm CR}, 
\end{equation} 

where $C_{\mathrm{CR}} \simeq 2^{n_r+2}-4n_r-4$  \cite{mottonen2004transformation}. 

We add a few remarks: 

\begin{itemize}
    \item Our estimates exclude the cost of preparing a unitary $V$ whose action on $|0\rangle_{n_b}$ yields $|b\rangle$. The possibility of efficient input state preparation is an open problem in QPE and algorithms that build on it. It could also occur in VQE when we employ multi-reference ans\"atze (for example, see Refs. \cite{zade2025quantum,chawla2025trapped}). 
    \item We also exclude the cost of extracting a molecular property as a feature of the solution ket, $|x\rangle \approx A^{-1}|b\rangle$ that HHL prepares. Let us consider the example of ground state energy calculation using the CSWAP test, whose inputs are $|b\rangle$ and $|x\rangle$ (output of HHL). The circuit requires $n_b$ CSWAP gates, and since each CSWAP is known to have a straightforward decomposition into 8 CX gates, the net CX gate count from the feature extraction module for obtaining ground state energies is $\approx 8 n_b$. We note that for positive definite $A$, the sign of the LCCSD correlation energy is guaranteed to be negative. Assuming that $A>0$ for the single reference LCCSD case (which is what we are considering here) \cite{peniel2026}, the squared overlap obtained from the CSWAP test, along with the appropriate normalization factors, suffices to recover the correlation energy. 
    \item We expect the logical $T$ count to scale similarly to that for QPE, that is, $\sim \frac{\ell_A}{\epsilon}\log(\frac{1}{\delta})$, where $\epsilon \sim 2^{-n_r}$ denotes the phase resolution and $\delta$ is the tolerance in synthesizing an arbitrary angle rotation gate. The CR module is relatively cheaper, since it involves $\sim 2^{n_r}$ rotation gates, yielding a $T$ count of $\sim \frac{1}{\epsilon}\log(\frac{1}{\delta})$. 
\end{itemize}

We now move to numerical estimates of CX counts and ground state energies. For the representative example of Li atom, we construct the CISD Hamiltonian
matrix using an all-electron ROHF reference and including all virtual orbitals. The resulting LCCSD $A$ matrix is obtained by removing the reference
determinant row and column from the CISD Hamiltonian matrix and subtracting the HF energy from its diagonal elements. Further, we embed $A$ into a matrix of dimension $D_A$ as $\widetilde{A}=A\oplus\lambda I_y$,
where $\lambda=\|A\|_2$ and $y=D_A-d_A$. The corresponding $\vec{b}$ is padded with zeros. Unlike in QPE-CASCI, where the electronic Hamiltonian is realized in its qubit form, the matrix that we simulate in HHL-LCCSD is the amplitude-encoded linear-system matrix $\widetilde{A}$. The number of Pauli terms $\ell_A$ in the decomposition of $\widetilde{A}$ is bounded by 

\begin{equation}
\ell_A \leq 4^{n_b}=D_A^2. 
\end{equation} 

For the LCCSD case, with a fixed number of occupied orbitals and an increasing number of virtual orbitals $n_v$, the matrix dimension scales as $\mathcal{O}(n_v^2)$, yielding an upper bound of $\mathcal{O}(n_v^4)$ on the number of Pauli terms. However, for the numerical estimates presented in Table \ref{tab:HHLresources} of Appendix C, we explicitly compute the Pauli decomposition of $\widetilde{A}$, using a coefficient threshold of $10^{-10}$. The CISD Hamiltonian matrix is obtained by modifying PySCF, from which we construct $A$. We find from Fig. \ref{fig:combined_resources}(d) that the results are very similar in terms of trends observed for QPE (and supporting data in Table \ref{tab:HHLresources} of Appendix C). We unsurprisingly find that the STO-6G and MSTO-11G bases yield identical Pauli-term counts and hence identical CX-gate estimates within our circuit model. However, MSTO-11G gives a substantially lower LCCSD ground state energy. Furthermore, within our circuit resource estimation framework, the estimated CX counts for MSTO-11G are lower than those for the 6-31G and cc-pVDZ bases by factors of approximately 33.2 and 166.4, respectively, while simultaneously yielding a lower FCI reference energy. 

We note that the resource estimates presented in this section are intended to illustrate circuit sizes across the considered bases, rather than to
establish algorithmic performance at a fixed
target energy accuracy. Furthermore, for QPE-CASCI and HHL-LCCSD, we assume first-order Trotterization with one Trotter step per evolution operator and fix the number of clock register qubits to 8. Also, we assume that the electronic Hamiltonian $H$ and
the padded LCCSD matrix $\widetilde{A}$ are
rescaled such that their spectral norms do not
exceed unity, and use unit evolution time. 
Thus, the same phase resolution need
not correspond to the same energy resolution
across different basis sets. 

We reiterate that the estimates exclude input state preparation, measurement costs, repeated circuit executions, and QEC overheads. In the case of HHL, the condition number of the $A$ matrix and the associated success probability also
need to be accounted for in a comprehensive
resource analysis. 

\section{Comparison with earlier works in literature} \label{lit} 

Now that we have discussed our results, it is in order to survey the existing related works in literature, and compare our work in relation to them. We note that earlier efforts towards refining and generating better single particle basis sets were not typically aimed at quantum computing applications, with very few exceptions. 

We also note a recent work by Hakim et al \cite{sto4}, where the authors optimize the exponents of Slater-type orbitals with the goal of improving electronic structure calculations within the quantum computing framework, while retaining the number of qubits. Although their motivation is related to ours, their work employs STOs directly, whereas we specifically focus on improving the STO-$k$G family of contracted Gaussian minimal basis sets. Our comparisons therefore concentrate on earlier developments of STO-$k$G and related optimized Gaussian minimal bases. 

We begin with the very first work that introduced the minimal basis sets \cite{hehre1969self}, and in particular, the basis sets STO-3G through STO-6G for H through F with the exception of He. They use the method of least squares to obtain the exponents and contraction coefficients. We find that our MSTO-$k$G results are consistently lower than their STO-$k$G and their STO (the actual Slater type orbitals) results. It is also important to note that the STO-$k$G results from Ref. \cite{hehre1969self} and those obtained using the basis sets from Basis Set Exchange database differ slightly, with the latter being lower than the former. In our data, when we make a comparison between STO-$k$G and MSTO-$k$G bases, we use the refined STO-$k$G values provided in the Basis Set Exchange database. 

The work by Tavouktsoglou and Huzinaga \cite{tavouktsoglou1980new} improve upon the STO-3G basis by using the method of least squares. We find that our MSTO-3G results are lower than those obtained in this work for all of the atoms that they consider: Li through F. 

\begin{figure}
    \centering
    \includegraphics[width=0.9\linewidth]{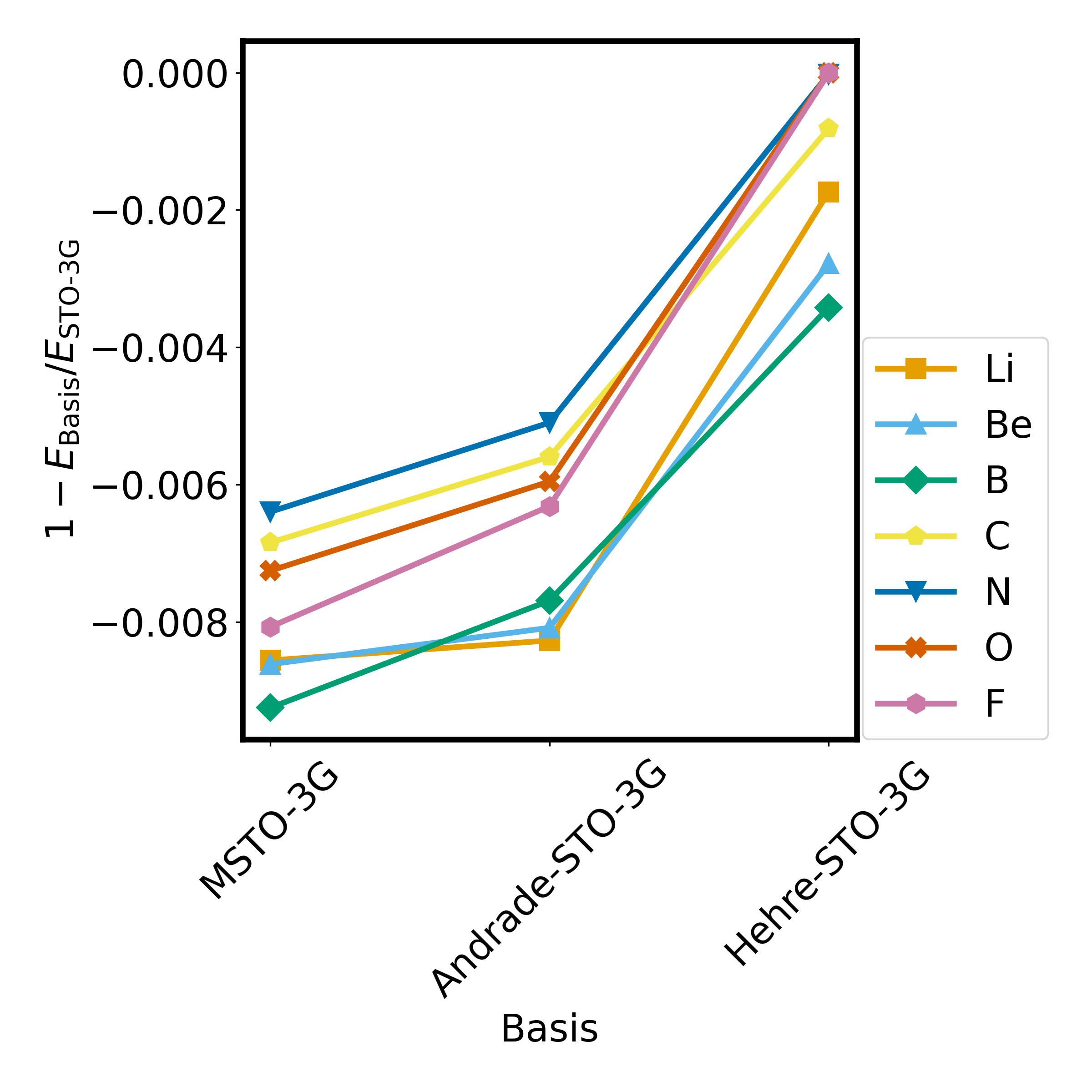}\\
    \includegraphics[width=0.9\linewidth]{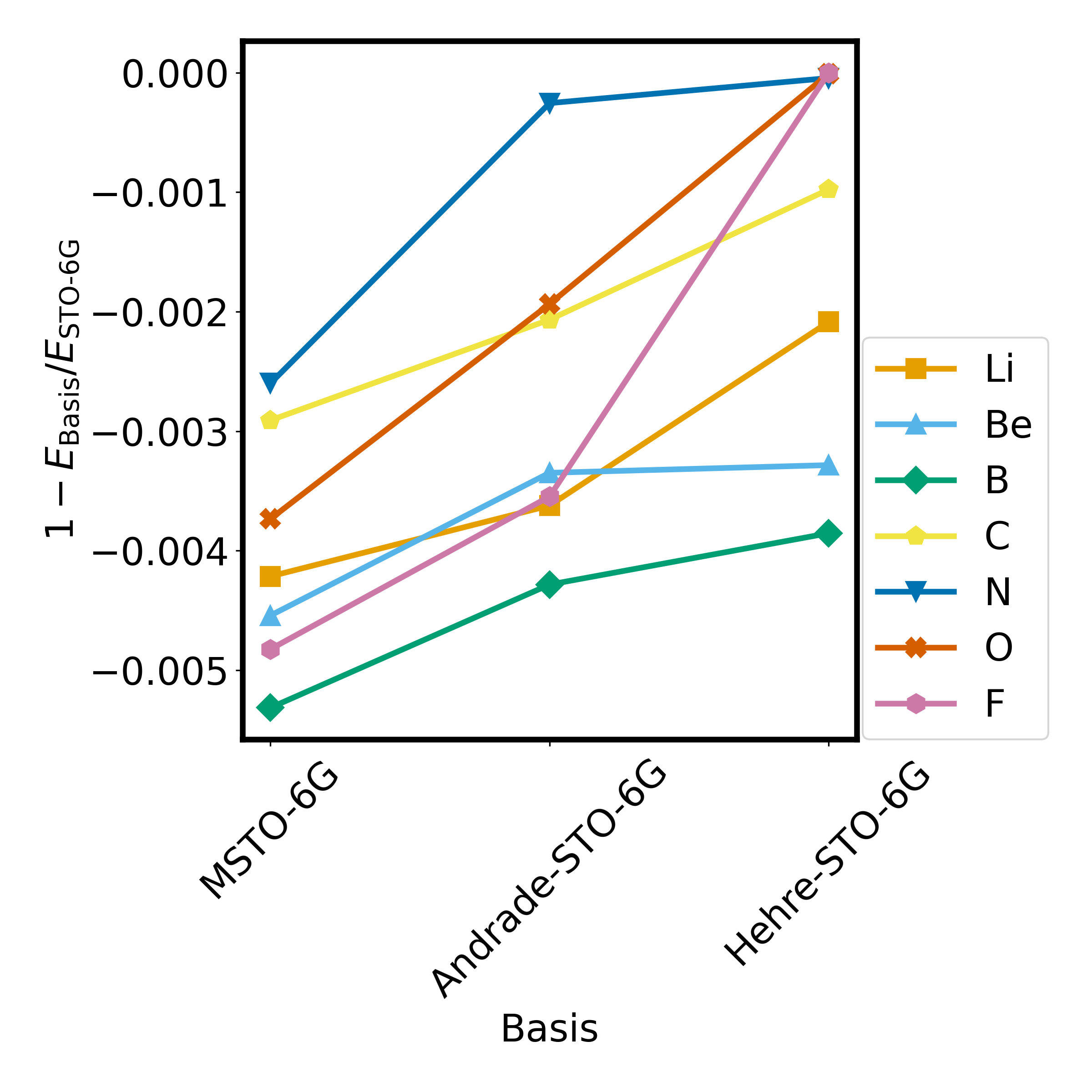}\\
    \caption{Relative changes in Hartree-Fock energies for Li through F, expressed as $1-E_{\mathrm{basis}}^{\mathrm{HF}}/E_{\mathrm{STO}\text{-}k\mathrm{G}}^{\mathrm{HF}}$, for $k=3$ (top) and $k=6$ (bottom), using our basis as well as those from Refs. \cite{hehre1969self} and \cite{de2008atomic}. The reference energy for each atom is the Hartree-Fock energy obtained with the corresponding standard STO-$k$G basis. Negative values indicate a lower energy than the reference. }
    \label{fig:EcompHehreAndrade}
\end{figure} 

We now discuss a work \cite{de2008atomic} that is partially similar in spirit to ours, where the authors optimize STO-3G and STO-6G basis sets for the first row elements of the periodic table, using a generalized simulated annealing based approach. They optimize the Hartree-Fock functional, and the parameters that are optimized are not only the exponents and contraction coefficients but also the coefficients of atomic orbitals in the linear combination of atomic orbitals (LCAO) approach. We deviate in that we use a genetic algorithm with aggressive optimization by beginning with existing STO series bases, and do so for STO-3G through 6G and extend it up to a cardinal number of 11. We find that our STO-3G and 6G and their STO-3G and 6G HF energies match exactly for all atoms, serving as a sanity check that our STO calculations are correct. At the STO-3G level, all of our MSTO HF energies are lower than theirs, except for the case of H, where our energies exactly match. Especially for larger systems considered, we obtain lower energies at $\sim$ 2 to 172 mHa level. At the STO-6G level, all of our MSTO energies are once again lower by $\sim$ 4 to 134 mHa than theirs, except for H, where they obtain slightly better quality (lower) energy than ours (differing at $\sim 3.4 \cdot 10^{-2}$  mHa level). We also note that we extend our basis set generation beyond $k$=6, up to 11. Fig. \ref{fig:EcompHehreAndrade} compares the energies obtained for atoms \ce{Li} through \ce{F} using our MSTO-3G and MSTO-6G bases with results from their 3G and 6G quality bases in Refs. \cite{hehre1969self} and \cite{de2008atomic}, as a visual indicator of the performance of our basis sets relative to these two works. 

In Ref. \cite{magalhaes2014gaussian}, the author uses both least squares fit and energy minimization approaches for the H atom for $k$ of 2, 3, and 5, and finds that the latter approach yields lower energies. The results from this work are consistent with those obtained using MSTO-$k$G basis for $k=2$, 3, and 5. However, the work is limited to only the H atom and up to $k=5$. 

Kapusta \textit{et al} \cite{Kapusta2018ReconstructionOS} have refined the existing STO-3G bases and generated their STO(0)-3G basis sets for principal quantum numbers from 1 to 7. They employ a minimized sum of the squared deviations approach to carry out their optimization. We note that a limitation of this approach is assigning the same exponents and contraction coefficients for any atom for a given orbital. We compare our HF MSTO-$3$G results and theirs using STO(0)-3G for the atoms that we consider, namely H, C, N, O, and F, and find that our ground state energies are consistently lower than theirs. In particular, we note that since an important focus of their work was getting better energies for the atoms C, N, O, and H as they are the main building blocks of several organic molecules, one could consider employing our basis sets for large quantum chemical calculations involving molecules built out of these atoms. However, their transferability to such systems cannot be established solely from the atomic ground state energies, and thus needs detailed analyses. Finally, recalling the limitation in capturing correlation effects for minimal type bases, we recommend employing the more expensive 6-31G basis sets for such computations whenever feasible computationally. Reoptimizing the 6-31G bases with our algorithm is a possibility, but is outside the scope of the current study. Preliminary investigations in that direction indicate that the 6-31G bases are already well-optimized, and improvements are of order 1 milliHartree or smaller. 

We finally turn our attention to a relatively recent work from 2023 \cite{kwon2023adaptive}, where the authors optimize the exponents and contraction coefficients of the STO-3G basis for every bond distance to obtain potential energy surfaces of the hydrogen molecule with the goal of executing VQE on an IBM superconducting qubit quantum computer. They restrict themselves to only the hydrogen molecule and $k$=3 in their study. We do not carry out an adaptive calculation in our work but is very much possible in principle in future studies. On this note, we also add that a very interesting extension of our work could be in the direction of developing modified mixed ramp Gaussian basis sets \cite{ireland2025fully,cox2020mixed}. 

\section{Conclusion}\label{sec:conclusion} 

In summary, we have designed our memetic algorithm (a genetic algorithm followed by aggressive refinement) to construct modified minimal basis sets with cardinal numbers 2 through 11 (MSTO-$k$G; $k=2$ through 11), for atoms H through F, excluding He. The MSTO basis sets are intended for quantum computing applications, where we are expected to be limited by qubit quality in the coming few years. Our work builds on the expectation that sufficiently reoptimized minimal basis sets between $k$ of 3 and 6 should yield better (lower) energies than their existing versions, and further exploits the fact that in their contracted form, minimal basis sets with increasing cardinal numbers retain the same number of spin orbitals for a given atom, and therefore the same qubit footprint in a quantum computing calculation, while yielding better (lower) ground state energies. 

We find that our MSTO-$k$G basis sets consistently outperform the conventional STO-$k$G basis sets in predicting atomic FCI ground state energies. For H, Li, and Be, we obtain lower FCI energies than those from the 6-31G basis, while for the remaining atoms, the FCI energies are higher than the corresponding 6-31G results. This is achieved with fewer spin orbitals: 10 instead of 18 for the atoms Li through F, and 2 instead of 4 for H. Thus, our approach illustrates the potential of classical basis set optimization for improving ground state energies within a restricted quantum resource budget. 

Of particular interest are our results for Li, for which the MSTO basis sets at higher $k$ yield FCI ground state energies lower than that obtained with the cc-pVQZ basis. Specifically, the best result for Li, obtained with MSTO-10G, is lower than the cc-pVQZ FCI energy by approximately 3.846 mHa and lies within 24.382 mHa of the benchmark Hylleraas ground state energy. For Be, the MSTO-11G basis yields a lower FCI energy than the 6-31G basis but does not
outperform the Dunning basis sets considered. On the other hand, the MSTO bases inherit the limitation of not having sufficient spin orbitals for the elements to the right of the periodic table (N, O, and F); although the base Hartree-Fock energies are better than their STO and 6-31G counterparts, the FCI correlation energies from MSTO bases are near zero and consequently, the total FCI energies are higher than 6-31G FCI energies. 

We also investigate the transferability of our atom-optimized basis sets to the H$_2$, Li$_2$, C$_2$, LiH, BeH, and BeH$_2$ molecules through potential energy curve calculations. While we obtain favourable results for some molecules, notably Li$_2$ and LiH, the results also reveal limitations for other systems, particularly H$_2$ and C$_2$ near equilibrium. These observations highlight that improvements in atomic total energies do not necessarily translate into improvements in molecular energy differences or properties. 

We carry out an illustrative resource estimation exercise for the VQE-UCCSD, QPE-CASCI, and HHL-LCCSD algorithms, comparing the estimated quantum resources across different basis sets alongside their classically computed FCI reference energies for the Li atom, taken as a best-case example. Within our resource model, the CX count scales as $\mathcal{O}(n_s^5)$ for one preparation of the VQE-UCCSD ansatz and for one QPE-CASCI circuit at fixed clock-register size. With HHL-LCCSD, for a fixed number of occupied orbitals and clock register qubits, the upper bound on the CX gate count within our estimation framework scales as $\mathcal{O}(n_v^4\log n_v)$, where $n_v$ denotes the number of virtual orbitals. Thus, the choice of single-particle basis set also significantly impacts the gate counts for HHL-LCCSD. For a fixed phase resolution $\epsilon$ and single-qubit rotation gate's synthesis tolerance $\delta$, the T count estimates for QPE-CASCI scale as $\mathcal{O}(n_s^4\epsilon^{-1}\log(1/\delta))$, while the corresponding upper bound for HHL-LCCSD scales as $\mathcal{O}(n_v^4\epsilon^{-1}\log(1/\delta))$ for a fixed number of occupied orbitals. Thus, the choice of basis set significantly (polynomial in spin orbitals) impacts the gate counts for VQE, QPE, and HHL applied to chemistry. We note that these estimates illustrate the impact of basis set choice on circuit resources and do not imply a comparison of end-to-end quantum computational costs at a fixed target energy accuracy. 

We then compare our methods and bases with related works in literature that generate modified minimal basis sets, and find that our basis sets generally yield lower atomic ground state energies than those reported in the earlier works considered, at the corresponding numbers of primitives per contraction. 

Given the usefulness of classical pre-processing to reduce quantum resources, and combined with the one-time nature of generating useful basis sets for this purpose with state-of-the-art optimization algorithms, we plan to extend in future works our basis sets to larger atomic systems with newer and more high performance classical optimization methods. With improved algorithms, we also plan to reoptimize some of the larger basis sets available in literature. 

\section*{Data availability} 

The codes that we developed are open-source, and can be accessed here: \url{https://github.com/subimal/MSTO-kG}. 

\begin{acknowledgments}
VSP and SD acknowledge Ms. Palak Chawla and Ms. Disha Shetty for very useful discussions on the quantum computational aspects of the problem, and also thank Aashna, Peniel, Tushti, Kalam, Suprava, and Fazil (in no particular order) for patiently going through the manuscript and giving their views. VSP acknowledges CRG grant (CRG/2023/002558). 
\end{acknowledgments}


\bibliographystyle{apsrev4-2}
\bibliography{gtobib}


\appendix

\begin{widetext}

\section{Computed MSTO-$k$G basis sets}\label{app:basis_data} 

\subsection{$k = 2$}
\begin{tabular}{lllllll}
\toprule
Atom & $n$ & $l$ & Shell type & Primitive  & Exponent & Contraction Coeff. \\
\midrule
H & 1 & 0 & s & 1 & 1.332500 & 0.335072 \\
H & 1 & 0 & s & 2 & 0.201530 & 1.002799 \\
\bottomrule
\end{tabular}

\begin{tabular}{lllllll}
\toprule
Atom & $n$ & $l$ & Shell type & Primitive  & Exponent & Contraction Coeff. \\
\midrule
Li & 1 & 0 & s & 1 & 10.443437 & 0.329496 \\
Li & 1 & 0 & s & 2 & 1.438243 & 1.009065 \\
Li & 2 & 0 & s & 1 & 0.018208 & 0.064413 \\
Li & 2 & 0 & s & 2 & 0.054956 & 0.272310 \\
Li & 2 & 1 & p & 1 & 5.366872 & 0.028000 \\
Li & 2 & 1 & p & 2 & 1.583844 & 0.031362 \\
\bottomrule
\end{tabular}

\begin{tabular}{lllllll}
\toprule
Atom & $n$ & $l$ & Shell type & Primitive  & Exponent & Contraction Coeff. \\
\midrule
Be & 1 & 0 & s & 1 & 19.490692 & 0.180891 \\
Be & 1 & 0 & s & 2 & 2.750151 & 0.563571 \\
Be & 2 & 0 & s & 1 & 0.042760 & 0.004905 \\
Be & 2 & 0 & s & 2 & 0.125839 & 0.015434 \\
Be & 2 & 1 & p & 1 & 0.522639 & 0.178367 \\
Be & 2 & 1 & p & 2 & 0.130491 & 0.459031 \\
\bottomrule
\end{tabular}

\begin{tabular}{lllllll}
\toprule
Atom & $n$ & $l$ & Shell type & Primitive  & Exponent & Contraction Coeff. \\
\midrule
B & 1 & 0 & s & 1 & 31.391771 & 0.362265 \\
B & 1 & 0 & s & 2 & 4.497927 & 1.138905 \\
B & 2 & 0 & s & 1 & 0.080070 & 0.016973 \\
B & 2 & 0 & s & 2 & 0.237878 & 0.037366 \\
B & 2 & 1 & p & 1 & 0.936604 & 0.115695 \\
B & 2 & 1 & p & 2 & 0.178634 & 0.280743 \\
\bottomrule
\end{tabular}

\begin{tabular}{lllllll}
\toprule
Atom & $n$ & $l$ & Shell type & Primitive  & Exponent & Contraction Coeff. \\
\midrule
C & 1 & 0 & s & 1 & 46.098325 & 0.191217 \\
C & 1 & 0 & s & 2 & 6.670982 & 0.604580 \\
C & 2 & 0 & s & 1 & 0.125152 & 0.297728 \\
C & 2 & 0 & s & 2 & 0.381227 & 0.598927 \\
C & 2 & 1 & p & 1 & 1.533198 & 0.194977 \\
C & 2 & 1 & p & 2 & 0.278145 & 0.460385 \\
\bottomrule
\end{tabular}

\begin{tabular}{lllllll}
\toprule
Atom & $n$ & $l$ & Shell type & Primitive  & Exponent & Contraction Coeff. \\
\midrule
N & 1 & 0 & s & 1 & 63.565489 & 0.309760 \\
N & 1 & 0 & s & 2 & 9.263207 & 0.982451 \\
N & 2 & 0 & s & 1 & 0.559701 & 0.176877 \\
N & 2 & 0 & s & 2 & 0.181843 & 0.097210 \\
N & 2 & 1 & p & 1 & 2.294462 & 0.542664 \\
N & 2 & 1 & p & 2 & 0.408574 & 1.259188 \\
\bottomrule
\end{tabular}

\begin{tabular}{lllllll}
\toprule
Atom & $n$ & $l$ & Shell type & Primitive  & Exponent & Contraction Coeff. \\
\midrule
O & 1 & 0 & s & 1 & 83.878966 & 0.129142 \\
O & 1 & 0 & s & 2 & 12.286765 & 0.410711 \\
O & 2 & 0 & s & 1 & 0.767709 & 0.146798 \\
O & 2 & 0 & s & 2 & 0.246543 & 0.080930 \\
O & 2 & 1 & p & 1 & 3.123926 & 0.269617 \\
O & 2 & 1 & p & 2 & 0.539346 & 0.609357 \\
\bottomrule
\end{tabular}

\begin{tabular}{lllllll}
\toprule
Atom & $n$ & $l$ & Shell type & Primitive  & Exponent & Contraction Coeff. \\
\midrule
F & 1 & 0 & s & 1 & 106.959844 & 0.207832 \\
F & 1 & 0 & s & 2 & 15.729023 & 0.661991 \\
F & 2 & 0 & s & 1 & 1.015465 & 0.043129 \\
F & 2 & 0 & s & 2 & 0.327844 & 0.025017 \\
F & 2 & 1 & p & 1 & 4.117196 & 0.219395 \\
F & 2 & 1 & p & 2 & 0.700375 & 0.489066 \\
\bottomrule
\end{tabular}

\subsection{$k = 3$}
\begin{tabular}{lllllll}
\toprule
Atom & $n$ & $l$ & Shell type & Primitive  & Exponent & Contraction Coeff. \\
\midrule
H & 1 & 0 & s & 1 & 4.492911 & 0.062664 \\
H & 1 & 0 & s & 2 & 0.680729 & 0.362123 \\
H & 1 & 0 & s & 3 & 0.151340 & 0.574683 \\
\bottomrule
\end{tabular}

\begin{tabular}{lllllll}
\toprule
Atom & $n$ & $l$ & Shell type & Primitive  & Exponent & Contraction Coeff. \\
\midrule
Li & 1 & 0 & s & 1 & 34.979819 & 0.077973 \\
Li & 1 & 0 & s & 2 & 5.191709 & 0.420472 \\
Li & 1 & 0 & s & 3 & 1.059549 & 0.702156 \\
Li & 2 & 0 & s & 1 & 0.739885 & -0.095207 \\
Li & 2 & 0 & s & 2 & 0.067459 & 0.659570 \\
Li & 2 & 0 & s & 3 & 0.026022 & 0.425545 \\
Li & 2 & 1 & p & 1 & 5.597350 & 0.254213 \\
Li & 2 & 1 & p & 2 & 0.012414 & 0.000163 \\
Li & 2 & 1 & p & 3 & 1.581591 & 0.273436 \\
\bottomrule
\end{tabular}

\begin{tabular}{lllllll}
\toprule
Atom & $n$ & $l$ & Shell type & Primitive  & Exponent & Contraction Coeff. \\
\midrule
Be & 1 & 0 & s & 1 & 65.687087 & 0.102922 \\
Be & 1 & 0 & s & 2 & 9.803806 & 0.569129 \\
Be & 1 & 0 & s & 3 & 2.042436 & 0.958605 \\
Be & 2 & 0 & s & 1 & 1.877925 & -0.078147 \\
Be & 2 & 0 & s & 2 & 0.156694 & 0.561264 \\
Be & 2 & 0 & s & 3 & 0.056552 & 0.394628 \\
Be & 2 & 1 & p & 1 & 1.024971 & 0.107598 \\
Be & 2 & 1 & p & 2 & 0.255234 & 0.514764 \\
Be & 2 & 1 & p & 3 & 0.089890 & 0.431360 \\
\bottomrule
\end{tabular}

\begin{tabular}{lllllll}
\toprule
Atom & $n$ & $l$ & Shell type & Primitive  & Exponent & Contraction Coeff. \\
\midrule
B & 1 & 0 & s & 1 & 105.903536 & 0.080683 \\
B & 1 & 0 & s & 2 & 15.835665 & 0.452640 \\
B & 1 & 0 & s & 3 & 3.340707 & 0.764567 \\
B & 2 & 0 & s & 1 & 3.519756 & -0.094484 \\
B & 2 & 0 & s & 2 & 0.287910 & 0.670508 \\
B & 2 & 0 & s & 3 & 0.096851 & 0.535122 \\
B & 2 & 1 & p & 1 & 2.576456 & 0.122402 \\
B & 2 & 1 & p & 2 & 0.520791 & 0.538773 \\
B & 2 & 1 & p & 3 & 0.130548 & 0.705790 \\
\bottomrule
\end{tabular}

\begin{tabular}{lllllll}
\toprule
Atom & $n$ & $l$ & Shell type & Primitive  & Exponent & Contraction Coeff. \\
\midrule
C & 1 & 0 & s & 1 & 155.296401 & 0.063022 \\
C & 1 & 0 & s & 2 & 23.288858 & 0.356559 \\
C & 1 & 0 & s & 3 & 4.947649 & 0.602982 \\
C & 2 & 0 & s & 1 & 5.743665 & -0.141755 \\
C & 2 & 0 & s & 2 & 0.458845 & 0.991060 \\
C & 2 & 0 & s & 3 & 0.150745 & 0.850078 \\
C & 2 & 1 & p & 1 & 4.151182 & 0.098060 \\
C & 2 & 1 & p & 2 & 0.848398 & 0.407291 \\
C & 2 & 1 & p & 3 & 0.200680 & 0.541462 \\
\bottomrule
\end{tabular}

\begin{tabular}{lllllll}
\toprule
Atom & $n$ & $l$ & Shell type & Primitive  & Exponent & Contraction Coeff. \\
\midrule
N & 1 & 0 & s & 1 & 213.361835 & 0.081482 \\
N & 1 & 0 & s & 2 & 32.067919 & 0.462794 \\
N & 1 & 0 & s & 3 & 6.854283 & 0.780701 \\
N & 2 & 0 & s & 1 & 8.122134 & -0.084313 \\
N & 2 & 0 & s & 2 & 0.692531 & 0.540436 \\
N & 2 & 0 & s & 3 & 0.225628 & 0.550794 \\
N & 2 & 1 & p & 1 & 6.061023 & 0.123704 \\
N & 2 & 1 & p & 2 & 1.248199 & 0.498304 \\
N & 2 & 1 & p & 3 & 0.289500 & 0.652939 \\
\bottomrule
\end{tabular}

\begin{tabular}{lllllll}
\toprule
Atom & $n$ & $l$ & Shell type & Primitive  & Exponent & Contraction Coeff. \\
\midrule
O & 1 & 0 & s & 1 & 282.356352 & 0.088788 \\
O & 1 & 0 & s & 2 & 42.456533 & 0.507544 \\
O & 1 & 0 & s & 3 & 9.100558 & 0.859179 \\
O & 2 & 0 & s & 1 & 11.337381 & -0.086138 \\
O & 2 & 0 & s & 2 & 0.911652 & 0.597201 \\
O & 2 & 0 & s & 3 & 0.289879 & 0.530326 \\
O & 2 & 1 & p & 1 & 8.110314 & 0.110116 \\
O & 2 & 1 & p & 2 & 1.676526 & 0.425669 \\
O & 2 & 1 & p & 3 & 0.373068 & 0.551263 \\
\bottomrule
\end{tabular}

\begin{tabular}{lllllll}
\toprule
Atom & $n$ & $l$ & Shell type & Primitive  & Exponent & Contraction Coeff. \\
\midrule
F & 1 & 0 & s & 1 & 359.860551 & 0.082573 \\
F & 1 & 0 & s & 2 & 54.172371 & 0.473389 \\
F & 1 & 0 & s & 3 & 11.647123 & 0.801526 \\
F & 2 & 0 & s & 1 & 14.794239 & -0.094705 \\
F & 2 & 0 & s & 2 & 1.189262 & 0.659799 \\
F & 2 & 0 & s & 3 & 0.373855 & 0.579782 \\
F & 2 & 1 & p & 1 & 10.492940 & 0.092888 \\
F & 2 & 1 & p & 2 & 2.178671 & 0.349654 \\
F & 2 & 1 & p & 3 & 0.476030 & 0.447176 \\
\bottomrule
\end{tabular}

\subsection{$k = 4$}
\begin{tabular}{lllllll}
\toprule
Atom & $n$ & $l$ & Shell type & Primitive  & Exponent & Contraction Coeff. \\
\midrule
H & 1 & 0 & s & 1 & 12.433109 & 0.013700 \\
H & 1 & 0 & s & 2 & 1.903457 & 0.093517 \\
H & 1 & 0 & s & 3 & 0.436273 & 0.317705 \\
H & 1 & 0 & s & 4 & 0.120787 & 0.326516 \\
\bottomrule
\end{tabular}

\begin{tabular}{lllllll}
\toprule
Atom & $n$ & $l$ & Shell type & Primitive  & Exponent & Contraction Coeff. \\
\midrule
Li & 1 & 0 & s & 1 & 97.695864 & 0.032939 \\
Li & 1 & 0 & s & 2 & 14.820762 & 0.218052 \\
Li & 1 & 0 & s & 3 & 3.266465 & 0.701671 \\
Li & 1 & 0 & s & 4 & 0.834229 & 0.786918 \\
Li & 2 & 0 & s & 1 & 0.730454 & -0.067421 \\
Li & 2 & 0 & s & 2 & 0.015888 & 0.071705 \\
Li & 2 & 0 & s & 3 & 0.034784 & 0.334934 \\
Li & 2 & 0 & s & 4 & 0.073672 & 0.349323 \\
Li & 2 & 1 & p & 1 & 5.997339 & 0.425393 \\
Li & 2 & 1 & p & 2 & 1.860719 & 0.675672 \\
Li & 2 & 1 & p & 3 & 26.457155 & 0.039499 \\
Li & 2 & 1 & p & 4 & 0.009098 & 0.000800 \\
\bottomrule
\end{tabular}

\begin{tabular}{lllllll}
\toprule
Atom & $n$ & $l$ & Shell type & Primitive  & Exponent & Contraction Coeff. \\
\midrule
Be & 1 & 0 & s & 1 & 187.952501 & 0.028393 \\
Be & 1 & 0 & s & 2 & 28.348333 & 0.192973 \\
Be & 1 & 0 & s & 3 & 6.280651 & 0.642124 \\
Be & 1 & 0 & s & 4 & 1.624882 & 0.732551 \\
Be & 2 & 0 & s & 1 & 1.850640 & -0.088707 \\
Be & 2 & 0 & s & 2 & 0.023583 & 0.041457 \\
Be & 2 & 0 & s & 3 & 0.161387 & 0.593620 \\
Be & 2 & 0 & s & 4 & 0.062840 & 0.446471 \\
Be & 2 & 1 & p & 1 & 2.727388 & 0.030689 \\
Be & 2 & 1 & p & 2 & 0.519016 & 0.194603 \\
Be & 2 & 1 & p & 3 & 0.190163 & 0.446290 \\
Be & 2 & 1 & p & 4 & 0.078765 & 0.285602 \\
\bottomrule
\end{tabular}

\begin{tabular}{lllllll}
\toprule
Atom & $n$ & $l$ & Shell type & Primitive  & Exponent & Contraction Coeff. \\
\midrule
B & 1 & 0 & s & 1 & 303.839820 & 0.023912 \\
B & 1 & 0 & s & 2 & 45.839141 & 0.163899 \\
B & 1 & 0 & s & 3 & 10.211701 & 0.552840 \\
B & 1 & 0 & s & 4 & 2.669101 & 0.634086 \\
B & 2 & 0 & s & 1 & 3.759129 & -0.081876 \\
B & 2 & 0 & s & 2 & 1.383510 & -0.028087 \\
B & 2 & 0 & s & 3 & 0.318470 & 0.599307 \\
B & 2 & 0 & s & 4 & 0.104570 & 0.604576 \\
B & 2 & 1 & p & 1 & 6.046818 & 0.034645 \\
B & 2 & 1 & p & 2 & 1.224802 & 0.198254 \\
B & 2 & 1 & p & 3 & 0.335538 & 0.508769 \\
B & 2 & 1 & p & 4 & 0.099653 & 0.429998 \\
\bottomrule
\end{tabular}

\begin{tabular}{lllllll}
\toprule
Atom & $n$ & $l$ & Shell type & Primitive  & Exponent & Contraction Coeff. \\
\midrule
C & 1 & 0 & s & 1 & 447.313496 & 0.026621 \\
C & 1 & 0 & s & 2 & 67.341312 & 0.183823 \\
C & 1 & 0 & s & 3 & 15.018155 & 0.627207 \\
C & 1 & 0 & s & 4 & 3.946006 & 0.714807 \\
C & 2 & 0 & s & 1 & 5.281819 & -0.071389 \\
C & 2 & 0 & s & 2 & 0.058683 & 0.035344 \\
C & 2 & 0 & s & 3 & 0.491311 & 0.437041 \\
C & 2 & 0 & s & 4 & 0.171114 & 0.436048 \\
C & 2 & 1 & p & 1 & 8.397557 & 0.060580 \\
C & 2 & 1 & p & 2 & 1.820185 & 0.309546 \\
C & 2 & 1 & p & 3 & 0.505859 & 0.707381 \\
C & 2 & 1 & p & 4 & 0.144746 & 0.592636 \\
\bottomrule
\end{tabular}

\begin{tabular}{lllllll}
\toprule
Atom & $n$ & $l$ & Shell type & Primitive  & Exponent & Contraction Coeff. \\
\midrule
N & 1 & 0 & s & 1 & 614.349623 & 0.025689 \\
N & 1 & 0 & s & 2 & 92.690354 & 0.177481 \\
N & 1 & 0 & s & 3 & 20.760638 & 0.607633 \\
N & 1 & 0 & s & 4 & 5.479167 & 0.692917 \\
N & 2 & 0 & s & 1 & 7.621857 & -0.089882 \\
N & 2 & 0 & s & 2 & 0.071787 & 0.054803 \\
N & 2 & 0 & s & 3 & 0.716923 & 0.539584 \\
N & 2 & 0 & s & 4 & 0.248686 & 0.567983 \\
N & 2 & 1 & p & 1 & 12.177783 & 0.036020 \\
N & 2 & 1 & p & 2 & 2.670970 & 0.183212 \\
N & 2 & 1 & p & 3 & 0.740773 & 0.403646 \\
N & 2 & 1 & p & 4 & 0.207078 & 0.340253 \\
\bottomrule
\end{tabular}

\begin{tabular}{lllllll}
\toprule
Atom & $n$ & $l$ & Shell type & Primitive  & Exponent & Contraction Coeff. \\
\midrule
O & 1 & 0 & s & 1 & 812.095834 & 0.026268 \\
O & 1 & 0 & s & 2 & 122.440542 & 0.182070 \\
O & 1 & 0 & s & 3 & 27.467276 & 0.626194 \\
O & 1 & 0 & s & 4 & 7.267088 & 0.714723 \\
O & 2 & 0 & s & 1 & 10.460464 & -0.090855 \\
O & 2 & 0 & s & 2 & 0.141754 & 0.061492 \\
O & 2 & 0 & s & 3 & 0.952670 & 0.583500 \\
O & 2 & 0 & s & 4 & 0.321901 & 0.530712 \\
O & 2 & 1 & p & 1 & 16.766357 & 0.045915 \\
O & 2 & 1 & p & 2 & 3.688268 & 0.235697 \\
O & 2 & 1 & p & 3 & 1.009334 & 0.507337 \\
O & 2 & 1 & p & 4 & 0.268049 & 0.446129 \\
\bottomrule
\end{tabular}

\begin{tabular}{lllllll}
\toprule
Atom & $n$ & $l$ & Shell type & Primitive  & Exponent & Contraction Coeff. \\
\midrule
F & 1 & 0 & s & 1 & 1024.126410 & 0.026796 \\
F & 1 & 0 & s & 2 & 154.966564 & 0.185088 \\
F & 1 & 0 & s & 3 & 34.893196 & 0.635767 \\
F & 1 & 0 & s & 4 & 9.269533 & 0.721131 \\
F & 2 & 0 & s & 1 & 13.573770 & -0.081181 \\
F & 2 & 0 & s & 2 & 0.122414 & 0.041086 \\
F & 2 & 0 & s & 3 & 1.271621 & 0.503645 \\
F & 2 & 0 & s & 4 & 0.424420 & 0.508406 \\
F & 2 & 1 & p & 1 & 22.081133 & 0.044035 \\
F & 2 & 1 & p & 2 & 4.867684 & 0.227192 \\
F & 2 & 1 & p & 3 & 1.320251 & 0.482604 \\
F & 2 & 1 & p & 4 & 0.341670 & 0.427171 \\
\bottomrule
\end{tabular}

\subsection{$k = 5$}
\begin{tabular}{lllllll}
\toprule
Atom & $n$ & $l$ & Shell type & Primitive  & Exponent & Contraction Coeff. \\
\midrule
H & 1 & 0 & s & 1 & 24.561782 & 0.015833 \\
H & 1 & 0 & s & 2 & 3.924283 & 0.105252 \\
H & 1 & 0 & s & 3 & 0.957988 & 0.416065 \\
H & 1 & 0 & s & 4 & 0.286151 & 0.905476 \\
H & 1 & 0 & s & 5 & 0.094807 & 0.590050 \\
\bottomrule
\end{tabular}

\begin{tabular}{lllllll}
\toprule
Atom & $n$ & $l$ & Shell type & Primitive  & Exponent & Contraction Coeff. \\
\midrule
Li & 1 & 0 & s & 1 & 232.706805 & 0.009798 \\
Li & 1 & 0 & s & 2 & 36.039372 & 0.068770 \\
Li & 1 & 0 & s & 3 & 8.326178 & 0.279340 \\
Li & 1 & 0 & s & 4 & 2.293087 & 0.637974 \\
Li & 1 & 0 & s & 5 & 0.686717 & 0.523345 \\
Li & 2 & 0 & s & 1 & 1.131744 & -0.043958 \\
Li & 2 & 0 & s & 2 & 0.355358 & -0.073511 \\
Li & 2 & 0 & s & 3 & 0.096858 & 0.215744 \\
Li & 2 & 0 & s & 4 & 0.085832 & 0.139570 \\
Li & 2 & 0 & s & 5 & 0.033043 & 0.574348 \\
Li & 2 & 1 & p & 1 & 1.458472 & 0.056637 \\
Li & 2 & 1 & p & 2 & 5.998030 & 0.188342 \\
Li & 2 & 1 & p & 3 & 1.772062 & 0.182682 \\
Li & 2 & 1 & p & 4 & 0.000804 & 0.000701 \\
Li & 2 & 1 & p & 5 & 0.000607 & -0.000571 \\
\bottomrule
\end{tabular}

\begin{tabular}{lllllll}
\toprule
Atom & $n$ & $l$ & Shell type & Primitive  & Exponent & Contraction Coeff. \\
\midrule
Be & 1 & 0 & s & 1 & 434.091838 & 0.009155 \\
Be & 1 & 0 & s & 2 & 67.177289 & 0.064447 \\
Be & 1 & 0 & s & 3 & 15.639776 & 0.263853 \\
Be & 1 & 0 & s & 4 & 4.357526 & 0.615606 \\
Be & 1 & 0 & s & 5 & 1.325476 & 0.501686 \\
Be & 2 & 0 & s & 1 & 6.374968 & -0.018018 \\
Be & 2 & 0 & s & 2 & 1.491138 & -0.139931 \\
Be & 2 & 0 & s & 3 & 0.564791 & 0.026597 \\
Be & 2 & 0 & s & 4 & 0.170667 & 0.720673 \\
Be & 2 & 0 & s & 5 & 0.062329 & 0.702130 \\
Be & 2 & 1 & p & 1 & 8.267417 & 0.017207 \\
Be & 2 & 1 & p & 2 & 1.410162 & 0.105174 \\
Be & 2 & 1 & p & 3 & 0.402551 & 0.484161 \\
Be & 2 & 1 & p & 4 & 0.146110 & 1.029627 \\
Be & 2 & 1 & p & 5 & 0.053272 & 0.234937 \\
\bottomrule
\end{tabular}

\begin{tabular}{lllllll}
\toprule
Atom & $n$ & $l$ & Shell type & Primitive  & Exponent & Contraction Coeff. \\
\midrule
B & 1 & 0 & s & 1 & 708.921534 & 0.008755 \\
B & 1 & 0 & s & 2 & 109.223926 & 0.062082 \\
B & 1 & 0 & s & 3 & 25.326119 & 0.259488 \\
B & 1 & 0 & s & 4 & 7.067193 & 0.612300 \\
B & 1 & 0 & s & 5 & 2.166124 & 0.494621 \\
B & 2 & 0 & s & 1 & 9.013960 & -0.016771 \\
B & 2 & 0 & s & 2 & 2.545165 & -0.084070 \\
B & 2 & 0 & s & 3 & 0.505581 & 0.128796 \\
B & 2 & 0 & s & 4 & 0.233939 & 0.490021 \\
B & 2 & 0 & s & 5 & 0.087398 & 0.344856 \\
B & 2 & 1 & p & 1 & 11.173515 & 0.016736 \\
B & 2 & 1 & p & 2 & 2.413354 & 0.101116 \\
B & 2 & 1 & p & 3 & 0.679748 & 0.354510 \\
B & 2 & 1 & p & 4 & 0.220221 & 0.589656 \\
B & 2 & 1 & p & 5 & 0.075165 & 0.315239 \\
\bottomrule
\end{tabular}

\begin{tabular}{lllllll}
\toprule
Atom & $n$ & $l$ & Shell type & Primitive  & Exponent & Contraction Coeff. \\
\midrule
C & 1 & 0 & s & 1 & 1007.888150 & 0.009996 \\
C & 1 & 0 & s & 2 & 156.503898 & 0.070181 \\
C & 1 & 0 & s & 3 & 36.465519 & 0.292900 \\
C & 1 & 0 & s & 4 & 10.261023 & 0.685502 \\
C & 1 & 0 & s & 5 & 3.174388 & 0.541710 \\
C & 2 & 0 & s & 1 & 12.719546 & -0.017893 \\
C & 2 & 0 & s & 2 & 3.498732 & -0.078070 \\
C & 2 & 0 & s & 3 & 0.966191 & 0.106778 \\
C & 2 & 0 & s & 4 & 0.357426 & 0.445108 \\
C & 2 & 0 & s & 5 & 0.126442 & 0.271942 \\
C & 2 & 1 & p & 1 & 15.657930 & 0.014928 \\
C & 2 & 1 & p & 2 & 3.417447 & 0.088593 \\
C & 2 & 1 & p & 3 & 0.989047 & 0.273952 \\
C & 2 & 1 & p & 4 & 0.314976 & 0.411186 \\
C & 2 & 1 & p & 5 & 0.102718 & 0.207661 \\
\bottomrule
\end{tabular}

\begin{tabular}{lllllll}
\toprule
Atom & $n$ & $l$ & Shell type & Primitive  & Exponent & Contraction Coeff. \\
\midrule
N & 1 & 0 & s & 1 & 1245.737030 & 0.010293 \\
N & 1 & 0 & s & 2 & 195.678174 & 0.070354 \\
N & 1 & 0 & s & 3 & 46.371297 & 0.283852 \\
N & 1 & 0 & s & 4 & 13.334402 & 0.628193 \\
N & 1 & 0 & s & 5 & 4.232987 & 0.461623 \\
N & 2 & 0 & s & 1 & 14.301413 & -0.026804 \\
N & 2 & 0 & s & 2 & 4.672186 & -0.075621 \\
N & 2 & 0 & s & 3 & 1.152511 & 0.178972 \\
N & 2 & 0 & s & 4 & 0.451010 & 0.494259 \\
N & 2 & 0 & s & 5 & 0.165098 & 0.267129 \\
N & 2 & 1 & p & 1 & 21.850124 & 0.021133 \\
N & 2 & 1 & p & 2 & 4.936508 & 0.122394 \\
N & 2 & 1 & p & 3 & 1.473998 & 0.358273 \\
N & 2 & 1 & p & 4 & 0.475944 & 0.531396 \\
N & 2 & 1 & p & 5 & 0.153194 & 0.300144 \\
\bottomrule
\end{tabular}

\begin{tabular}{lllllll}
\toprule
Atom & $n$ & $l$ & Shell type & Primitive  & Exponent & Contraction Coeff. \\
\midrule
O & 1 & 0 & s & 1 & 1622.509400 & 0.011464 \\
O & 1 & 0 & s & 2 & 255.457870 & 0.078649 \\
O & 1 & 0 & s & 3 & 60.488003 & 0.317980 \\
O & 1 & 0 & s & 4 & 17.429605 & 0.701919 \\
O & 1 & 0 & s & 5 & 5.546655 & 0.503621 \\
O & 2 & 0 & s & 1 & 18.724456 & -0.027624 \\
O & 2 & 0 & s & 2 & 6.541395 & -0.074259 \\
O & 2 & 0 & s & 3 & 1.623506 & 0.153709 \\
O & 2 & 0 & s & 4 & 0.678723 & 0.490992 \\
O & 2 & 0 & s & 5 & 0.239409 & 0.320295 \\
O & 2 & 1 & p & 1 & 27.108991 & 0.022531 \\
O & 2 & 1 & p & 2 & 6.120533 & 0.130840 \\
O & 2 & 1 & p & 3 & 1.798277 & 0.358561 \\
O & 2 & 1 & p & 4 & 0.559099 & 0.482238 \\
O & 2 & 1 & p & 5 & 0.172072 & 0.243054 \\
\bottomrule
\end{tabular}

\begin{tabular}{lllllll}
\toprule
Atom & $n$ & $l$ & Shell type & Primitive  & Exponent & Contraction Coeff. \\
\midrule
F & 1 & 0 & s & 1 & 2171.499400 & 0.009929 \\
F & 1 & 0 & s & 2 & 339.437935 & 0.068891 \\
F & 1 & 0 & s & 3 & 80.092451 & 0.283092 \\
F & 1 & 0 & s & 4 & 23.017945 & 0.641438 \\
F & 1 & 0 & s & 5 & 7.270549 & 0.481901 \\
F & 2 & 0 & s & 1 & 19.803230 & -0.043438 \\
F & 2 & 0 & s & 2 & 6.491131 & -0.086320 \\
F & 2 & 0 & s & 3 & 2.435812 & 0.177980 \\
F & 2 & 0 & s & 4 & 0.867524 & 0.559989 \\
F & 2 & 0 & s & 5 & 0.297995 & 0.324309 \\
F & 2 & 1 & p & 1 & 37.821166 & 0.021173 \\
F & 2 & 1 & p & 2 & 8.717346 & 0.122611 \\
F & 2 & 1 & p & 3 & 2.627408 & 0.341328 \\
F & 2 & 1 & p & 4 & 0.832228 & 0.488497 \\
F & 2 & 1 & p & 5 & 0.249685 & 0.309942 \\
\bottomrule
\end{tabular}

\subsection{$k = 6$}
\begin{tabular}{lllllll}
\toprule
Atom & $n$ & $l$ & Shell type & Primitive  & Exponent & Contraction Coeff. \\
\midrule
H & 1 & 0 & s & 1 & 42.431848 & 0.005984 \\
H & 1 & 0 & s & 2 & 6.527242 & 0.043765 \\
H & 1 & 0 & s & 3 & 1.527203 & 0.194622 \\
H & 1 & 0 & s & 4 & 0.448575 & 0.523163 \\
H & 1 & 0 & s & 5 & 0.153500 & 0.610899 \\
H & 1 & 0 & s & 6 & 0.057351 & 0.131116 \\
\bottomrule
\end{tabular}

\begin{tabular}{lllllll}
\toprule
Atom & $n$ & $l$ & Shell type & Primitive  & Exponent & Contraction Coeff. \\
\midrule
Li & 1 & 0 & s & 1 & 357.319648 & 0.005192 \\
Li & 1 & 0 & s & 2 & 56.415381 & 0.035738 \\
Li & 1 & 0 & s & 3 & 13.458931 & 0.151421 \\
Li & 1 & 0 & s & 4 & 3.901728 & 0.404018 \\
Li & 1 & 0 & s & 5 & 1.262537 & 0.565961 \\
Li & 1 & 0 & s & 6 & 0.443288 & 0.210397 \\
Li & 2 & 0 & s & 1 & 0.454459 & -0.021872 \\
Li & 2 & 0 & s & 2 & 1.224863 & -0.046715 \\
Li & 2 & 0 & s & 3 & 0.514853 & -0.066434 \\
Li & 2 & 0 & s & 4 & 0.027102 & 0.259290 \\
Li & 2 & 0 & s & 5 & 0.072389 & 0.724723 \\
Li & 2 & 0 & s & 6 & 0.027653 & 0.346335 \\
Li & 2 & 1 & p & 1 & 15.086516 & 0.097310 \\
Li & 2 & 1 & p & 2 & 1.540652 & 0.254062 \\
Li & 2 & 1 & p & 3 & 1.322475 & 0.137116 \\
Li & 2 & 1 & p & 4 & 1.402183 & 0.068783 \\
Li & 2 & 1 & p & 5 & 4.264866 & 0.543726 \\
Li & 2 & 1 & p & 6 & 0.000797 & 0.000020 \\
\bottomrule
\end{tabular}

\begin{tabular}{lllllll}
\toprule
Atom & $n$ & $l$ & Shell type & Primitive  & Exponent & Contraction Coeff. \\
\midrule
Be & 1 & 0 & s & 1 & 628.105510 & 0.005689 \\
Be & 1 & 0 & s & 2 & 99.611392 & 0.038544 \\
Be & 1 & 0 & s & 3 & 24.211405 & 0.160504 \\
Be & 1 & 0 & s & 4 & 7.160029 & 0.428508 \\
Be & 1 & 0 & s & 5 & 2.362672 & 0.590722 \\
Be & 1 & 0 & s & 6 & 0.848907 & 0.211013 \\
Be & 2 & 0 & s & 1 & 0.079822 & -0.004596 \\
Be & 2 & 0 & s & 2 & 2.644312 & -0.043432 \\
Be & 2 & 0 & s & 3 & 0.913087 & -0.059581 \\
Be & 2 & 0 & s & 4 & 0.218189 & 0.264613 \\
Be & 2 & 0 & s & 5 & 0.059969 & 0.409491 \\
Be & 2 & 0 & s & 6 & 0.132252 & 0.261167 \\
Be & 2 & 1 & p & 1 & 13.255033 & 0.004452 \\
Be & 2 & 1 & p & 2 & 1.888868 & 0.044621 \\
Be & 2 & 1 & p & 3 & 0.058765 & 0.078267 \\
Be & 2 & 1 & p & 4 & 0.396103 & 0.288892 \\
Be & 2 & 1 & p & 5 & 0.141375 & 0.489599 \\
Be & 2 & 1 & p & 6 & 0.061225 & 0.058752 \\
\bottomrule
\end{tabular}

\begin{tabular}{lllllll}
\toprule
Atom & $n$ & $l$ & Shell type & Primitive  & Exponent & Contraction Coeff. \\
\midrule
B & 1 & 0 & s & 1 & 1191.995380 & 0.003906 \\
B & 1 & 0 & s & 2 & 187.930922 & 0.027353 \\
B & 1 & 0 & s & 3 & 44.719153 & 0.120466 \\
B & 1 & 0 & s & 4 & 12.983995 & 0.345787 \\
B & 1 & 0 & s & 5 & 4.260820 & 0.519448 \\
B & 1 & 0 & s & 6 & 1.515436 & 0.220620 \\
B & 2 & 0 & s & 1 & 20.786505 & -0.004123 \\
B & 2 & 0 & s & 2 & 4.377161 & -0.045476 \\
B & 2 & 0 & s & 3 & 1.679385 & -0.051531 \\
B & 2 & 0 & s & 4 & 0.410009 & 0.260743 \\
B & 2 & 0 & s & 5 & 0.171020 & 0.403959 \\
B & 2 & 0 & s & 6 & 0.074160 & 0.194810 \\
B & 2 & 1 & p & 1 & 2.265859 & 0.002587 \\
B & 2 & 1 & p & 2 & 6.263301 & 0.023501 \\
B & 2 & 1 & p & 3 & 1.406857 & 0.107740 \\
B & 2 & 1 & p & 4 & 0.475123 & 0.246898 \\
B & 2 & 1 & p & 5 & 0.181495 & 0.328380 \\
B & 2 & 1 & p & 6 & 0.067564 & 0.157649 \\
\bottomrule
\end{tabular}

\begin{tabular}{lllllll}
\toprule
Atom & $n$ & $l$ & Shell type & Primitive  & Exponent & Contraction Coeff. \\
\midrule
C & 1 & 0 & s & 1 & 1487.729180 & 0.005473 \\
C & 1 & 0 & s & 2 & 237.875623 & 0.037129 \\
C & 1 & 0 & s & 3 & 57.569062 & 0.158239 \\
C & 1 & 0 & s & 4 & 16.957780 & 0.434670 \\
C & 1 & 0 & s & 5 & 5.638156 & 0.596224 \\
C & 1 & 0 & s & 6 & 2.016810 & 0.204718 \\
C & 2 & 0 & s & 1 & 33.245310 & -0.004243 \\
C & 2 & 0 & s & 2 & 6.707808 & -0.043769 \\
C & 2 & 0 & s & 3 & 2.558931 & -0.042958 \\
C & 2 & 0 & s & 4 & 0.622313 & 0.257876 \\
C & 2 & 0 & s & 5 & 0.242229 & 0.374496 \\
C & 2 & 0 & s & 6 & 0.099892 & 0.138412 \\
C & 2 & 1 & p & 1 & 30.671811 & 0.007197 \\
C & 2 & 1 & p & 2 & 6.571588 & 0.051820 \\
C & 2 & 1 & p & 3 & 1.932930 & 0.188690 \\
C & 2 & 1 & p & 4 & 0.690723 & 0.381574 \\
C & 2 & 1 & p & 5 & 0.264982 & 0.460064 \\
C & 2 & 1 & p & 6 & 0.098706 & 0.240541 \\
\bottomrule
\end{tabular}

\begin{tabular}{lllllll}
\toprule
Atom & $n$ & $l$ & Shell type & Primitive  & Exponent & Contraction Coeff. \\
\midrule
N & 1 & 0 & s & 1 & 1878.144370 & 0.005565 \\
N & 1 & 0 & s & 2 & 300.246020 & 0.037801 \\
N & 1 & 0 & s & 3 & 72.678832 & 0.159332 \\
N & 1 & 0 & s & 4 & 21.469328 & 0.425158 \\
N & 1 & 0 & s & 5 & 7.156883 & 0.542742 \\
N & 1 & 0 & s & 6 & 2.519045 & 0.152024 \\
N & 2 & 0 & s & 1 & 33.545042 & -0.007573 \\
N & 2 & 0 & s & 2 & 8.085021 & -0.057971 \\
N & 2 & 0 & s & 3 & 2.704251 & -0.042758 \\
N & 2 & 0 & s & 4 & 0.953890 & 0.259456 \\
N & 2 & 0 & s & 5 & 0.402114 & 0.350974 \\
N & 2 & 0 & s & 6 & 0.172071 & 0.242262 \\
N & 2 & 1 & p & 1 & 49.349058 & 0.006128 \\
N & 2 & 1 & p & 2 & 10.732785 & 0.046997 \\
N & 2 & 1 & p & 3 & 3.118240 & 0.184026 \\
N & 2 & 1 & p & 4 & 1.078864 & 0.391515 \\
N & 2 & 1 & p & 5 & 0.395848 & 0.478936 \\
N & 2 & 1 & p & 6 & 0.140051 & 0.255522 \\
\bottomrule
\end{tabular}

\begin{tabular}{lllllll}
\toprule
Atom & $n$ & $l$ & Shell type & Primitive  & Exponent & Contraction Coeff. \\
\midrule
O & 1 & 0 & s & 1 & 2459.678270 & 0.005909 \\
O & 1 & 0 & s & 2 & 393.874043 & 0.039799 \\
O & 1 & 0 & s & 3 & 96.232398 & 0.166137 \\
O & 1 & 0 & s & 4 & 28.753679 & 0.442812 \\
O & 1 & 0 & s & 5 & 9.662396 & 0.574660 \\
O & 1 & 0 & s & 6 & 3.427536 & 0.169654 \\
O & 2 & 0 & s & 1 & 39.246022 & -0.006918 \\
O & 2 & 0 & s & 2 & 11.094114 & -0.052302 \\
O & 2 & 0 & s & 3 & 4.654038 & -0.035522 \\
O & 2 & 0 & s & 4 & 1.241235 & 0.270447 \\
O & 2 & 0 & s & 5 & 0.508299 & 0.387728 \\
O & 2 & 0 & s & 6 & 0.208151 & 0.198744 \\
O & 2 & 1 & p & 1 & 82.246307 & 0.003753 \\
O & 2 & 1 & p & 2 & 16.290624 & 0.034649 \\
O & 2 & 1 & p & 3 & 4.581324 & 0.144492 \\
O & 2 & 1 & p & 4 & 1.522467 & 0.321960 \\
O & 2 & 1 & p & 5 & 0.521785 & 0.396872 \\
O & 2 & 1 & p & 6 & 0.172017 & 0.211401 \\
\bottomrule
\end{tabular}

\begin{tabular}{lllllll}
\toprule
Atom & $n$ & $l$ & Shell type & Primitive  & Exponent & Contraction Coeff. \\
\midrule
F & 1 & 0 & s & 1 & 3101.056010 & 0.005660 \\
F & 1 & 0 & s & 2 & 498.931583 & 0.037859 \\
F & 1 & 0 & s & 3 & 122.468326 & 0.157729 \\
F & 1 & 0 & s & 4 & 36.651305 & 0.421556 \\
F & 1 & 0 & s & 5 & 12.384611 & 0.543017 \\
F & 1 & 0 & s & 6 & 4.432168 & 0.164265 \\
F & 2 & 0 & s & 1 & 46.705201 & -0.006812 \\
F & 2 & 0 & s & 2 & 12.974718 & -0.047678 \\
F & 2 & 0 & s & 3 & 4.296215 & -0.024341 \\
F & 2 & 0 & s & 4 & 1.609904 & 0.220640 \\
F & 2 & 0 & s & 5 & 0.703264 & 0.248378 \\
F & 2 & 0 & s & 6 & 0.306749 & 0.197017 \\
F & 2 & 1 & p & 1 & 74.831193 & 0.006442 \\
F & 2 & 1 & p & 2 & 17.107923 & 0.044945 \\
F & 2 & 1 & p & 3 & 5.191097 & 0.167523 \\
F & 2 & 1 & p & 4 & 1.754160 & 0.345113 \\
F & 2 & 1 & p & 5 & 0.591643 & 0.391395 \\
F & 2 & 1 & p & 6 & 0.192520 & 0.181364 \\
\bottomrule
\end{tabular}

\subsection{$k = 7$}
\begin{tabular}{lllllll}
\toprule
Atom & $n$ & $l$ & Shell type & Primitive  & Exponent & Contraction Coeff. \\
\midrule
H & 1 & 0 & s & 1 & 36.215701 & 0.004636 \\
H & 1 & 0 & s & 2 & 5.945957 & 0.029019 \\
H & 1 & 0 & s & 3 & 1.672836 & 0.086907 \\
H & 1 & 0 & s & 4 & 0.372023 & 0.302611 \\
H & 1 & 0 & s & 5 & 0.145877 & 0.346923 \\
H & 1 & 0 & s & 6 & 0.058664 & 0.083617 \\
H & 1 & 0 & s & 7 & 0.794523 & 0.107805 \\
\bottomrule
\end{tabular}

\begin{tabular}{lllllll}
\toprule
Atom & $n$ & $l$ & Shell type & Primitive  & Exponent & Contraction Coeff. \\
\midrule
Li & 1 & 0 & s & 1 & 343.220752 & 0.005357 \\
Li & 1 & 0 & s & 2 & 54.985322 & 0.036420 \\
Li & 1 & 0 & s & 3 & 13.360790 & 0.148893 \\
Li & 1 & 0 & s & 4 & 3.986245 & 0.383807 \\
Li & 1 & 0 & s & 5 & 0.774205 & 0.155673 \\
Li & 1 & 0 & s & 6 & 0.414568 & 0.157633 \\
Li & 1 & 0 & s & 7 & 1.394985 & 0.478386 \\
Li & 2 & 0 & s & 1 & 0.009102 & 0.030398 \\
Li & 2 & 0 & s & 2 & 1.137499 & -0.047816 \\
Li & 2 & 0 & s & 3 & 0.386588 & -0.096878 \\
Li & 2 & 0 & s & 4 & 0.076804 & 0.247828 \\
Li & 2 & 0 & s & 5 & 0.031196 & 0.569284 \\
Li & 2 & 0 & s & 6 & 0.069971 & 0.271056 \\
Li & 2 & 0 & s & 7 & 0.165559 & 0.064038 \\
Li & 2 & 1 & p & 1 & 21.555617 & 0.066067 \\
Li & 2 & 1 & p & 2 & 1.679468 & 0.326949 \\
Li & 2 & 1 & p & 3 & 1.543082 & 0.320567 \\
Li & 2 & 1 & p & 4 & 0.005101 & -0.000992 \\
Li & 2 & 1 & p & 5 & 1.733977 & 0.009134 \\
Li & 2 & 1 & p & 6 & 0.083200 & 0.006066 \\
Li & 2 & 1 & p & 7 & 5.024132 & 0.612641 \\
\bottomrule
\end{tabular}

\begin{tabular}{lllllll}
\toprule
Atom & $n$ & $l$ & Shell type & Primitive  & Exponent & Contraction Coeff. \\
\midrule
Be & 1 & 0 & s & 1 & 656.902472 & 0.006197 \\
Be & 1 & 0 & s & 2 & 104.195338 & 0.042957 \\
Be & 1 & 0 & s & 3 & 24.749232 & 0.185009 \\
Be & 1 & 0 & s & 4 & 7.175726 & 0.503304 \\
Be & 1 & 0 & s & 5 & 2.331965 & 0.696914 \\
Be & 1 & 0 & s & 6 & 0.808728 & 0.234208 \\
Be & 1 & 0 & s & 7 & 0.265758 & -0.004638 \\
Be & 2 & 0 & s & 1 & 16.118838 & -0.002583 \\
Be & 2 & 0 & s & 2 & 4.259903 & -0.021703 \\
Be & 2 & 0 & s & 3 & 1.378626 & -0.090348 \\
Be & 2 & 0 & s & 4 & 0.220948 & 0.218324 \\
Be & 2 & 0 & s & 5 & 0.132218 & 0.400425 \\
Be & 2 & 0 & s & 6 & 0.059409 & 0.388960 \\
Be & 2 & 0 & s & 7 & 0.022203 & 0.022587 \\
Be & 2 & 1 & p & 1 & 10.364116 & 0.010060 \\
Be & 2 & 1 & p & 2 & 2.559345 & 0.061442 \\
Be & 2 & 1 & p & 3 & 0.563026 & 0.374122 \\
Be & 2 & 1 & p & 4 & 0.110633 & 0.124752 \\
Be & 2 & 1 & p & 5 & 0.071562 & 0.358074 \\
Be & 2 & 1 & p & 6 & 0.069727 & 0.119979 \\
Be & 2 & 1 & p & 7 & 0.192969 & 1.024303 \\
\bottomrule
\end{tabular}

\begin{tabular}{lllllll}
\toprule
Atom & $n$ & $l$ & Shell type & Primitive  & Exponent & Contraction Coeff. \\
\midrule
B & 1 & 0 & s & 1 & 895.059309 & 0.006588 \\
B & 1 & 0 & s & 2 & 11.051797 & 0.446943 \\
B & 1 & 0 & s & 3 & 146.019232 & 0.042622 \\
B & 1 & 0 & s & 4 & 36.492356 & 0.171727 \\
B & 1 & 0 & s & 5 & 3.736563 & 0.596800 \\
B & 1 & 0 & s & 6 & 1.245423 & 0.078461 \\
B & 1 & 0 & s & 7 & 1.433852 & 0.127030 \\
B & 2 & 0 & s & 1 & 1.292182 & -0.006817 \\
B & 2 & 0 & s & 2 & 1.844326 & -0.076663 \\
B & 2 & 0 & s & 3 & 4.959091 & -0.051761 \\
B & 2 & 0 & s & 4 & 0.420044 & 0.323967 \\
B & 2 & 0 & s & 5 & 0.103630 & 0.471972 \\
B & 2 & 0 & s & 6 & 0.063599 & 0.064601 \\
B & 2 & 0 & s & 7 & 0.224102 & 0.422335 \\
B & 2 & 1 & p & 1 & 25.847767 & 0.004318 \\
B & 2 & 1 & p & 2 & 4.339640 & 0.047770 \\
B & 2 & 1 & p & 3 & 0.343152 & 0.376045 \\
B & 2 & 1 & p & 4 & 0.036593 & 0.051667 \\
B & 2 & 1 & p & 5 & 0.116828 & 0.448116 \\
B & 2 & 1 & p & 6 & 0.388055 & 0.084825 \\
B & 2 & 1 & p & 7 & 1.105075 & 0.198175 \\
\bottomrule
\end{tabular}

\begin{tabular}{lllllll}
\toprule
Atom & $n$ & $l$ & Shell type & Primitive  & Exponent & Contraction Coeff. \\
\midrule
C & 1 & 0 & s & 1 & 1194.188930 & 0.006589 \\
C & 1 & 0 & s & 2 & 194.061751 & 0.042889 \\
C & 1 & 0 & s & 3 & 48.028224 & 0.171972 \\
C & 1 & 0 & s & 4 & 14.522873 & 0.431488 \\
C & 1 & 0 & s & 5 & 4.975868 & 0.514307 \\
C & 1 & 0 & s & 6 & 1.910933 & 0.135474 \\
C & 1 & 0 & s & 7 & 1.293437 & 0.011825 \\
C & 2 & 0 & s & 1 & 23.978838 & -0.010574 \\
C & 2 & 0 & s & 2 & 3.884708 & -0.048191 \\
C & 2 & 0 & s & 3 & 5.758997 & -0.055569 \\
C & 2 & 0 & s & 4 & 0.127766 & 0.388892 \\
C & 2 & 0 & s & 5 & 0.758987 & 0.103245 \\
C & 2 & 0 & s & 6 & 0.126331 & 0.011332 \\
C & 2 & 0 & s & 7 & 0.383645 & 0.617531 \\
C & 2 & 1 & p & 1 & 27.268783 & 0.018832 \\
C & 2 & 1 & p & 2 & 6.580807 & 0.096484 \\
C & 2 & 1 & p & 3 & 2.645286 & 0.222211 \\
C & 2 & 1 & p & 4 & 0.451507 & 0.750437 \\
C & 2 & 1 & p & 5 & 0.215191 & 0.862582 \\
C & 2 & 1 & p & 6 & 0.083276 & 0.399621 \\
C & 2 & 1 & p & 7 & 1.061166 & 0.639316 \\
\bottomrule
\end{tabular}

\begin{tabular}{lllllll}
\toprule
Atom & $n$ & $l$ & Shell type & Primitive  & Exponent & Contraction Coeff. \\
\midrule
N & 1 & 0 & s & 1 & 1751.597230 & 0.005324 \\
N & 1 & 0 & s & 2 & 272.872768 & 0.037455 \\
N & 1 & 0 & s & 3 & 64.619601 & 0.159329 \\
N & 1 & 0 & s & 4 & 18.897945 & 0.401234 \\
N & 1 & 0 & s & 5 & 6.763559 & 0.354623 \\
N & 1 & 0 & s & 6 & 4.345101 & 0.137402 \\
N & 1 & 0 & s & 7 & 1.721191 & 0.058985 \\
N & 2 & 0 & s & 1 & 21.833277 & -0.012670 \\
N & 2 & 0 & s & 2 & 5.913574 & -0.078740 \\
N & 2 & 0 & s & 3 & 0.466277 & -0.036636 \\
N & 2 & 0 & s & 4 & 0.166338 & 0.186849 \\
N & 2 & 0 & s & 5 & 0.549578 & 0.521859 \\
N & 2 & 0 & s & 6 & 0.206457 & 0.135290 \\
N & 2 & 0 & s & 7 & 1.469769 & 0.071018 \\
N & 2 & 1 & p & 1 & 43.663429 & 0.009504 \\
N & 2 & 1 & p & 2 & 11.371124 & 0.051490 \\
N & 2 & 1 & p & 3 & 3.716008 & 0.195534 \\
N & 2 & 1 & p & 4 & 1.320422 & 0.474218 \\
N & 2 & 1 & p & 5 & 0.486797 & 0.635050 \\
N & 2 & 1 & p & 6 & 0.141396 & 0.316219 \\
N & 2 & 1 & p & 7 & 0.245582 & 0.175968 \\
\bottomrule
\end{tabular}

\begin{tabular}{lllllll}
\toprule
Atom & $n$ & $l$ & Shell type & Primitive  & Exponent & Contraction Coeff. \\
\midrule
O & 1 & 0 & s & 1 & 2573.046670 & 0.006303 \\
O & 1 & 0 & s & 2 & 414.765540 & 0.042092 \\
O & 1 & 0 & s & 3 & 102.059410 & 0.175262 \\
O & 1 & 0 & s & 4 & 30.883247 & 0.470288 \\
O & 1 & 0 & s & 5 & 10.595279 & 0.639573 \\
O & 1 & 0 & s & 6 & 4.036538 & 0.237306 \\
O & 1 & 0 & s & 7 & 1.795940 & 0.013175 \\
O & 2 & 0 & s & 1 & 29.485037 & -0.011436 \\
O & 2 & 0 & s & 2 & 10.768628 & -0.092094 \\
O & 2 & 0 & s & 3 & 6.447004 & -0.043033 \\
O & 2 & 0 & s & 4 & 0.201562 & 0.209527 \\
O & 2 & 0 & s & 5 & 0.359231 & 0.403368 \\
O & 2 & 0 & s & 6 & 0.314721 & 0.221043 \\
O & 2 & 0 & s & 7 & 0.974912 & 0.807275 \\
O & 2 & 1 & p & 1 & 69.366016 & 0.006947 \\
O & 2 & 1 & p & 2 & 15.070592 & 0.053516 \\
O & 2 & 1 & p & 3 & 4.637408 & 0.186280 \\
O & 2 & 1 & p & 4 & 1.878070 & 0.315694 \\
O & 2 & 1 & p & 5 & 0.300433 & 0.485544 \\
O & 2 & 1 & p & 6 & 0.094442 & 0.102858 \\
O & 2 & 1 & p & 7 & 0.820574 & 0.483771 \\
\bottomrule
\end{tabular}

\begin{tabular}{lllllll}
\toprule
Atom & $n$ & $l$ & Shell type & Primitive  & Exponent & Contraction Coeff. \\
\midrule
F & 1 & 0 & s & 1 & 2924.910650 & 0.005546 \\
F & 1 & 0 & s & 2 & 473.005150 & 0.036708 \\
F & 1 & 0 & s & 3 & 116.797193 & 0.149967 \\
F & 1 & 0 & s & 4 & 35.608673 & 0.386072 \\
F & 1 & 0 & s & 5 & 12.302986 & 0.480870 \\
F & 1 & 0 & s & 6 & 4.679761 & 0.151294 \\
F & 1 & 0 & s & 7 & 1.401899 & 0.007168 \\
F & 2 & 0 & s & 1 & 6.743232 & -0.018544 \\
F & 2 & 0 & s & 2 & 19.338766 & -0.046747 \\
F & 2 & 0 & s & 3 & 7.425120 & -0.058363 \\
F & 2 & 0 & s & 4 & 2.422353 & 0.144688 \\
F & 2 & 0 & s & 5 & 0.000103 & -0.000027 \\
F & 2 & 0 & s & 6 & 0.305892 & 0.387313 \\
F & 2 & 0 & s & 7 & 0.933029 & 0.626171 \\
F & 2 & 1 & p & 1 & 61.985534 & 0.009229 \\
F & 2 & 1 & p & 2 & 14.399508 & 0.059677 \\
F & 2 & 1 & p & 3 & 4.540860 & 0.188529 \\
F & 2 & 1 & p & 4 & 1.870009 & 0.236241 \\
F & 2 & 1 & p & 5 & 0.526223 & 0.282588 \\
F & 2 & 1 & p & 6 & 0.207625 & 0.204702 \\
F & 2 & 1 & p & 7 & 1.029405 & 0.227162 \\
\bottomrule
\end{tabular}

\subsection{$k = 8$}
\begin{tabular}{lllllll}
\toprule
Atom & $n$ & $l$ & Shell type & Primitive  & Exponent & Contraction Coeff. \\
\midrule
H & 1 & 0 & s & 1 & 57.360124 & 0.010589 \\
H & 1 & 0 & s & 2 & 8.858612 & 0.075640 \\
H & 1 & 0 & s & 3 & 0.573397 & 0.145787 \\
H & 1 & 0 & s & 4 & 0.697915 & 0.620006 \\
H & 1 & 0 & s & 5 & 0.241767 & 1.575516 \\
H & 1 & 0 & s & 6 & 0.705874 & 0.139903 \\
H & 1 & 0 & s & 7 & 0.088118 & 0.938748 \\
H & 1 & 0 & s & 8 & 2.154795 & 0.319647 \\
\bottomrule
\end{tabular}

\begin{tabular}{lllllll}
\toprule
Atom & $n$ & $l$ & Shell type & Primitive  & Exponent & Contraction Coeff. \\
\midrule
Li & 1 & 0 & s & 1 & 421.892873 & 0.009975 \\
Li & 1 & 0 & s & 2 & 65.887804 & 0.069748 \\
Li & 1 & 0 & s & 3 & 15.832399 & 0.294670 \\
Li & 1 & 0 & s & 4 & 3.211819 & 0.040798 \\
Li & 1 & 0 & s & 5 & 4.929907 & 0.667753 \\
Li & 1 & 0 & s & 6 & 0.450908 & 0.504495 \\
Li & 1 & 0 & s & 7 & 2.564102 & 0.429413 \\
Li & 1 & 0 & s & 8 & 1.233569 & 1.204849 \\
Li & 2 & 0 & s & 1 & 0.346352 & -0.094191 \\
Li & 2 & 0 & s & 2 & 0.754829 & -0.112499 \\
Li & 2 & 0 & s & 3 & 1.987442 & -0.012769 \\
Li & 2 & 0 & s & 4 & 0.078373 & 0.123304 \\
Li & 2 & 0 & s & 5 & 0.030563 & 1.058465 \\
Li & 2 & 0 & s & 6 & 0.084378 & 0.439866 \\
Li & 2 & 0 & s & 7 & 2.969524 & -0.012975 \\
Li & 2 & 0 & s & 8 & 0.082152 & 0.304985 \\
Li & 2 & 1 & p & 1 & 0.182820 & 0.028823 \\
Li & 2 & 1 & p & 2 & 0.058139 & 0.000448 \\
Li & 2 & 1 & p & 3 & 0.003788 & -0.000101 \\
Li & 2 & 1 & p & 4 & 0.010807 & 0.001095 \\
Li & 2 & 1 & p & 5 & 0.000369 & -0.000958 \\
Li & 2 & 1 & p & 6 & 0.002838 & 0.000356 \\
Li & 2 & 1 & p & 7 & 1.793771 & 2.362083 \\
Li & 2 & 1 & p & 8 & 6.341340 & 1.658003 \\
\bottomrule
\end{tabular}

\begin{tabular}{lllllll}
\toprule
Atom & $n$ & $l$ & Shell type & Primitive  & Exponent & Contraction Coeff. \\
\midrule
Be & 1 & 0 & s & 1 & 1.886943 & 0.271926 \\
Be & 1 & 0 & s & 2 & 617.363676 & 0.002630 \\
Be & 1 & 0 & s & 3 & 5.994324 & 0.251085 \\
Be & 1 & 0 & s & 4 & 5.359786 & -0.025928 \\
Be & 1 & 0 & s & 5 & 93.888423 & 0.019227 \\
Be & 1 & 0 & s & 6 & 0.541312 & 0.048853 \\
Be & 1 & 0 & s & 7 & 21.626925 & 0.084649 \\
Be & 1 & 0 & s & 8 & 0.131021 & -0.019203 \\
Be & 2 & 0 & s & 1 & 1.756364 & -0.013903 \\
Be & 2 & 0 & s & 2 & 0.001614 & 0.000120 \\
Be & 2 & 0 & s & 3 & 0.011854 & -0.001460 \\
Be & 2 & 0 & s & 4 & 0.016135 & 0.004015 \\
Be & 2 & 0 & s & 5 & 0.002236 & -0.000107 \\
Be & 2 & 0 & s & 6 & 0.175458 & 0.029201 \\
Be & 2 & 0 & s & 7 & 0.062406 & 0.075316 \\
Be & 2 & 0 & s & 8 & 0.163880 & 0.057947 \\
Be & 2 & 1 & p & 1 & 1.242689 & 0.003834 \\
Be & 2 & 1 & p & 2 & 0.017768 & 0.000810 \\
Be & 2 & 1 & p & 3 & 0.274942 & 0.021229 \\
Be & 2 & 1 & p & 4 & 0.002541 & 0.001483 \\
Be & 2 & 1 & p & 5 & 0.101042 & 0.004445 \\
Be & 2 & 1 & p & 6 & 0.002724 & -0.000465 \\
Be & 2 & 1 & p & 7 & 0.004232 & -0.001254 \\
Be & 2 & 1 & p & 8 & 0.093609 & 0.015460 \\
\bottomrule
\end{tabular}

\begin{tabular}{lllllll}
\toprule
Atom & $n$ & $l$ & Shell type & Primitive  & Exponent & Contraction Coeff. \\
\midrule
B & 1 & 0 & s & 1 & 979.220174 & 0.007804 \\
B & 1 & 0 & s & 2 & 150.870702 & 0.055934 \\
B & 1 & 0 & s & 3 & 35.319511 & 0.242085 \\
B & 1 & 0 & s & 4 & 10.046125 & 0.635492 \\
B & 1 & 0 & s & 5 & 2.867956 & 0.568970 \\
B & 1 & 0 & s & 6 & 3.999577 & 0.201235 \\
B & 1 & 0 & s & 7 & 0.123791 & 0.628966 \\
B & 1 & 0 & s & 8 & 0.482255 & 0.430230 \\
B & 2 & 0 & s & 1 & 17.786375 & -0.009850 \\
B & 2 & 0 & s & 2 & 4.492607 & -0.067078 \\
B & 2 & 0 & s & 3 & 1.669305 & -0.099985 \\
B & 2 & 0 & s & 4 & 0.413626 & 0.411199 \\
B & 2 & 0 & s & 5 & 0.102139 & 0.532458 \\
B & 2 & 0 & s & 6 & 0.081821 & 0.001312 \\
B & 2 & 0 & s & 7 & 0.180878 & 0.409102 \\
B & 2 & 0 & s & 8 & 0.475335 & 0.039965 \\
B & 2 & 1 & p & 1 & 8.336493 & 0.029957 \\
B & 2 & 1 & p & 2 & 8.540898 & 0.005458 \\
B & 2 & 1 & p & 3 & 0.637511 & 0.113611 \\
B & 2 & 1 & p & 4 & 0.150859 & 0.796235 \\
B & 2 & 1 & p & 5 & 0.357337 & 0.105503 \\
B & 2 & 1 & p & 6 & 0.049024 & 0.192300 \\
B & 2 & 1 & p & 7 & 1.775569 & 0.198508 \\
B & 2 & 1 & p & 8 & 0.480516 & 0.453278 \\
\bottomrule
\end{tabular}

\begin{tabular}{lllllll}
\toprule
Atom & $n$ & $l$ & Shell type & Primitive  & Exponent & Contraction Coeff. \\
\midrule
C & 1 & 0 & s & 1 & 1935.542100 & 0.009644 \\
C & 1 & 0 & s & 2 & 308.793054 & 0.067021 \\
C & 1 & 0 & s & 3 & 73.034477 & 0.307667 \\
C & 1 & 0 & s & 4 & 20.884887 & 0.942475 \\
C & 1 & 0 & s & 5 & 6.695430 & 1.553847 \\
C & 1 & 0 & s & 6 & 0.051813 & 0.000818 \\
C & 1 & 0 & s & 7 & 1.947105 & 0.719484 \\
C & 1 & 0 & s & 8 & 0.238017 & 0.524677 \\
C & 2 & 0 & s & 1 & 35.187413 & 0.000411 \\
C & 2 & 0 & s & 2 & 3.280029 & -0.402695 \\
C & 2 & 0 & s & 3 & 0.056207 & 0.002141 \\
C & 2 & 0 & s & 4 & 1.108019 & 0.128010 \\
C & 2 & 0 & s & 5 & 0.112037 & 0.592702 \\
C & 2 & 0 & s & 6 & 0.139744 & 0.051578 \\
C & 2 & 0 & s & 7 & 1.041333 & 0.540821 \\
C & 2 & 0 & s & 8 & 0.286833 & 2.125882 \\
C & 2 & 1 & p & 1 & 1.512565 & 0.004466 \\
C & 2 & 1 & p & 2 & 19.246413 & 0.038565 \\
C & 2 & 1 & p & 3 & 4.371958 & 0.230821 \\
C & 2 & 1 & p & 4 & 0.122502 & 0.941409 \\
C & 2 & 1 & p & 5 & 0.699632 & 0.287368 \\
C & 2 & 1 & p & 6 & 0.018183 & 0.036358 \\
C & 2 & 1 & p & 7 & 1.309885 & 0.717066 \\
C & 2 & 1 & p & 8 & 0.363818 & 1.355376 \\
\bottomrule
\end{tabular}

\begin{tabular}{lllllll}
\toprule
Atom & $n$ & $l$ & Shell type & Primitive  & Exponent & Contraction Coeff. \\
\midrule
N & 1 & 0 & s & 1 & 1878.683780 & 0.005811 \\
N & 1 & 0 & s & 2 & 297.849168 & 0.039956 \\
N & 1 & 0 & s & 3 & 71.805053 & 0.168733 \\
N & 1 & 0 & s & 4 & 21.297556 & 0.446982 \\
N & 1 & 0 & s & 5 & 7.122089 & 0.557259 \\
N & 1 & 0 & s & 6 & 2.701200 & 0.155555 \\
N & 1 & 0 & s & 7 & 0.652792 & 0.027782 \\
N & 1 & 0 & s & 8 & 0.185818 & 0.003580 \\
N & 2 & 0 & s & 1 & 0.023102 & -0.005820 \\
N & 2 & 0 & s & 2 & 9.020485 & -0.118642 \\
N & 2 & 0 & s & 3 & 2.788727 & -0.107685 \\
N & 2 & 0 & s & 4 & 0.316192 & 0.385519 \\
N & 2 & 0 & s & 5 & 0.382367 & 0.655418 \\
N & 2 & 0 & s & 6 & 0.135416 & 0.368596 \\
N & 2 & 0 & s & 7 & 0.506011 & -0.002559 \\
N & 2 & 0 & s & 8 & 0.949504 & 0.642842 \\
N & 2 & 1 & p & 1 & 19.748264 & 0.055500 \\
N & 2 & 1 & p & 2 & 8.129135 & 0.051984 \\
N & 2 & 1 & p & 3 & 3.491695 & 0.401659 \\
N & 2 & 1 & p & 4 & 0.593652 & 0.282318 \\
N & 2 & 1 & p & 5 & 0.483376 & 0.412560 \\
N & 2 & 1 & p & 6 & 0.076239 & 0.206603 \\
N & 2 & 1 & p & 7 & 1.148110 & 0.889266 \\
N & 2 & 1 & p & 8 & 0.244241 & 0.964370 \\
\bottomrule
\end{tabular}

\begin{tabular}{lllllll}
\toprule
Atom & $n$ & $l$ & Shell type & Primitive  & Exponent & Contraction Coeff. \\
\midrule
O & 1 & 0 & s & 1 & 2050.376470 & 0.007187 \\
O & 1 & 0 & s & 2 & 336.608199 & 0.046225 \\
O & 1 & 0 & s & 3 & 84.270010 & 0.183324 \\
O & 1 & 0 & s & 4 & 26.072692 & 0.443769 \\
O & 1 & 0 & s & 5 & 8.668950 & 0.449690 \\
O & 1 & 0 & s & 6 & 3.126057 & 0.120443 \\
O & 1 & 0 & s & 7 & 0.289788 & 0.662240 \\
O & 1 & 0 & s & 8 & 0.960094 & 0.720026 \\
O & 2 & 0 & s & 1 & 16.212248 & -0.029172 \\
O & 2 & 0 & s & 2 & 10.735611 & -0.131827 \\
O & 2 & 0 & s & 3 & 5.153579 & -0.061122 \\
O & 2 & 0 & s & 4 & 1.328291 & 0.171560 \\
O & 2 & 0 & s & 5 & 0.287162 & 0.890111 \\
O & 2 & 0 & s & 6 & 0.283139 & 0.284953 \\
O & 2 & 0 & s & 7 & 0.476486 & 0.000880 \\
O & 2 & 0 & s & 8 & 0.901631 & 1.173436 \\
O & 2 & 1 & p & 1 & 47.846486 & 0.018417 \\
O & 2 & 1 & p & 2 & 11.985199 & 0.100991 \\
O & 2 & 1 & p & 3 & 3.969366 & 0.340489 \\
O & 2 & 1 & p & 4 & 0.144380 & 0.340147 \\
O & 2 & 1 & p & 5 & 0.678441 & 0.211572 \\
O & 2 & 1 & p & 6 & 0.301000 & 0.145453 \\
O & 2 & 1 & p & 7 & 0.455334 & 0.643587 \\
O & 2 & 1 & p & 8 & 1.423602 & 0.699513 \\
\bottomrule
\end{tabular}

\begin{tabular}{lllllll}
\toprule
Atom & $n$ & $l$ & Shell type & Primitive  & Exponent & Contraction Coeff. \\
\midrule
F & 1 & 0 & s & 1 & 3386.120700 & 0.006014 \\
F & 1 & 0 & s & 2 & 545.486408 & 0.040340 \\
F & 1 & 0 & s & 3 & 134.173685 & 0.168956 \\
F & 1 & 0 & s & 4 & 40.566453 & 0.463858 \\
F & 1 & 0 & s & 5 & 13.771119 & 0.628601 \\
F & 1 & 0 & s & 6 & 4.740271 & 0.263775 \\
F & 1 & 0 & s & 7 & 0.951284 & 0.232965 \\
F & 1 & 0 & s & 8 & 0.287062 & 0.115580 \\
F & 2 & 0 & s & 1 & 5.839263 & -0.005497 \\
F & 2 & 0 & s & 2 & 14.022904 & -0.134460 \\
F & 2 & 0 & s & 3 & 5.859306 & -0.030912 \\
F & 2 & 0 & s & 4 & 1.678833 & 0.447212 \\
F & 2 & 0 & s & 5 & 0.723841 & 0.179907 \\
F & 2 & 0 & s & 6 & 0.223872 & 0.292970 \\
F & 2 & 0 & s & 7 & 0.471435 & 0.530757 \\
F & 2 & 0 & s & 8 & 0.930295 & 0.497480 \\
F & 2 & 1 & p & 1 & 72.861610 & 0.011447 \\
F & 2 & 1 & p & 2 & 16.074749 & 0.083357 \\
F & 2 & 1 & p & 3 & 2.273626 & 0.351930 \\
F & 2 & 1 & p & 4 & 5.111718 & 0.249185 \\
F & 2 & 1 & p & 5 & 0.070307 & 0.049031 \\
F & 2 & 1 & p & 6 & 0.010721 & 0.000324 \\
F & 2 & 1 & p & 7 & 0.307624 & 0.534361 \\
F & 2 & 1 & p & 8 & 0.952243 & 0.635310 \\
\bottomrule
\end{tabular}

\subsection{$k = 9$}
\begin{tabular}{lllllll}
\toprule
Atom & $n$ & $l$ & Shell type & Primitive  & Exponent & Contraction Coeff. \\
\midrule
H & 1 & 0 & s & 1 & 66.397548 & 0.008338 \\
H & 1 & 0 & s & 2 & 10.149185 & 0.061116 \\
H & 1 & 0 & s & 3 & 0.249153 & 0.235949 \\
H & 1 & 0 & s & 4 & 0.794520 & 0.448464 \\
H & 1 & 0 & s & 5 & 0.225253 & 1.107420 \\
H & 1 & 0 & s & 6 & 0.806379 & 0.235587 \\
H & 1 & 0 & s & 7 & 0.087305 & 0.867243 \\
H & 1 & 0 & s & 8 & 2.483953 & 0.259263 \\
H & 1 & 0 & s & 9 & 0.413336 & 0.419480 \\
\bottomrule
\end{tabular}

\begin{tabular}{lllllll}
\toprule
Atom & $n$ & $l$ & Shell type & Primitive  & Exponent & Contraction Coeff. \\
\midrule
Li & 1 & 0 & s & 1 & 0.011363 & 0.256536 \\
Li & 1 & 0 & s & 2 & 0.641313 & -1.525312 \\
Li & 1 & 0 & s & 3 & 15.993439 & 0.057106 \\
Li & 1 & 0 & s & 4 & 0.029647 & 7.404341 \\
Li & 1 & 0 & s & 5 & 0.029612 & 0.393366 \\
Li & 1 & 0 & s & 6 & 0.028809 & 0.448355 \\
Li & 1 & 0 & s & 7 & 0.072790 & 9.374863 \\
Li & 1 & 0 & s & 8 & 175.847312 & 0.005491 \\
Li & 1 & 0 & s & 9 & 0.011996 & 0.216768 \\
Li & 2 & 0 & s & 1 & 1.510707 & 12.102813 \\
Li & 2 & 0 & s & 2 & 0.481769 & 4.117196 \\
Li & 2 & 0 & s & 3 & 0.029662 & 0.469861 \\
Li & 2 & 0 & s & 4 & 0.682128 & 2.950108 \\
Li & 2 & 0 & s & 5 & 0.087231 & 0.561270 \\
Li & 2 & 0 & s & 6 & 4.423732 & 8.145436 \\
Li & 2 & 0 & s & 7 & 59.935483 & 0.750028 \\
Li & 2 & 0 & s & 8 & 370.549484 & 0.113669 \\
Li & 2 & 0 & s & 9 & 14.695858 & 3.094047 \\
Li & 2 & 1 & p & 1 & 14.309611 & 0.325934 \\
Li & 2 & 1 & p & 2 & 0.068540 & -0.027000 \\
Li & 2 & 1 & p & 3 & 7.823817 & 0.009849 \\
Li & 2 & 1 & p & 4 & 3.627798 & -0.089347 \\
Li & 2 & 1 & p & 5 & 0.142801 & -0.042128 \\
Li & 2 & 1 & p & 6 & 0.088389 & 0.045283 \\
Li & 2 & 1 & p & 7 & 4.430907 & 1.647322 \\
Li & 2 & 1 & p & 8 & 0.640898 & 0.207518 \\
Li & 2 & 1 & p & 9 & 1.714672 & 1.547356 \\
\bottomrule
\end{tabular}

\begin{tabular}{lllllll}
\toprule
Atom & $n$ & $l$ & Shell type & Primitive  & Exponent & Contraction Coeff. \\
\midrule
Be & 1 & 0 & s & 1 & 611.309439 & 0.008527 \\
Be & 1 & 0 & s & 2 & 0.624954 & 0.161003 \\
Be & 1 & 0 & s & 3 & 94.986364 & 0.060176 \\
Be & 1 & 0 & s & 4 & 3.229133 & 0.120667 \\
Be & 1 & 0 & s & 5 & 1.900250 & 0.798600 \\
Be & 1 & 0 & s & 6 & 0.090549 & 0.015221 \\
Be & 1 & 0 & s & 7 & 22.357310 & 0.259372 \\
Be & 1 & 0 & s & 8 & 0.507804 & 0.010308 \\
Be & 1 & 0 & s & 9 & 6.444248 & 0.663903 \\
Be & 2 & 0 & s & 1 & 4.465355 & -0.059653 \\
Be & 2 & 0 & s & 2 & 0.056202 & -0.167939 \\
Be & 2 & 0 & s & 3 & 1.331428 & -0.245457 \\
Be & 2 & 0 & s & 4 & 0.068524 & 1.019335 \\
Be & 2 & 0 & s & 5 & 0.198545 & 1.068622 \\
Be & 2 & 0 & s & 6 & 0.098364 & 0.667237 \\
Be & 2 & 0 & s & 7 & 0.240874 & 0.006101 \\
Be & 2 & 0 & s & 8 & 0.034695 & 0.274937 \\
Be & 2 & 0 & s & 9 & 0.039079 & 0.008474 \\
Be & 2 & 1 & p & 1 & 0.121105 & 0.046869 \\
Be & 2 & 1 & p & 2 & 0.035296 & 0.053536 \\
Be & 2 & 1 & p & 3 & 2.056997 & 0.060357 \\
Be & 2 & 1 & p & 4 & 0.021393 & 0.014362 \\
Be & 2 & 1 & p & 5 & 0.116840 & 0.069334 \\
Be & 2 & 1 & p & 6 & 0.006096 & -0.004061 \\
Be & 2 & 1 & p & 7 & 0.844951 & 0.048824 \\
Be & 2 & 1 & p & 8 & 0.109946 & 0.573207 \\
Be & 2 & 1 & p & 9 & 0.317325 & 0.544455 \\
\bottomrule
\end{tabular}

\begin{tabular}{lllllll}
\toprule
Atom & $n$ & $l$ & Shell type & Primitive  & Exponent & Contraction Coeff. \\
\midrule
B & 1 & 0 & s & 1 & 775.940317 & 0.007908 \\
B & 1 & 0 & s & 2 & 124.730600 & 0.052383 \\
B & 1 & 0 & s & 3 & 30.452505 & 0.211618 \\
B & 1 & 0 & s & 4 & 9.027329 & 0.519359 \\
B & 1 & 0 & s & 5 & 2.881014 & 0.484591 \\
B & 1 & 0 & s & 6 & 0.307805 & 0.162751 \\
B & 1 & 0 & s & 7 & 0.115726 & 0.803441 \\
B & 1 & 0 & s & 8 & 1.023049 & 0.036758 \\
B & 1 & 0 & s & 9 & 0.406167 & 0.446557 \\
B & 2 & 0 & s & 1 & 3.142152 & -0.002691 \\
B & 2 & 0 & s & 2 & 4.000426 & -0.096008 \\
B & 2 & 0 & s & 3 & 2.000779 & -0.082904 \\
B & 2 & 0 & s & 4 & 0.415419 & 0.112786 \\
B & 2 & 0 & s & 5 & 0.235630 & 0.064254 \\
B & 2 & 0 & s & 6 & 0.084380 & 0.177212 \\
B & 2 & 0 & s & 7 & 0.119384 & 0.846280 \\
B & 2 & 0 & s & 8 & 0.355154 & 0.634490 \\
B & 2 & 0 & s & 9 & 0.166601 & 0.111560 \\
B & 2 & 1 & p & 1 & 42.545682 & 0.001546 \\
B & 2 & 1 & p & 2 & 6.633165 & 0.028044 \\
B & 2 & 1 & p & 3 & 2.585709 & 0.041071 \\
B & 2 & 1 & p & 4 & 0.076519 & 0.161462 \\
B & 2 & 1 & p & 5 & 0.342688 & 0.000894 \\
B & 2 & 1 & p & 6 & 0.093505 & 0.211677 \\
B & 2 & 1 & p & 7 & 0.943828 & 0.241549 \\
B & 2 & 1 & p & 8 & 0.547669 & 0.032054 \\
B & 2 & 1 & p & 9 & 0.270370 & 0.548872 \\
\bottomrule
\end{tabular}

\begin{tabular}{lllllll}
\toprule
Atom & $n$ & $l$ & Shell type & Primitive  & Exponent & Contraction Coeff. \\
\midrule
C & 1 & 0 & s & 1 & 1066.069100 & 0.011969 \\
C & 1 & 0 & s & 2 & 166.093230 & 0.083659 \\
C & 1 & 0 & s & 3 & 2.052613 & 0.050999 \\
C & 1 & 0 & s & 4 & 11.279091 & 0.824023 \\
C & 1 & 0 & s & 5 & 39.120476 & 0.347334 \\
C & 1 & 0 & s & 6 & 3.595335 & 0.319724 \\
C & 1 & 0 & s & 7 & 1.807489 & 0.036742 \\
C & 1 & 0 & s & 8 & 0.174701 & -0.014630 \\
C & 1 & 0 & s & 9 & 3.784566 & 0.386289 \\
C & 2 & 0 & s & 1 & 5.756476 & -0.068825 \\
C & 2 & 0 & s & 2 & 4.931187 & -0.064823 \\
C & 2 & 0 & s & 3 & 0.007215 & -0.031266 \\
C & 2 & 0 & s & 4 & 0.139940 & 0.761736 \\
C & 2 & 0 & s & 5 & 1.112330 & -0.015306 \\
C & 2 & 0 & s & 6 & 0.005518 & 0.002680 \\
C & 2 & 0 & s & 7 & 0.461065 & 0.982403 \\
C & 2 & 0 & s & 8 & 0.191100 & 0.012181 \\
C & 2 & 0 & s & 9 & 0.945243 & 0.004789 \\
C & 2 & 1 & p & 1 & 13.021071 & -0.004323 \\
C & 2 & 1 & p & 2 & 15.767797 & 0.064865 \\
C & 2 & 1 & p & 3 & 3.993967 & 0.279747 \\
C & 2 & 1 & p & 4 & 0.393416 & 0.796770 \\
C & 2 & 1 & p & 5 & 0.209343 & 0.589933 \\
C & 2 & 1 & p & 6 & 0.055758 & 0.189750 \\
C & 2 & 1 & p & 7 & 0.138433 & 0.825958 \\
C & 2 & 1 & p & 8 & 1.360687 & 0.782205 \\
C & 2 & 1 & p & 9 & 0.572093 & 0.814409 \\
\bottomrule
\end{tabular}

\begin{tabular}{lllllll}
\toprule
Atom & $n$ & $l$ & Shell type & Primitive  & Exponent & Contraction Coeff. \\
\midrule
N & 1 & 0 & s & 1 & 1736.915230 & 0.007053 \\
N & 1 & 0 & s & 2 & 284.402618 & 0.045798 \\
N & 1 & 0 & s & 3 & 71.344937 & 0.183229 \\
N & 1 & 0 & s & 4 & 22.061050 & 0.462890 \\
N & 1 & 0 & s & 5 & 8.020781 & 0.443273 \\
N & 1 & 0 & s & 6 & 3.163662 & 0.241639 \\
N & 1 & 0 & s & 7 & 0.199798 & 0.764996 \\
N & 1 & 0 & s & 8 & 0.714970 & 0.847771 \\
N & 1 & 0 & s & 9 & 0.421315 & 0.222484 \\
N & 2 & 0 & s & 1 & 21.282954 & -0.003225 \\
N & 2 & 0 & s & 2 & 7.357421 & -0.301210 \\
N & 2 & 0 & s & 3 & 1.511586 & 0.002142 \\
N & 2 & 0 & s & 4 & 1.011041 & 0.114304 \\
N & 2 & 0 & s & 5 & 0.230272 & 0.685415 \\
N & 2 & 0 & s & 6 & 0.348368 & 0.215028 \\
N & 2 & 0 & s & 7 & 0.694240 & 1.018700 \\
N & 2 & 0 & s & 8 & 0.636048 & 0.835676 \\
N & 2 & 0 & s & 9 & 0.185425 & 0.948374 \\
N & 2 & 1 & p & 1 & 23.002377 & 0.017961 \\
N & 2 & 1 & p & 2 & 25.333866 & 0.035776 \\
N & 2 & 1 & p & 3 & 5.421593 & 0.332118 \\
N & 2 & 1 & p & 4 & 0.708259 & 0.627992 \\
N & 2 & 1 & p & 5 & 0.306702 & 0.331424 \\
N & 2 & 1 & p & 6 & 0.055830 & 0.010769 \\
N & 2 & 1 & p & 7 & 0.140756 & 0.717807 \\
N & 2 & 1 & p & 8 & 1.622379 & 0.948942 \\
N & 2 & 1 & p & 9 & 0.423722 & 0.962336 \\
\bottomrule
\end{tabular}%

\subsection{$k = 10$}
\begin{tabular}{lllllll}
\toprule
Atom & $n$ & $l$ & Shell type & Primitive  & Exponent & Contraction Coeff. \\
\midrule
H & 1 & 0 & s & 1 & 0.023408 & 0.018613 \\
H & 1 & 0 & s & 2 & 47.451794 & 0.011323 \\
H & 1 & 0 & s & 3 & 8.632545 & 0.060140 \\
H & 1 & 0 & s & 4 & 0.266933 & 1.278035 \\
H & 1 & 0 & s & 5 & 0.080237 & 0.026456 \\
H & 1 & 0 & s & 6 & 0.070984 & 0.248660 \\
H & 1 & 0 & s & 7 & 0.495950 & 0.089100 \\
H & 1 & 0 & s & 8 & 0.787069 & 0.674741 \\
H & 1 & 0 & s & 9 & 0.114002 & 0.727133 \\
H & 1 & 0 & s & 10 & 2.425374 & 0.230988 \\
\bottomrule
\end{tabular}%

\subsection{$k = 11$}
%

\newpage
\section{Comparison of basis set energies}

\begin{table*}[h]
\caption{Comparison of FCI energies (in Ha) for full active space $E(FCI)$ versus active space of 10 spin orbitals $E(asFCI)$. We employed the cc-pVDZ basis sets for these calculations. }\label{tab:Ecor_elements}
\setlength{\tabcolsep}{12pt} 
%

\end{table*}


\begin{table*}[h!]
        \centering
    \caption{Comparison of basis set energies (Ha) for \ce{H}. }\label{tab:En_H}
    \setlength{\tabcolsep}{10pt}
    %
    \end{table*}

\begin{table*}[h!]
    \caption{Comparison of basis set energies (Ha) for Li. The Hylleraas result is not quite FCI, but is placed under that column for convenience. }\label{tab:En_Li}
    \setlength{\tabcolsep}{10pt}
    %

    
    \end{table*}

\begin{table*}[h!]
        \centering
    \caption{Comparison of basis set energies for Be. }\label{tab:En_Be}
    \setlength{\tabcolsep}{10pt}
    %

    \end{table*}

\begin{table*}[h!]
    \caption{Comparison of basis set energies (Ha) for B. }\label{tab:En_B}
    \setlength{\tabcolsep}{10pt}
    %

    \end{table*}

\begin{table*}[h!]
    \caption{Comparison of basis set energies (Ha) for C. }\label{tab:En_C}
    \setlength{\tabcolsep}{10pt}
    %

    \end{table*}

\begin{table*}[h!]

        \centering
    \caption{Comparison of basis set energies (Ha) for N. }\label{tab:En_N}
    \setlength{\tabcolsep}{10pt}
    %

    \end{table*}

\begin{table*}[h!]
    \caption{Comparison of basis set energies (Ha) for O. }\label{tab:En_O}
    \setlength{\tabcolsep}{10pt}
    %

    \end{table*}

\begin{table*}[h!]
    \caption{Comparison of basis set energies (Ha) for F. }\label{tab:En_F}
    \setlength{\tabcolsep}{10pt}
    %

    \end{table*} 

\end{widetext}

\clearpage 

\begin{widetext}

\section{Comparison of quantum resources across basis sets} \label{app:res}

\begin{table*}[h!]
\centering
\caption{Illustrative resource requirements for a VQE-UCCSD calculation of the ground state energy of the Li atom. We choose STO-6G and MSTO-11G as representative minimal basis sets. The FCI energies are classically computed reference values rather than the outputs of the quantum algorithm. } 
\setlength{\tabcolsep}{12pt} 
%
\label{tab:VQEresources}
\end{table*}

\begin{table*}[h!]
\centering
\caption{Qubit counts and CX gate estimates for different basis sets in a QPE-CASCI calculation of the ground state energy of the Li atom. The number of QPE clock register qubits is set to 8. JWT stands for Jordan-Wigner transformation. }
\setlength{\tabcolsep}{12pt}
%
\label{tab:QPEresources}
\end{table*}

\begin{table*}[h!]
\centering
\caption{Qubit counts and CX gate estimates for HHL-LCCSD calculations of the ground state energy
of the Li atom. The number of QPE clock register
qubits is set to 8. The LCCSD energies are obtained
by solving the linear systems classically, and the FCI energies are provided
as reference values. }
\setlength{\tabcolsep}{12pt}
%
 
\label{tab:HHLresources}
\end{table*} 

\begin{table}[]
    \centering
    \caption{Symmetries and basis combinations that we studied (for di- and triatomic cases) of the molecules considered. }
    \label{tab:molsyms}
    %

\end{table}

\end{widetext}

\end{document}